\documentclass[reprint,showpacs,prd,preprintnumbers,amssymb,nofootinbib,superscriptaddress,aps]{revtex4-1}
\usepackage{graphicx,floatflt,amssymb,rotate,epsfig}
\usepackage[utf8]{inputenc}
\usepackage[inline]{enumitem}
\usepackage{mathrsfs}
\usepackage{amssymb}
\usepackage{amsmath}
\usepackage{color}
\usepackage{graphicx}
\usepackage{subfig}
\usepackage{multirow}
\usepackage{mathtools}
\usepackage{float}
\usepackage{pstricks}
\usepackage[toc,page]{appendix}
\usepackage{pifont}
\usepackage{braket}
\usepackage{bbold}
\usepackage{hyperref}

\usepackage{relsize}
\def\Babar{{\mbox{\slshape B\kern-0.1em{\smaller A}\kern-0.1em B\kern-0.1em{\smaller A\kern-0.2em R}}}}

\newcommand{\ba}{\begin{array}}
	\newcommand{\ea}{\end{array}}
\def\beq{\begin{equation}}
\def\eeq{\end{equation}}
\def\bea{\begin{eqnarray}}
\def\eea{\end{eqnarray}}
\def\nn{\nonumber}

\def\roughly#1{\mathrel{\raise.3ex\hbox
		{$#1$\kern-.75em\lower1ex\hbox{$\sim$}}}}

\def\sla#1{\raise.15ex\hbox{$/$}\kern-.57em #1}% Feynman slash

\def\bd{B_d^0}

\def\order{\lower 1.8ex \hbox{\LARGE\~{}}}

\def\bd0tau{B\to D \tau\nu_{\tau}}

\def\be {\begin{equation}}
\def\ee {\end{equation}}

\usepackage[normalem]{ulem}

\definecolor{darkgreen}{cmyk}{1,0,1,0.4}

\definecolor{pink}{cmyk}{0.4,1,0.3,0}

\def\com2#1{\textcolor{red}{\it{#1}}}
\begin{document}
	
	%opening
	\title{Exhaustive Model Selection in $b \to s \ell \ell$ Decays: \\
		Pitting Cross-Validation against AIC$_c$}
	
	\author{Srimoy Bhattacharya}
	\email{bhattacharyasrimoy@gmail.com}
	\affiliation{The Institute of Mathematical Sciences, C.I.T Campus, Taramani, Chennai 600 113, India}

	\author{Aritra Biswas}
	\email{iluvnpur@gmail.com}
	\affiliation{Indian Institute of Technology, North Guwahati, Guwahati 781039, Assam, India }
	
	\author{Soumitra Nandi}
	\email{soumitra.nandi@iitg.ernet.in}
	\affiliation{Indian Institute of Technology, North Guwahati, Guwahati 781039, Assam, India }
	
	\author{Sunando Kumar Patra}
	\email{sunando.patra@gmail.com}
	\affiliation{Indian Institute of Technology, North Guwahati, Guwahati 781039, Assam, India }

	\begin{abstract}  
		In the light of recent data, we study the new physics effects in the exclusive $b \to s \ell^+\ell^-$ decays from a model independent perspective. Different combinations of the dimension six effective operators along with their respective Wilson coefficients are chosen for the analysis. To find out the operator or sets of operators that can best explain the available data in this channel, we simultaneously apply popular model selection tools like cross-validation and the information theoretic approach like Akaike Information Criterion (AIC). There are one, two, and three-operator scenarios which survive the test and a left-handed quark current with vector muon coupling is common among them. This is also the only surviving one-operator scenario. Best-fit values and correlations of the new Wilson coefficients are supplied for all the selected scenarios. We find that the angular observables play the dominant role in the model selection procedure. We also note that while a left-handed quark current with axial-vector muon coupling is the only one-operator scenario able to explain the ratios $R_{K^{(*)}}$ ($R_{K^*}$ for $q^2\in [ 0.045, 1.1] {\rm GeV}^2$ in particular), there are also a couple of two operator scenarios that can simultaneously explain the measured $R_{K^{(*)}}$.
	\end{abstract}   
	
	\maketitle
	
%%%%%%%%%%%%%%%%%%%%%%%%%%%%%%%%%%%%%%%%%%%%%%%%%	
\section{Introduction}\label{sec:intro}
%%%%%%%%%%%%%%%%%%%%%%%%%%%%%%%%%%%%%%%%%%%%%%%%%

Decays involving $b \to s \ell\ell$ transitions are suppressed in the standard model (SM). These decay modes are potentially sensitive to new physics effects. Whether the contributions appear at the tree or the loop level depends on the type of the new physics (NP). A lot of attention, both experimental and theoretical, have been given to $B\to K^{(*)}\mu^+\mu^-$ decays in the last couple of years. There are several angular observables associated with these decays, which are potentially sensitive to the NP effects and are measured by LHCb \cite{Aaij:2015oid,Aaboud:2018krd}. A couple of them have shown discrepancies with their respective SM predictions \cite{DescotesGenon:2012zf,Descotes-Genon:2013vna,Horgan:2013pva,Straub:2015ica}. However, these angular observables are not free from hadronic uncertainties and it is fairly possible that the observed discrepancies are due to poorly known hadronic effects, e.g., see \cite{Ciuchini:2015qxb} for details. Furthermore, these modes offer theoretically clean observables like 
\begin{equation}
 R_H = \frac{\int_{\it q_{min}^2}^{\it q_{max}^2} \frac{d\Gamma(B\to H\mu^+\mu^-)}{dq^2}}{\int_{\it q_{min}^2}^{\it q_{max}^2} \frac{d\Gamma(B\to H e^+ e^-)}{dq^2}}
\end{equation}
where $H$ is either $K$ or $K^*$ meson and $q^2$ is the dilepton squared mass. These ratios are useful to test lepton flavor universality violation (LFUV) and within appropriately chosen ranges of $q^2$, these observables can be predicted very precisely in the SM; see \cite{Hiller:2003js,Bordone:2016gaq} for details. The SM predictions are, respectively, $R(K) = 1.0004 (8)$, and  
\begin{equation}
 R_{K^*} = 
 \begin{cases}
   0.920 \pm 0.007,\ \ \text{$q^2 \in [0.045,1.1]$ {\rm {GeV}$^2$}},\\
   0.996 \pm 0.002,\ \ \text{$q^2 \in [1.1,6]$ {\rm {GeV}$^2$}}\,.
 \end{cases}
\end{equation}
The LHCb collaboration has measured~\cite{Aaij:2017vbb,Aaij:2019wad}
\begin{equation}
 R_K = 0.846^{+0.060 \> + 0.016}_{-0.054 \> -0.014}, \  \text{$q^2 \in [1.1,6]$ {\it {\rm GeV}$^2$}},
\end{equation}
and 
\begin{equation}
 R_{K^*} = 
 \begin{cases}
   0.660^{+0.110}_{- 0.070} \pm 0.024,\  \text{$q^2 \in [0.045,1.1]$ {\it {\rm GeV}$^2$}},\\
   0.685^{+0.113}_{- 0.069} \pm 0.047,\  \text{$q^2 \in [1.1,6]$ {\it {\rm GeV}$^2$}}.
 \end{cases}
\end{equation}
We will use the notation $R^{Low}_{K^*}$ and $R^{Central}_{K^*}$ from now on to represent $R_{K^*}$ for values of $q^2$ in $[0.045,1.1]{\rm GeV}^2$ and $[1.1,6]{\rm GeV}^2$, respectively. Very recently Belle has measured the observables $R_{K^{(*)}}$ and the measured values are given by \cite{Abdesselam:2019wac}
\begin{equation}
 R_{K^*} = 
 \begin{cases}
   0.52^{+0.36}_{- 0.26} \pm 0.05,\  \text{$q^2 \in [0.045,1.1]$ {\it {\rm GeV}$^2$}},\\
   0.96^{+0.45}_{- 0.29} \pm 0.11,\  \text{$q^2 \in [1.1,6]$ {\it {\rm GeV}$^2$}}.
 \end{cases}
\end{equation}
and \cite{Abdesselam:2019lab}
\begin{equation}
 R_K = 0.98^{+0.27}_{-0.23} \pm 0.06, \  \text{$q^2 \in [1 , 6]$ {\it {\rm GeV}$^2$}},
\end{equation}
These new measurements have larger uncertainties compared to those from LHCb, but the results are consistent with each other. Belle has also measured separate ratios like $R_{K^{*}}^0$ and $R_{K^{*}}^+$, but the associated uncertainties are quite large at the moment. On the whole, the deviation between data and SM predictions stand at the level of 2.5 to 3 $\sigma$. Future measurements of these ratios with enough statistical significance would have the potential to discover NP unambiguously. 

The observed discrepancies can be explained in various NP models. Different types of new physics interactions (like vector, scalar etc.) with different Lorentz structures may contribute to these decays and explain the data. A lot of work has already been done and it is a difficult task to quote all of them. We are more interested, in the present work, in a model independent analysis. There are a few related analyses available in the literature, which mainly focus on considering one or two operators at a time \cite{Aebischer:2019mlg,Alok:2019ufo,Capdevila:2017bsm,Arbey:2018ics,Descotes-Genon:2013wba,Altmannshofer:2013foa,Beaujean:2013soa,Hurth:2013ssa,Altmannshofer:2014rta,Altmannshofer:2008dz,Egede:2010zc,Das:2012kz,Hurth:2014vma,Ciuchini:2019usw,Bobeth:2011nj,Geng:2017svp,Ciuchini:2017mik,Hiller:2003js,Kowalska:2019ley}. 

%Capdevila:2018jhy

In this article, we have done a model independent analysis of the NP affecting the $b\to s \ell^+\ell^-$ decay modes. The operator basis is exactly the same as that given in Ref.~\cite{Altmannshofer:2008dz}. We have considered all possible combinations of these operators and categorized them as independent `model's (scenarios). There are several models capable of describing the observed data and one is thus confronted with the problem of model selection. 

In short, the problem of model selection is as follows: any model, used to represent certain observation, will almost never be exact; chances are, that some information will be lost due to the choice of that particular model  \cite{Box:1976, box2005statistics}. Choosing a simplistic model with too few parameters can involve making unrealistically simple assumptions and lead to high bias, poor prediction, and consequent missed opportunities for insight. While simplistic models are not flexible enough to describe the sample or the population well, a model with a larger number of parameters can fit the observed data very well. Does this make it a better model? With too many parameters, we face the possibility of just fitting the noise in the data and losing sight of the important trends. The most general problem in model selection is thus the optimization of the parameters required to explain certain observation \cite{Burnham, Geisser:1979}. The motivation: there must be a happy medium somewhere. 

The most generally applicable, powerful, and reliable method for model comparison (also computationally expensive) is `cross-validation'\cite{Andrae:2010gh}, which, in addition to testing the predictive power of the model, minimizes the bias and variance together by minimizing the mean-squared-error (MSE). On the other hand, penalized-likelihood information criteria, such as the Akaike Information Criterion (AIC) \cite{akaike}, and the Bayesian Information Criterion (BIC; more aptly named as Schwarz Information Criterion) are widely used for model selection. AIC estimates the relative amount of information lost by a given model: the less the information lost by a model, the higher the quality of that model. For a detailed discussion on various model selection procedures and their relative performance with respect to cross-validation, ref. \cite{arlot2010}. In our earlier publications \cite{Bhattacharya:2016zcw,Bhattacharya:2018kig}, we have used these criteria in the context of NP model selections in $b \to c\tau \nu_{\tau}$ decays. 

Very recently, in a Bayesian analysis of $b\to s\ell^+\ell^-$ decays \cite{Ciuchini:2019usw}, an information criterion has been used. They have shown the use of a criterion closely related to DIC (Deviance Information Criterion; the definition by Gelman et al \cite{2013arXiv1307.5928G}) and BPIC (Bayesian Predictive Information Criterion \cite{AndoBiomet}) for model selection, which is not only ideal for samples from a Markov Chain Monte Carlo but is also asymptotically equivalent \cite{claeskens_hjort_2008} to natural model-robust version of AIC.

AIC is easy to calculate in a frequentist analysis, which is not the case for Bayesian analyses. The main difference between that analysis and ours is that they created the model hierarchy by defining $\Delta IC = IC_{SM} - IC_{NP}$. As a result, the quality of a model is determined with respect to the SM, whereas in our case the best model is picked up first and the hierarchy is defined with respect to that. 
Still, similar to our findings, they have found that the case $\Delta C_9$ ($C^{NP}_{9,~\mu}$ in their paper) provides the optimal outcome for $B\to K^* \ell \ell$ transitions if we consider only the `Moments' data for the angular observables, in addition to the new LFUV data. 

In the present analysis, we use both AIC and cross-validation to pin down the best possible model(s), and find out how one can use both procedures in tandem to glean the most out of the data at hand.

The article is organized as follows: in section \ref{sec:backdrop}, we discuss the present experimental and theoretical status of the observables used in this analysis. Section \ref{sec:method} discusses the detailed methodology of the statistical analysis, as well as model selection. We present our results in section \ref{sec:res} and in section \ref{sec:summ} we summarize.

%%%%%%%%%%%%%%%%%%%%%%%%%%%%%%%%%%%%%%%%%%%%%%%%%	
\section{Backdrop}\label{sec:backdrop}
\subsection{Experimental}\label{sec:backexp}
%%%%%%%%%%%%%%%%%%%%%%%%%%%%%%%%%%%%%%%%%%%%%%%%%

We list the experimental results used in this analysis and the corresponding references below:

\begin{itemize}
	\item Binned data on the angular observables related to the $B^0\to K^{*0}\mu^+\mu^-$ decays have been obtained from refs.~\cite{Aaij:2015oid} (LHCb) and~\cite{Aaboud:2018krd} (ATLAS)\footnote{We refrain from using the very old (2012) CDF data available from the public note~\cite{cdf2012cdf}
		%  CDF Collaboration, “Precise Measurements of Exclusive b → sμ + μ − Decay Amplitudes Using the Full CDF Data Set”, CDF public note 10894.
		on the angular observables in the $B\to K^{(*)}\mu^+\mu^-$ decays.}.
	\item Binned data for the differential branching fraction for $B^0\to K^{*0}\mu^+\mu^-$ have been obtained from ref.~\cite{Aaij:2016flj} and that for $B^+\to K^{*+}\mu^+\mu^-$ from~\cite{Aaij:2014pli}. Both of these are LHCb references.
	\item Binned data on the angular observables for $B^+\to K^+\mu^+\mu^-$ ($A_{FB}$ and $F_H$) have been taken from ref.~\cite{Sirunyan:2018jll} (CMS).
	\item Binned data on the differential branching fraction for $B^+\to K^+\mu^+\mu^-$ and $B^0\to K^0\mu^+\mu^-$ reported by LHCb have been taken from ref.~\cite{Aaij:2014pli}.
	\item Binned data on the angular observables for $B_s\to\phi\mu^+\mu^-$ (LHCb) have been taken from ref.~\cite{Aaij:2015esa}.
	\item The lepton flavor universality violating (LFUV) observables $R_{K^{*}}$, both for the low and central bin, have been obtained from ref.~\cite{Aaij:2017vbb} (LHCb). We also include the recent measurements on these observables (for the same bins) from Belle~\cite{Abdesselam:2019wac}. The old $R_{K}$ data from LHCb has been taken from ref.~\cite{Aaij:2014ora}. The updated result on the same has also been included~\cite{Aaij:2019wad}.
	\item The experimental result for the branching ratio (BR) corresponding to $B_s\to\mu^+\mu^-$ has been taken from~\cite{hflav} which is the average of the measured values by CMS, ATLAS and LHCb. The value is given by 
	\begin{equation}
	 Br(B_s \to \mu\mu) = \left(3.1 \pm 0.6\right)\times 10^{-9}\,.
	\end{equation}
	The decay constant is taken from ref. \cite{Charles:2013aka,Aoki:2019cca}
	\begin{equation}
         f_{B_s} = 0.2284 \pm 0.0037 \rm~ GeV\,.%\ \ B_{B_s} = 1.327 \pm 0.034.
	\end{equation}

\end{itemize}  
All numerical uncertainties quoted in this analysis, unless otherwise specified, denotes the 1-$\sigma$ (68\% c.l.) range.

A few words regarding the data on the angular observables due to LHCb taken from ref.~\cite{Aaij:2015oid} is in order at this point. LHCb has provided the data corresponding to the angular observables in bins of $q^2$ ($q=p_{\mu^+}+p_{\mu^-}$, $p_{\mu}$ being the four-momentum of muon) by performing two separate analyses. The more commonly used dataset in the community is that due to the ``Method of Moments''%, where the number of data is high, owing to the $q^2$ bin size being small
. The angular observables in this case are determined by using a principal moment analysis of the angular distribution without carrying out any angular fit to the data~\cite{Beaujean:2015xea,Gratrex:2015hna}. These moments are continuous functions of $q^2$. The statistical uncertainties for these angular moments are estimated using a bootstrapping technique~\cite{efron1979} and confidence intervals are defined such that they include the $16^{th}$–$84^{th}$ percentiles of the bootstrap distribution of the observables. The other method termed the ``Maximum likelihood fit'' involves an unbinned maximum likelihood fit to the invariant mass $m(\mu^+\mu^-(K^*\to)K^+\pi^-)$ and the three decay angles $\cos\theta_l$, $\cos\theta_K$ and $\phi$ in each $q^2$ bin, where:
\begin{itemize}
	\item $\theta_l$ is the angle between the $\mu^+$ ($\mu^-$) and the direction opposite to that of the $B^0$ ($\bar{B}^0$) in the rest frame of the dimuon system,
	\item $\theta_K$ is the angle between the direction of the $K^+$ ($K^-$) and the $B^0$ ($\bar{B}^0$) in the rest frame of the $K^{*0}$ ($\bar{K}^{*0}$) where the $K^{*0}$ meson is reconstructed through the decay $K^{*0}\to K^+\pi^-$, and
	\item $\phi$ is the angle between the plane defined by the dimuon pair and the plane defined by the $K$ and the $\pi$ in the $B^0$ ($\bar{B}^0$) rest frame.
\end{itemize}
%In order to ensure correct coverage for the uncertainties of the angular observables, the Feldman-Cousins method~\cite{Feldman:1997qc} is used with nuisance parameters treated according to the plug-in method~\cite{Bodhisattva:2009uba}.
%To keep the precision of the data in each bin of the similar order as the ``Method of Moments'', the bin sizes are variable over the $q^2$ range and in general larger than the ``Method of Moments''.
The bin sizes corresponding to the maximum likelihood analysis are larger than those for the method of moments. This is done since there is a dearth of statistics, and an increase in the bin-size renders the precision comparable with that for the method of moments. With increased number of events in the future, an unbinned likelihood analysis will become the norm, but at the present precision level, the `Moments' data is at least equally dependable, if not more. To examine and point to any fundamental difference between these two data-sets in presence of NP models, we use both these sets as separate cases in our analysis. To the best of our knowledge, this is the first global $b\to s\ell\ell$ analysis that takes both of these datasets into account.

Apart from classifying the data according to whether it corresponds to the ``Likelihood'' or the ``Moment'' method for the angular observables, we have also prepared separate datasets which we call:
\begin{itemize}
	\item ``Old'' dataset, containing the (previous) estimates for the LFUV $R_{K}$ and $R_{K^*}$ ratios from refs.~\cite{Aaij:2014ora, Aaij:2017vbb} respectively, and
	\item ``New'' dataset, where the old estimate for $R(K)$ ~\cite{Aaij:2014ora} by LHCb is replaced by the new one~\cite{Aaij:2019wad}, while both the previous~\cite{Aaij:2017vbb} and the current~\cite{Abdesselam:2019wac} estimates for the $R_{K^*}$ ratio have been included. We also include the most recent measurement for $R_K$ due to Belle~\cite{Abdesselam:2019lab}.
\end{itemize}
We should mention here that we have only taken the ``low bins'' ($q^2\leq 6$ GeV$^2$) corresponding to the experimental data referred to above. This is done so that we can avoid the region around the $J/\psi$ resonance (the ``broad charmonium'' region) since a trustworthy theoretical estimate for this region is challenging. Hence, we do not include the $R_{K^*}$ data from the low-recoil region provided by the recent Belle measurement from~\cite{Abdesselam:2019wac}. We take care of the systematic and statistical correlations separately in the data as and when they have been reported.

\begin{table}[tbp]
	\centering
	\begin{tabular}{ccc}
		\hline
		\hline
		Observable & SM prediction & Measurement  \\
		& $\times 10^{5}$ & $\times 10^{5}$ \\
		\hline
		$\text{BR}(B\to X_s\gamma)_{E_\gamma>1.6\,\text{GeV}}$ & $33.6\pm 2.6$~\cite{Misiak:2015xwa} & $32.7 \pm 1.4$~\cite{Misiak:2017bgg} \\
		$\text{BR}(B^+\to K^*\gamma)$ & $3.51 \pm 0.78$ & $4.21 \pm 0.18$~\cite{Amhis:2014hma} \\
		$\text{BR}(B^0\to K^*\gamma)$ & $3.49 \pm 0.78$ & $4.33 \pm 0.15$~\cite{Amhis:2014hma} \\
		$\overline{\text{BR}}(B_s\to \phi\gamma)$ & $4.33 \pm 0.77$ & $3.5 \pm 0.4$\cite{Aaij:2012ita,Dutta:2014sxo} \\
		\hline
		\hline
	\end{tabular}
	\caption{SM values and experimental world averages of inclusive and exclusive $b\to s\gamma$ observables used in our analysis.}
	\label{tab:rad}
\end{table}

\begin{figure*}[t]
	\caption{\small Correlations between $R_K$ and $R_{K^*}$ for different single-operator NP scenarios. The arrows indicate the increasing values of the WCs from $-2$ to $+2$. All the experimental data are considered within their $1 \sigma$ ranges.}
	\label{fig:corpltrkrkst1d}
	\centering
	\subfloat[]{\includegraphics[width=0.45\textwidth]{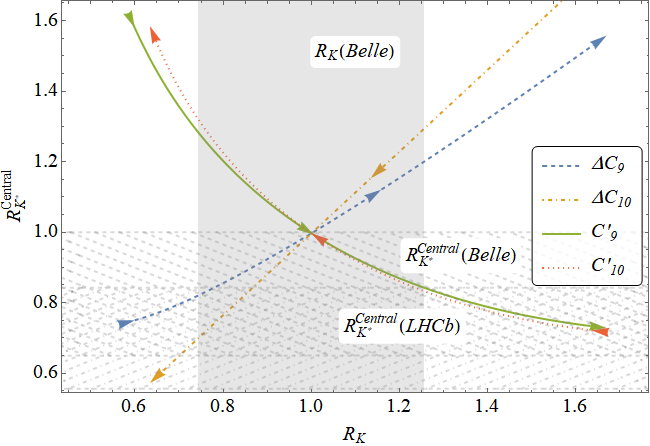}\label{fig:rkrkstcen1p}}~
	\subfloat[]{\includegraphics[width=0.45\textwidth]{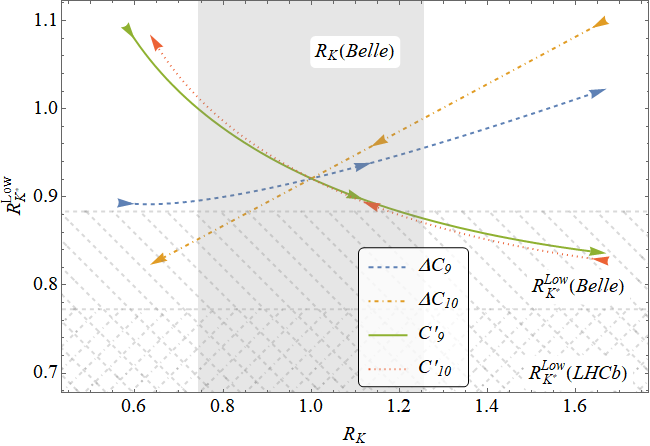}\label{fig:rkrkstLow1p}}
\end{figure*}

\begin{figure*}[htbp]
	\caption{Figure \ref{fig:rkrkstCenDc9c9pr} to figure \ref{fig:rkrkstLowDc9Dc10} shows the correlations between $R_K$ and $R_{K^*}$ in different NP scenarios. The constraints on $\Delta C_{10}$ and $C'_{10}$ from the measured value of $Br(B_s\to \mu\mu)$ can be inferred from figure \ref{fig:Bs2mmplt}.}
	\label{fig:corpltrkrkst2d}
	\centering
	\subfloat[]{\includegraphics[width=0.32\textwidth]{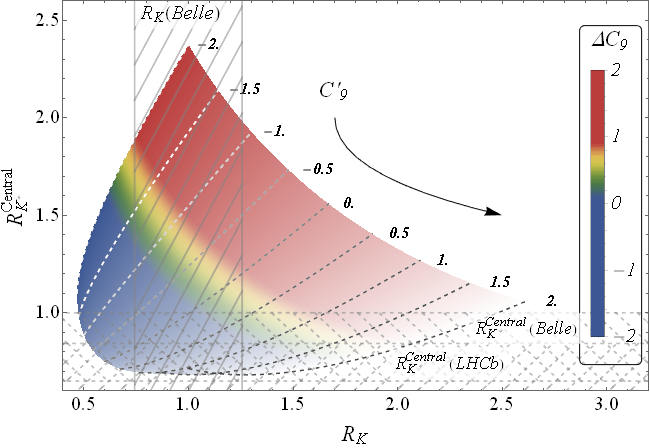}\label{fig:rkrkstCenDc9c9pr}}~
	\subfloat[]{\includegraphics[width=0.32\textwidth]{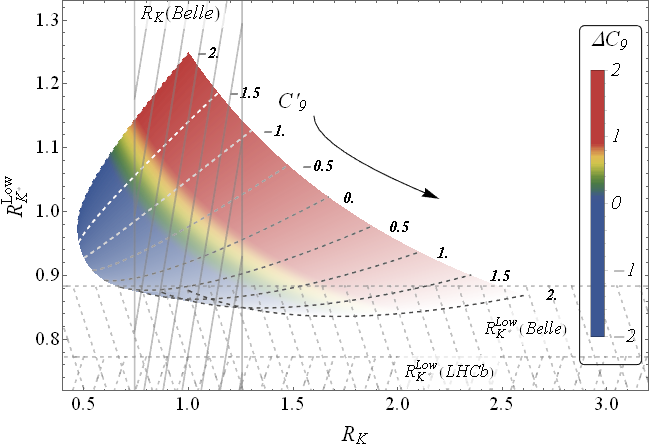}\label{fig:rkrkstLowDc9c9pr}}~
	\subfloat[]{\includegraphics[width=0.32\textwidth]{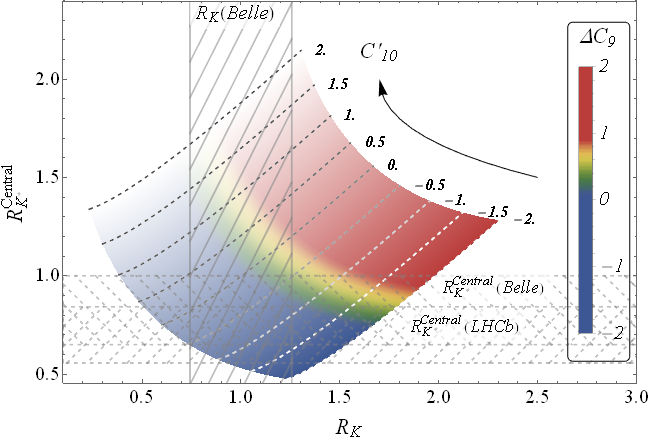}\label{fig:rkrkstCenDc9c10pr}}\\
	\subfloat[]{\includegraphics[width=0.32\textwidth]{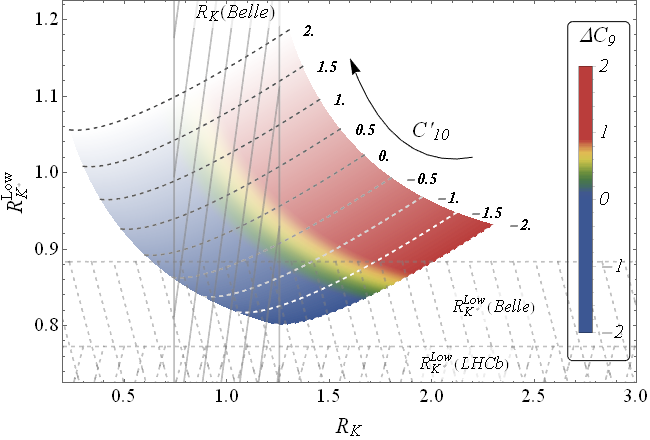}\label{fig:rkrkstLowDc9c10pr}}~
	\subfloat[]{\includegraphics[width=0.32\textwidth]{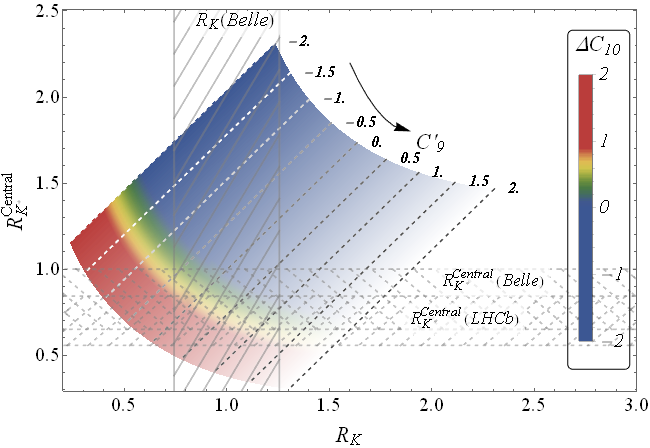}\label{fig:rkrkstCenDc10c9pr}}~
	\subfloat[]{\includegraphics[width=0.32\textwidth]{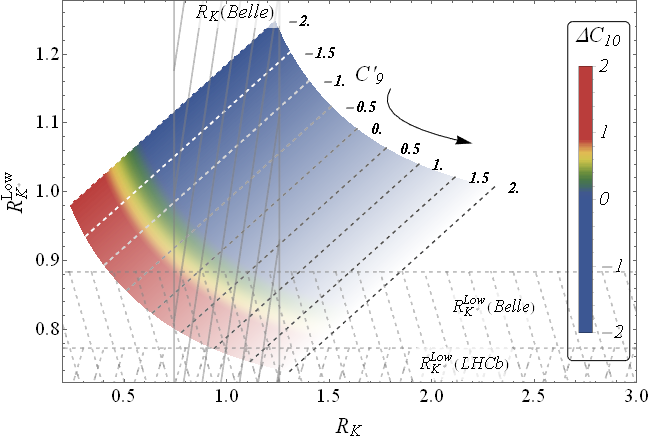}\label{fig:rkrkstLowDc10c9pr}}\\
	\subfloat[]{\includegraphics[width=0.32\textwidth]{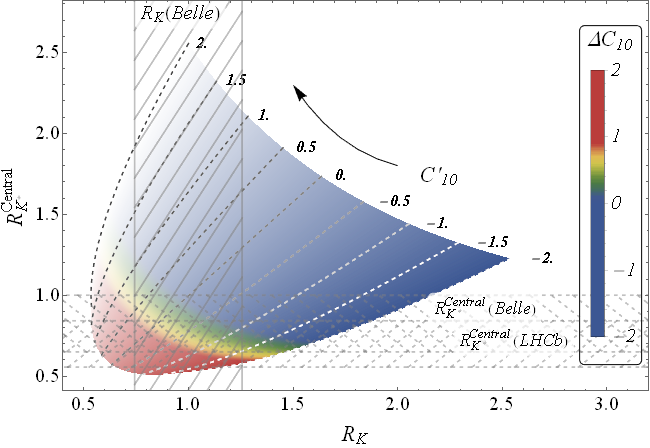}\label{fig:rkrkstCenDc10c10pr}}~
	\subfloat[]{\includegraphics[width=0.32\textwidth]{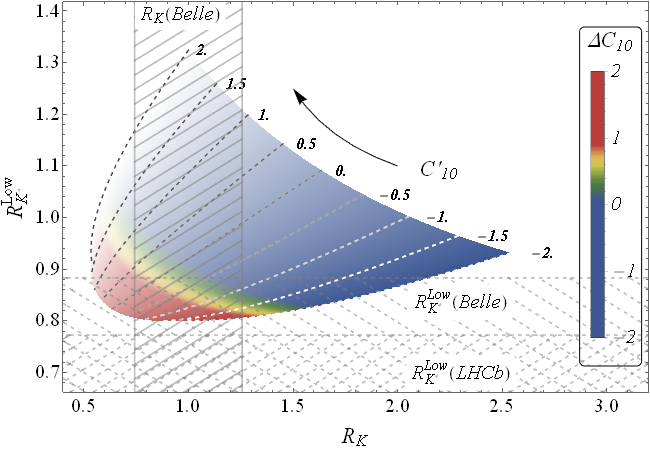}\label{fig:rkrkstLowDc10c10pr}}~
	\subfloat[]{\includegraphics[width=0.32\textwidth]{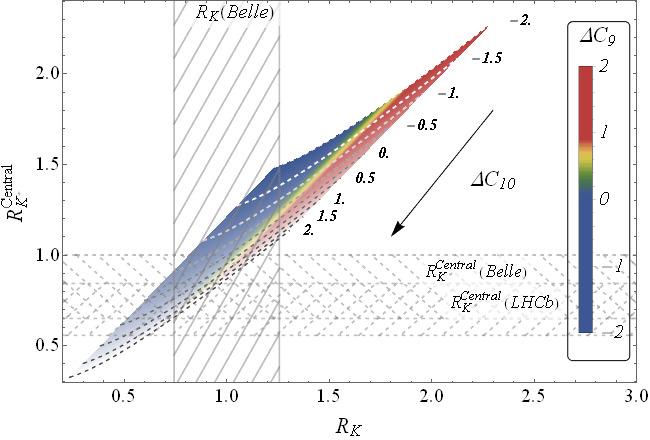}\label{fig:rkrkstCenDc9Dc10}}\\
	\subfloat[]{\includegraphics[width=0.32\textwidth]{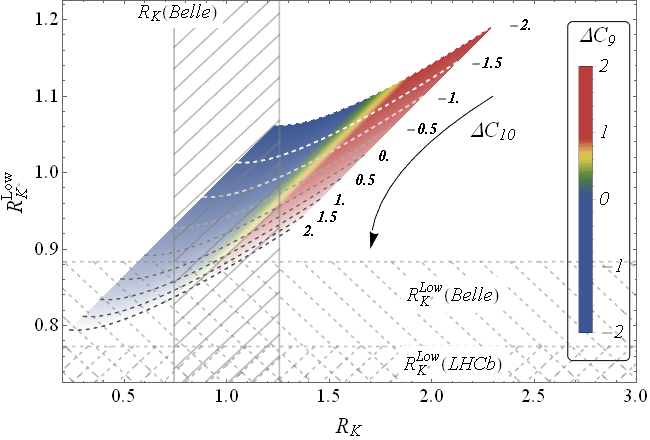}\label{fig:rkrkstLowDc9Dc10}}~~~
	\subfloat[]{\includegraphics[width=0.32\textwidth]{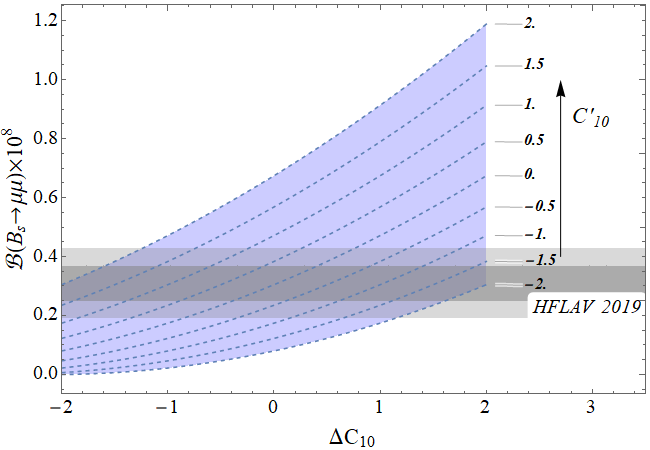}\label{fig:Bs2mmplt}}
\end{figure*}

%%%%%%%%%%%%%%%%%%%%%%%%%%%%%%%%%%%%%%%%%%%%%%%%%
\subsection{Theoretical}\label{sec:backtheo}
%%%%%%%%%%%%%%%%%%%%%%%%%%%%%%%%%%%%%%%%%%%%%%%%%
    
The effective Hamiltonian and the operator basis for exclusive $b\to s \mu^+\mu^-$ decays are taken from \cite{Altmannshofer:2008dz, Altmannshofer:2014rta} and is written as:
\begin{equation} \label{eq:Heff}
{\cal H}_{eff} = - \frac{4\,G_F}{\sqrt{2}}\left(
\lambda_t {\cal H}_{eff}^{(t)} + \lambda_u {\cal
	H}_{eff}^{(u)}\right)
\end{equation}
with the CKM combination $\lambda_i=V_{ib}V_{is}^*$ and
\begin{align}
	\nn {\cal H}_{eff}^{(t)} & = C_1 \mathcal O_1^c + C_2 \mathcal O_2^c +  \sum_{i=3}^{6} C_i \mathcal O_i + \\
	\nn &\quad \sum_{i=7,8,9,10,P,S} (C_i \mathcal O_i + C'_i \mathcal
	O'_i)\,, \\
	{\cal H}_{eff}^{(u)} & = C_1 (\mathcal O_1^c-\mathcal O_1^u)  + C_2(\mathcal O_2^c-\mathcal
	O_2^u)\,.
\end{align}
We consider NP effects in the following operators:
{\small
\begin{align}
\nn {\mathcal{O}}_{7} &= \frac{e}{g^2} m_b (\bar{s} \sigma_{\mu \nu} P_R b) F^{\mu \nu}, ~~~
{\mathcal{O}}_{7}^\prime = \frac{e}{g^2} m_b (\bar{s} \sigma_{\mu \nu} P_L b) F^{\mu \nu} ,\\
\nn {\mathcal{O}}_{9} &= \frac{e^2}{g^2} (\bar{s} \gamma_{\mu} P_L b)(\bar{\mu} \gamma^\mu \mu), ~~~ {\mathcal{O}}_{9}^\prime = \frac{e^2}{g^2} (\bar{s} \gamma_{\mu} P_R b)(\bar{\mu} \gamma^\mu \mu), \\
\nn {\mathcal{O}}_{10} &=\frac{e^2}{g^2} (\bar{s}  \gamma_{\mu} P_L b)(  \bar{\mu} \gamma^\mu \gamma_5 \mu), ~~~ {\mathcal{O}}_{10}^\prime =\frac{e^2}{g^2} (\bar{s}  \gamma_{\mu} P_R b)(  \bar{\mu} \gamma^\mu \gamma_5 \mu), \\
\nn {\mathcal{O}}_{S} &=\frac{e^2}{16\pi^2} m_b (\bar{s} P_R b)(  \bar{\mu} \mu), ~~~ {\mathcal{O}}_{S}^\prime =\frac{e^2}{16\pi^2} m_b (\bar{s} P_L b)(  \bar{\mu} \mu) ,\\
{\mathcal{O}}_{P} &=\frac{e^2}{16\pi^2} m_b (\bar{s} P_R b)(  \bar{\mu} \gamma_5 \mu) , ~~~ {\mathcal{O}}_{P}^\prime =\frac{e^2}{16\pi^2} m_b (\bar{s} P_L b)(  \bar{\mu} \gamma_5 \mu)\,.
\end{align}
}%

The NP contributions to operators $\mathcal{O}_{9,10}$ is given by $\Delta C_{9,10}$. In these decays, when the final state contains a vector meson, one can construct various helicity amplitudes. These helicity amplitudes are used to form angular coefficients which are relevant in defining the CP-symmetric and asymmetric observables measured by the different experimental collaborations. The details about various transversity amplitudes and the respective angular coefficients can be obtained from \cite{Altmannshofer:2008dz}. The two major components that go into the formation of the helicity amplitudes are the Wilson coefficients (WC) of different operators and the form factors which are defined as the hadronic matrix elements of various operators. We follow ref.~\cite{Straub:2015ica} for the form factors in $B\to K^*$ and $B_s \to \phi$ decays \footnote{Although the latest LCSR estimates for the $B\to K^*$ decays is calculated in ref.~\cite{Gubernari:2018wyi}, it does not include the corresponding $B_s\to\phi$ matrix elements. Hence we refrain from using these results.}.

For the $B\to K$ sector we closely follow the methodology communicated in ref.~\cite{Bobeth:2007dw}. This includes expressing the differential decay distribution in terms of a polynomial in $\cos\theta$, where $\theta$ denotes the angle between the direction of motion of the parent $B$ meson and the positively charged lepton in the dilepton center of mass frame. The coefficients of these terms can then be expressed as combinations of the corresponding WC and form factors. For the form factors, we use the results from ref.~\cite{Altmannshofer:2014rta}, where the authors perform a combined fit to the lattice computation in ref.~\cite{Bouchard:2013pna} as well as LCSR predictions at $q^2=0$~\cite{Ball:2004ye,Bartsch:2009qp}, using the parametrization and conventions of~\cite{Bouchard:2013pna}. The method is described in details in the appendix of ref.~\cite{Buras:2014fpa}.

We also take care of the correlations among these form factor elements as reported in these references, in order to propagate them to form the theoretical correlations and errors for the corresponding observables.

Since our aim is to perform a global model selection based on the plethora of available $b\to s\ell\ell$ data discussed in sec.~\ref{sec:backexp}, there is a possibility that amongst the selected models the operator with $C_{7}^\prime$ as coefficient may appear as a plausible solution. Such an operator is also relevant for the radiative decays like inclusive and exclusive $b\to s\gamma$. For such scenarios, we have checked whether parameter spaces which are allowed by $b\to sll$ data are also allowed by the inclusive $B\to X_s\gamma$ measurement, alongwith the branching ratios for the three exclusive radiative modes $B^+\to K^{*+}\gamma$, $B^0\to K^{*0}\gamma$ and the time integrated $\overline{\text{BR}}(B_s\to\phi\gamma)$  \footnote{We refrain from using measurements for the CP asymmetries since our NP Wilson coefficients are taken to be real, thus excluding the possibility of CP violation in NP}. The definitions and formulae for these modes are taken from ref.~\cite{Paul:2016urs}. We provide the experimental values and the SM estimates used in our analysis in Table.~\ref{tab:rad}. The corresponding theoretical (for $B\to X_s\gamma$) and experimental references are provided therein. Our SM values are consistent with the estimates of ref.~\cite{Paul:2016urs}, within $1~\sigma$.

\section{NP and current data on $R_K$ and $R_{K^*}$}

Before pursuing a detailed discussion on model selection, let us look for the NP effects in $b\to s \mu\mu$ decays, only in the light of recently updated measurements on $R_K$ and $R_{K^*}$ in this section, focusing on the measurements of $R_{K^*}$ in the low $q^2$ bins. There is some discrepancy between this particular data and the corresponding predicted value in the SM. However, one needs to remember that the angular observables are not free from hadronic uncertainties. In this part of the study, we do not consider any of the angular observables, neither do we carry out any fit to data. We simply check the dependencies of  $R^{Low}_{K^*}$, $R^{Central}_{K^*}$, and $R_K$ on various WCs in one and two operator scenarios. We do not include $C_S$, $C_P$, $C'_S$, and $C'_P$, since the corresponding operators by themselves, or combinations including such operators, are unable to explain the observed data in $R^{Low}_{K^*}$. These WCs also suffer from tight constraints due to $B_s \to \mu\mu$ decays \cite{Alonso:2014csa}. Also, the new electromagnetic dipole operator ${\cal O}'_7$ alone would not be able to explain the observed data on $R_K^{(*)}$ and the branching fractions in the above mentioned radiative decays simultaneously. Hence, we have not considered the effects of this operator in this part of the analysis.

The results of our analysis are presented in figures \ref{fig:corpltrkrkst1d} and \ref{fig:corpltrkrkst2d}. In single operator scenarios, the correlations between all the above-mentioned observables are shown in figures \ref{fig:rkrkstcen1p} and \ref{fig:rkrkstLow1p}. It would be difficult to explain the observed data for $R^{Low}_{K^*}$ within their $1 \sigma$ ranges will be difficult in the single-operator scenarios. Although the allowed region is tightly constrained, ${\cal O}_{10}$ (with WC $\Delta C_{10}$) is the only operator that can simultaneously explain all the data on $R_{K^{(*)}}$ except $R^{Low}_{K^*}$ from LHCb. The required value of $\Delta C_{10}$ lies in between 0.5 and 1.5, which is consistent with the measured value of $Br(B_s\to \mu\mu)$ within its 2-$\sigma$ range for detail see figure \ref{fig:Bs2mmplt}. However, there are several candidates in the two operator scenarios that could explain all the data simultaneously. Among various possible combinations, the highly probable scenarios are the operators with the WCs $[\Delta C_9, C'_{10}]$,  $[\Delta C_{10}, C'_{9}]$, $[\Delta C_{10}, C'_{10}]$, and $[\Delta C_{10}, \Delta C_{9}]$. The other possible scenario $[\Delta C_9, C'_{9}]$ is less favored but allowed by the data. Also, the allowed values of $\Delta C_{10}$ and/or $C'_{10}$ can explain $Br(B_s\to \mu\mu)$ within its 1-$\sigma$ range, wherever applicable; see figure \ref{fig:Bs2mmplt} for details. In fig. \ref{fig:Bs2mmplt}, we have shown the variations of $Br(B_s\to \mu\mu)$ w.r.t. the parameters $[\Delta C_{10}, C'_{10}]$. However, there are scenarios where only $\Delta C_{10}$ or $C'_{10}$ appears. In such cases, depending on the scenario, one needs to look at the plot with either $\Delta C_{10} =0 $ or $C'_{10} = 0$. To conclude this section, we would like to mention that the three or more operator scenarios could also be relevant to explain the present data on $R_{K^{(*)}}$ simultaneously. The take home message is that simultaneous contributions from various operators are required for a simultaneous explanation of the $R_{K^{(*)}}$ data alone. The results of this section will be useful for a better understanding of the results in the following section.

\begin{figure*}[t]
	\caption{\small For the fit with `New Data', indices of competing scenarios with $\Delta\text{AIC}_c \leq 4$ in the MSE$_{\text{X-val}}$ vs. $w^{\Delta\text{AIC}_c}_i$ plane. We break the plane in four regions, with the one in right-bottom being the best one. Models in this region are chosen as the best ones from these two criteria and the labels are colored blue. Wilson coefficients contained in these models are shown in-box. For comparison, indices for models picked up by the criterion $\Delta\text{AIC}_c \leq 2$ are framed. For details, check sec. \ref{modelselection}.}
	\label{fig:ModselNew}
	\centering
	\subfloat[New Data (Moments)]{\includegraphics[width=\columnwidth]{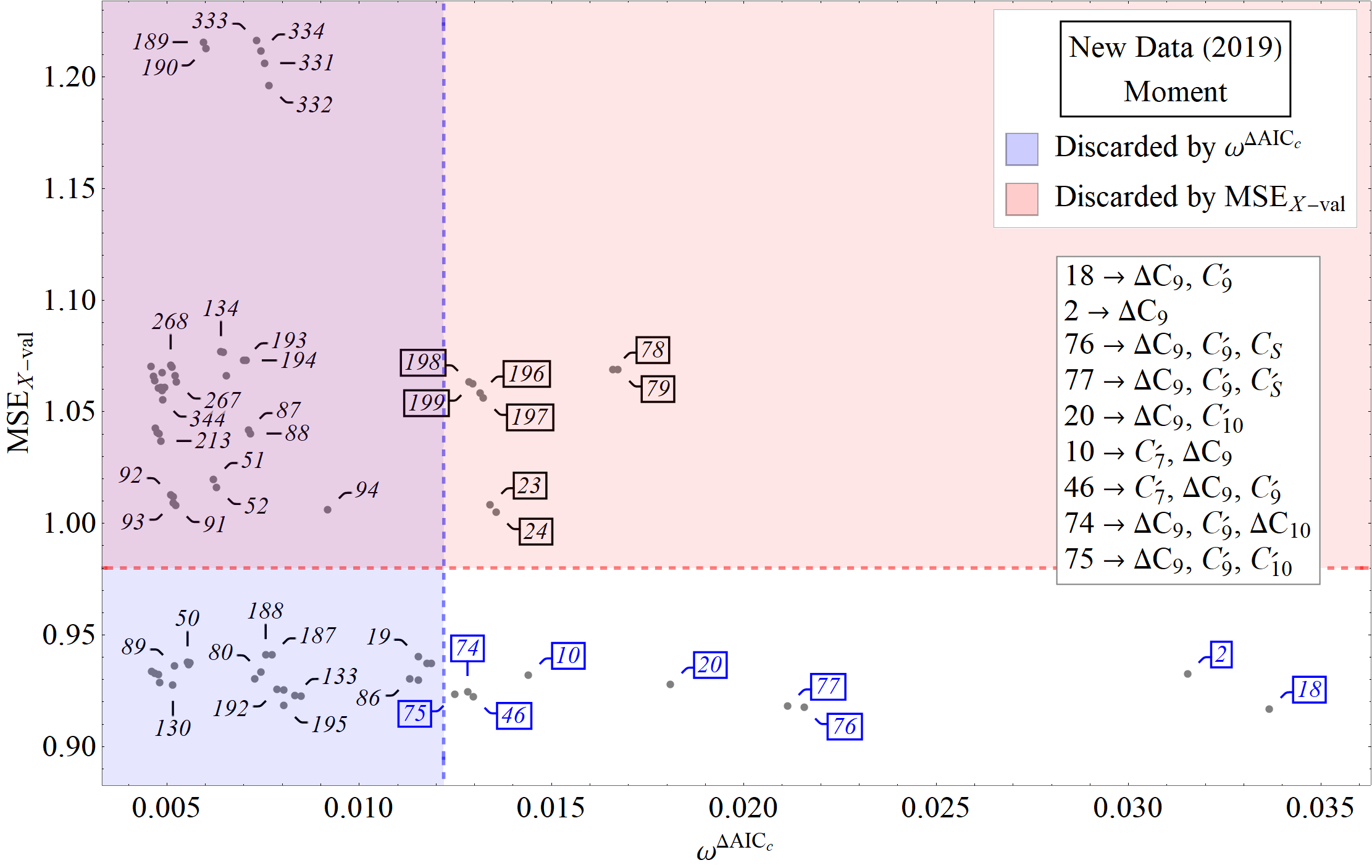}
		\label{fig:ModselNewMom}}~~
	\subfloat[New Data (Likelihood)]{\includegraphics[width=\columnwidth]{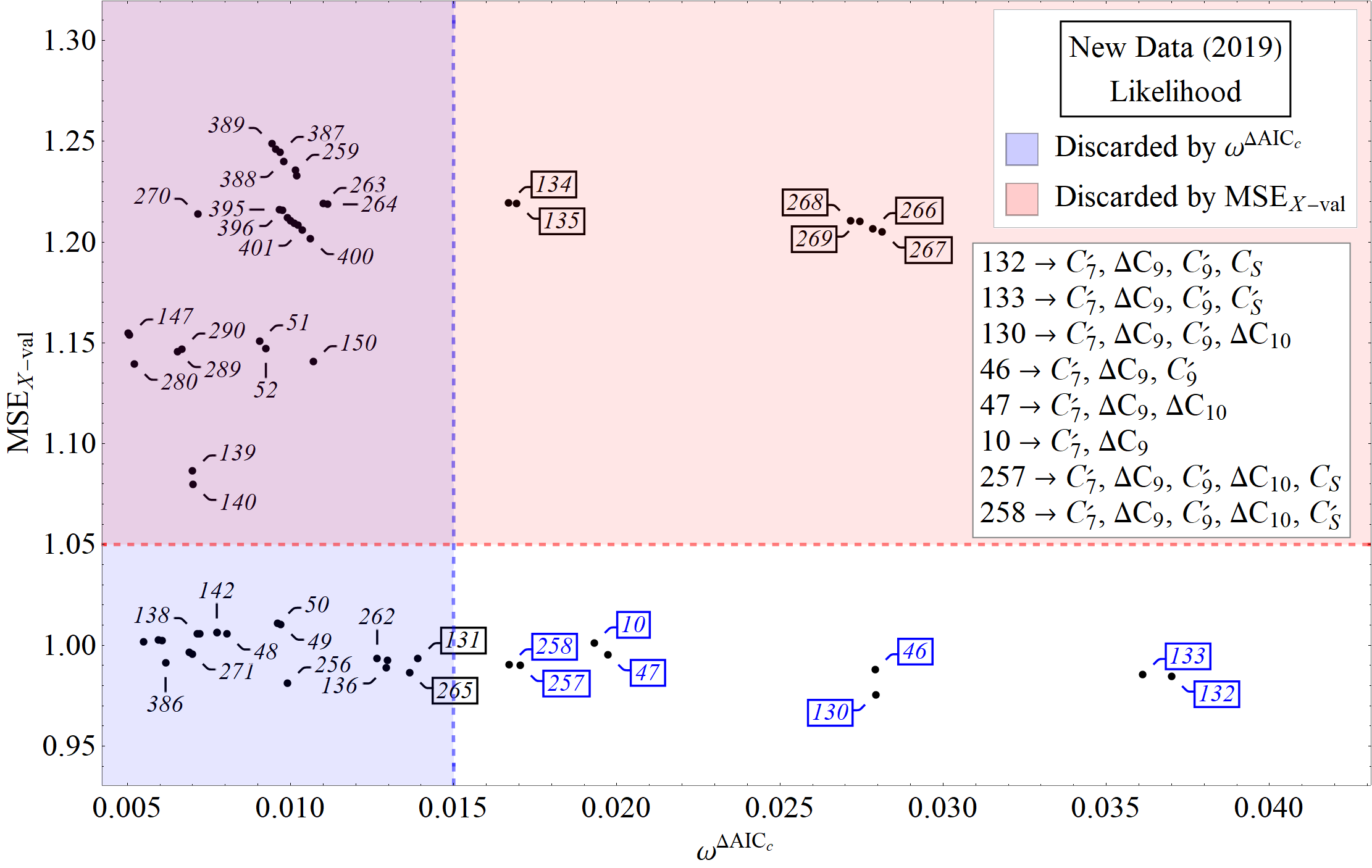}
		\label{fig:ModselNewLike}}
\end{figure*}

\begin{figure*}[t]
	\caption{\small Same as fig. \ref{fig:ModselOld}, but for the fit with `Old Data'.}
	\label{fig:ModselOld}
	\centering
	\subfloat[Old data (Moments)]{\includegraphics[width=\columnwidth]{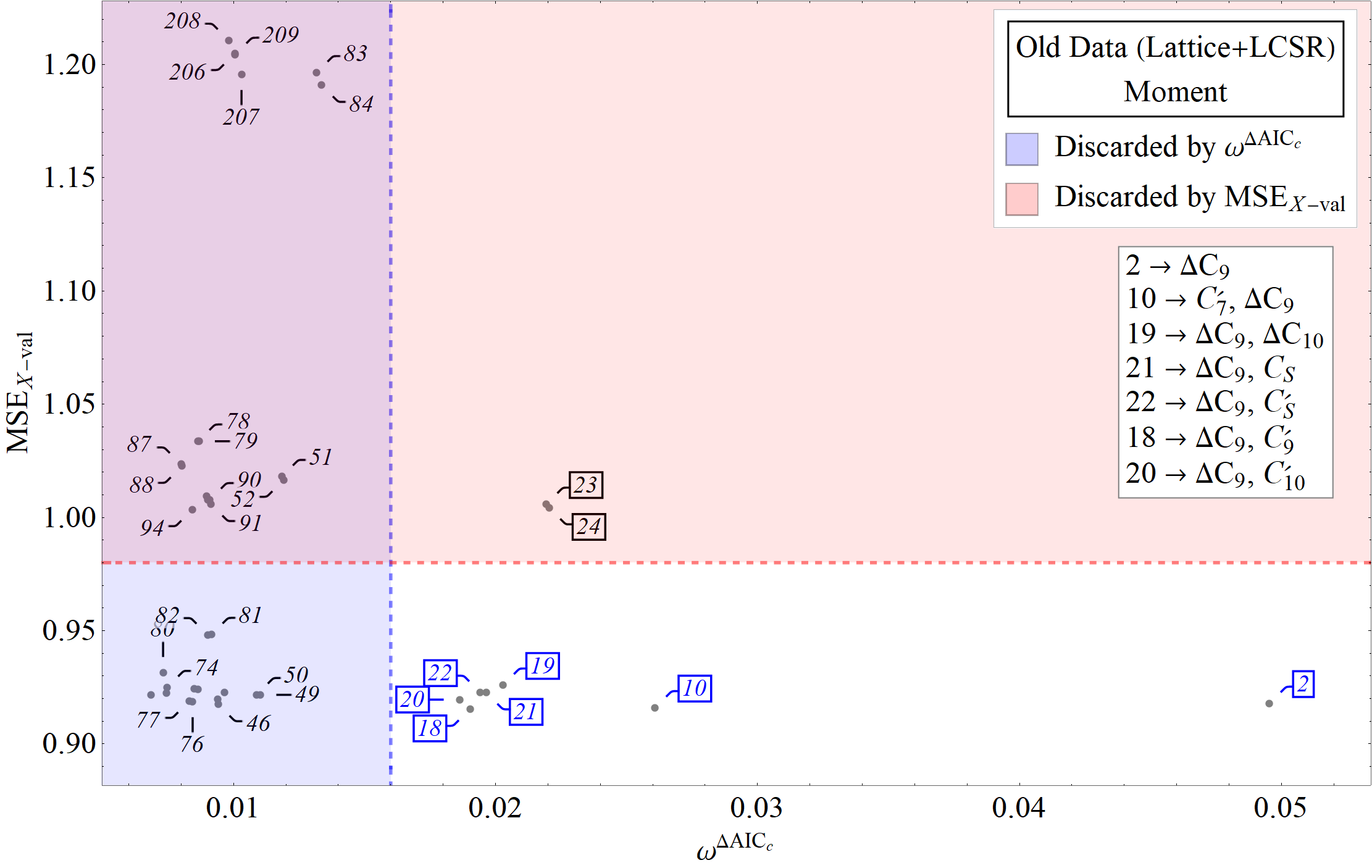}
		\label{fig:ModselOldMom}}~~
	\subfloat[Old data (Likelihood)]{\includegraphics[width=\columnwidth]{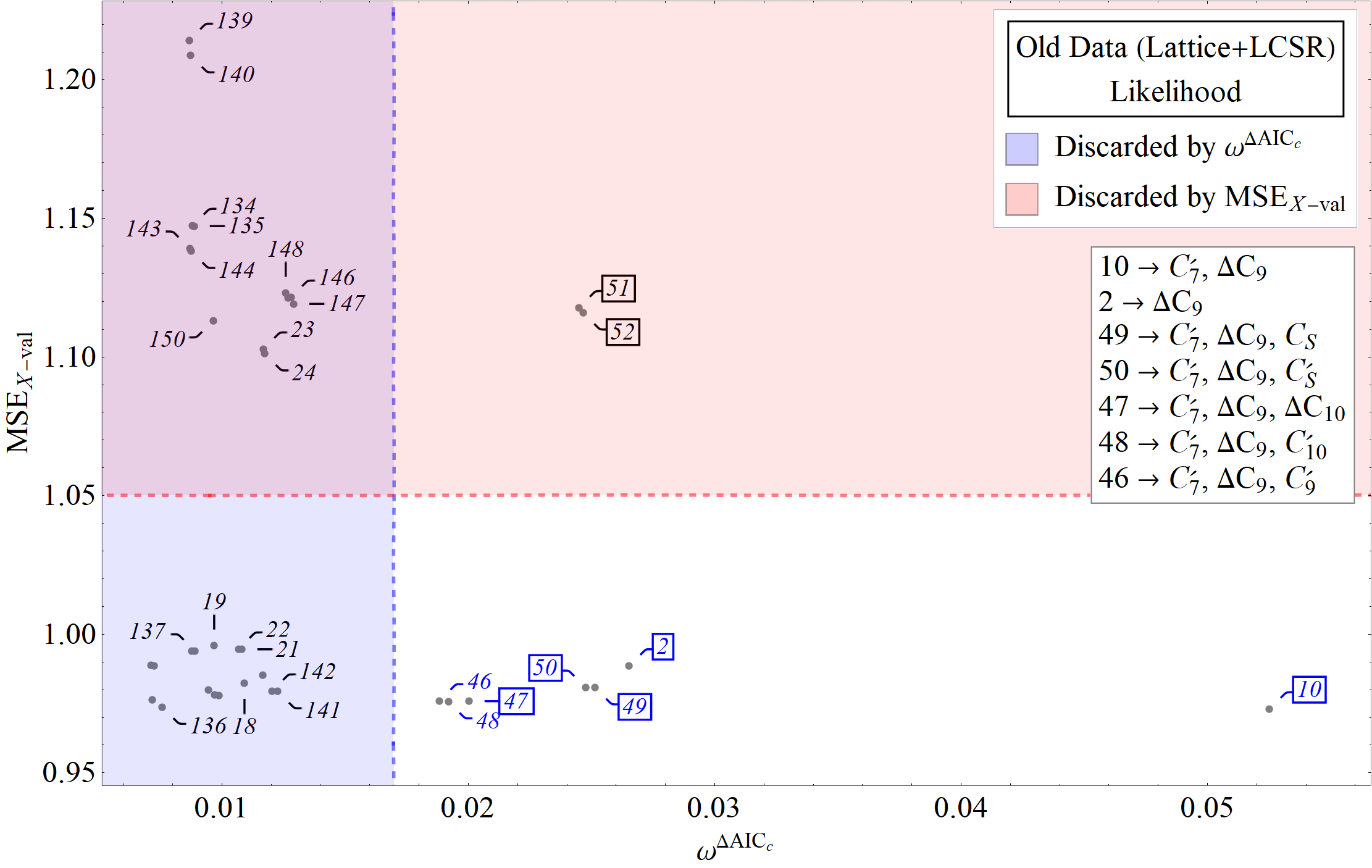}
		\label{fig:ModselOldLike}}
\end{figure*}

\begin{table*}[tbp]
	\small
	\caption{Fit-qualities, model selection criteria, parameter estimates and effects on radiative decays for the `best' selected models with the `New' data-set, with the `Moments' estimate of the angular observables. Selected models are obtained from fig. \ref{fig:ModselNewMom}. Last four columns showcase the deviations (in units of `$\sigma$') between the experimental value of the radiative decays and the corresponding value obtained with the fit results.}
	\begin{tabular}{*{10}{c}}
		\hline
		\hline
		$\text{Model}$  &  $\left.\chi _{\text{Min}}^2\right/$  &  $\text{p-val}$  &  $\omega ^{\text{$\Delta $AIC}_c}$  &  $\text{MSE}_{X-\text{val}}$  &  $\text{Parameter}$  &  \multicolumn{4}{c|}{Deviation in $\sigma$} \\
		\cline{7-10}
		$\text{Index}$  &  $\text{DOF}$  &  $\text{($\%$)}$  &  $\text{($\%$)}$  &  $\text{}$  &  $\text{Values}$  &  $\text{B$\to $}X_s\gamma$  &  $B^+\to K^*\gamma$  &  $\text{$\Delta $B}^0\to K^*\gamma$  &  $\text{$\Delta $B}_s\to \phi \gamma$  \\
		\hline
		$18$  &  $\text{250.28/256}$  &  $58.9$  &  $3.4$  &  $0.917$  &  $\begin{array}{l}
		\Delta C_9\to \text{-1.13$\pm $0.13} \\
		C'_{9}\to \text{0.25$\pm $0.17} \\
		\end{array}$  &  $-$  &  $-$  &  $-$  &  $-$  \\
		\cline{6-6}
		$2$  &  $\text{252.44/257}$  &  $56.9$  &  $3.2$  &  $0.933$  &  $\begin{array}{l}
		\Delta C_9\to \text{-1.12$\pm $0.13} \\
		\end{array}$  &  $-$  &  $-$  &  $-$  &  $-$  \\
		\cline{6-6}
		$76$  &  $\text{249.12/255}$  &  $59.2$  &  $2.2$  &  $0.918$  &  $\begin{array}{l}
		\Delta C_9\to \text{-1.18$\pm $0.14} \\
		C'_{9}\to \text{0.34$\pm $0.19} \\
		C_S\to \text{-0.035$\pm $0.016} \\
		\end{array}$  &  $-$  &  $-$  &  $-$  &  $-$  \\
		\cline{6-6}
		$77$  &  $\text{249.16/255}$  &  $59.1$  &  $2.1$  &  $0.918$  &  $\begin{array}{l}
		\Delta C_9\to \text{-1.18$\pm $0.14} \\
		C'_{9}\to \text{0.34$\pm $0.19} \\
		C'_S\to \text{0.035$\pm $0.016} \\
		\end{array}$  &  $-$  &  $-$  &  $-$  &  $-$  \\
		\cline{6-6}
		$20$  &  $\text{251.52/256}$  &  $56.7$  &  $1.8$  &  $0.928$  &  $\begin{array}{l}
		\Delta C_9\to \text{-1.15$\pm $0.14} \\
		C'_{10}\to \text{-0.1$\pm $0.104} \\
		\end{array}$  &  $-$  &  $-$  &  $-$  &  $-$  \\
		\cline{6-6}
		$10$  &  $\text{251.97/256}$  &  $55.9$  &  $1.4$  &  $0.932$  &  $\begin{array}{l}
		C'_7\to \text{0.01$\pm $0.015} \\
		\Delta C_9\to \text{-1.15$\pm $0.14} \\
		\end{array}$  &  $0.31$  &  $-0.87$  &  $-1.06$  &  $1.22$  \\
		\cline{6-6}
		$46$  &  $\text{250.14/255}$  &  $57.4$  &  $1.3$  &  $0.922$  &  $\begin{array}{l}
		C'_7\to \text{0.0058$\pm $0.0155} \\
		\Delta C_9\to \text{-1.15$\pm $0.14} \\
		C'_{9}\to \text{0.24$\pm $0.18} \\
		\end{array}$  &  $0.3$  &  $-0.87$  &  $-1.06$  &  $1.22$  \\
		\cline{6-6}
		$74$  &  $\text{250.16/255}$  &  $57.4$  &  $1.3$  &  $0.925$  &  $\begin{array}{l}
		\Delta C_9\to \text{-1.16$\pm $0.15} \\
		C'_{9}\to \text{0.26$\pm $0.17} \\
		\Delta C_{10}\to \text{-0.041$\pm $0.118} \\
		\end{array}$  &  $-$  &  $-$  &  $-$  &  $-$  \\
		\cline{6-6}
		$75$  &  $\text{250.21/255}$  &  $57.3$  &  $1.2$  &  $0.923$  &  $\begin{array}{l}
		\Delta C_9\to \text{-1.12$\pm $0.14} \\
		C'_{9}\to \text{0.3$\pm $0.26} \\
		C'_{10}\to \text{0.04$\pm $0.157} \\
		\end{array}$  &  $-$  &  $-$  &  $-$  &  $-$  \\
		\hline
		\hline
	\end{tabular}
	\label{tab:NewMom}
\end{table*}

\begin{table*}[tbp]
\small
\caption{Same as the table \ref{tab:NewMom}, but with the `Likelihood' estimate of the angular observables. Selected models are obtained from fig. \ref{fig:ModselNewLike}.}
\begin{tabular}{*{10}{c}}
\hline
\hline
$\text{Model}$  &  $\left.\chi _{\text{Min}}^2\right/$  &  $\text{p-val}$  &  $\omega ^{\text{$\Delta $AIC}_c}$  &  $\text{MSE}_{X-\text{val}}$  &  $\text{Parameter}$  &  \multicolumn{4}{c|}{Deviation in $\sigma$} \\
\cline{7-10}
$\text{Index}$  &  $\text{DOF}$  &  $\text{($\%$)}$  &  $\text{($\%$)}$  &  $\text{}$  &  $\text{Values}$  &  $\text{B$\to $}X_s\gamma$  &  $B^+\to K^*\gamma$  &  $\text{$\Delta $B}^0\to K^*\gamma$  &  $\text{$\Delta $B}_s\to \phi \gamma$  \\
\hline
$132$  &  $\text{217.02/210}$  &  $35.5$  &  $3.7$  &  $0.985$  &  $\begin{array}{l}
C'_7\to \text{0.04$\pm $0.015} \\
\Delta C_9\to \text{-1.39$\pm $0.13} \\
C'_9\to \text{0.45$\pm $0.2} \\
C_S\to \text{-0.042$\pm $0.013} \\
\end{array}$  &  $0.44$  &  $-0.82$  &  $-1.02$  &  $1.27$  \\
\cline{6-6}
$133$  &  $\text{217.07/210}$  &  $35.4$  &  $3.6$  &  $0.986$  &  $\begin{array}{l}
C'_7\to \text{0.04$\pm $0.015} \\
\Delta C_9\to \text{-1.39$\pm $0.13} \\
C'_9\to \text{0.45$\pm $0.2} \\
C'_S\to \text{0.042$\pm $0.013} \\
\end{array}$  &  $0.44$  &  $-0.82$  &  $-1.02$  &  $1.27$  \\
\cline{6-6}
$130$  &  $\text{217.58/210}$  &  $34.5$  &  $2.8$  &  $0.976$  &  $\begin{array}{l}
C'_7\to \text{0.044$\pm $0.015} \\
\Delta C_9\to \text{-1.42$\pm $0.14} \\
C'_9\to \text{0.32$\pm $0.19} \\
\Delta C_{10}\to \text{-0.16$\pm $0.11} \\
\end{array}$  &  $0.47$  &  $-0.81$  &  $-1.01$  &  $1.28$  \\
\cline{6-6}
$46$  &  $\text{219.66/211}$  &  $32.7$  &  $2.8$  &  $0.988$  &  $\begin{array}{l}
C'_7\to \text{0.04$\pm $0.015} \\
\Delta C_9\to \text{-1.34$\pm $0.13} \\
C'_9\to \text{0.33$\pm $0.2} \\
\end{array}$  &  $0.44$  &  $-0.82$  &  $-1.02$  &  $1.27$  \\
\cline{6-6}
$47$  &  $\text{220.36/211}$  &  $31.5$  &  $2.$  &  $0.995$  &  $\begin{array}{l}
C'_7\to \text{0.048$\pm $0.015} \\
\Delta C_9\to \text{-1.43$\pm $0.15} \\
\Delta C_{10}\to \text{-0.16$\pm $0.11} \\
\end{array}$  &  $0.5$  &  $-0.8$  &  $-1.$  &  $1.29$  \\
\cline{6-6}
$10$  &  $\text{222.46/212}$  &  $29.7$  &  $1.9$  &  $1.001$  &  $\begin{array}{l}
C'_7\to \text{0.043$\pm $0.015} \\
\Delta C_9\to \text{-1.33$\pm $0.13} \\
\end{array}$  &  $0.46$  &  $-0.81$  &  $-1.01$  &  $1.28$  \\
\cline{6-6}
$257$  &  $\text{216.47/209}$  &  $34.7$  &  $1.7$  &  $0.99$  &  $\begin{array}{l}
C'_7\to \text{0.042$\pm $0.015} \\
\Delta C_9\to \text{-1.42$\pm $0.14} \\
C'_9\to \text{0.41$\pm $0.21} \\
\Delta C_{10}\to \text{-0.091$\pm $0.123} \\
C_S\to \text{-0.036$\pm $0.017} \\
\end{array}$  &  $0.46$  &  $-0.82$  &  $-1.01$  &  $1.28$  \\
\cline{6-6}
$258$  &  $\text{216.51/209}$  &  $34.6$  &  $1.7$  &  $0.99$  &  $\begin{array}{l}
C'_7\to \text{0.042$\pm $0.015} \\
\Delta C_9\to \text{-1.42$\pm $0.14} \\
C'_9\to \text{0.41$\pm $0.21} \\
\Delta C_{10}\to \text{-0.092$\pm $0.123} \\
C'_S\to \text{0.036$\pm $0.017} \\
\end{array}$  &  $0.46$  &  $-0.82$  &  $-1.01$  &  $1.28$  \\
\cline{6-6}
$131$  &  $\text{218.98/210}$  &  $32.1$  &  $1.4$  &  $0.993$  &  $\begin{array}{l}
C'_7\to \text{0.04$\pm $0.015} \\
\Delta C_9\to \text{-1.32$\pm $0.13} \\
C'_9\to \text{0.47$\pm $0.25} \\
\text{$C_{10}^{\prime}$}\to \text{0.11$\pm $0.14} \\
\end{array}$  &  $0.44$  &  $-0.82$  &  $-1.02$  &  $1.27$  \\
\cline{6-6}
$265$  &  $\text{216.92/209}$  &  $33.9$  &  $1.4$  &  $0.986$  &  $\begin{array}{l}
C'_7\to \text{0.04$\pm $0.015} \\
\Delta C_9\to \text{-1.39$\pm $0.13} \\
C'_9\to \text{0.44$\pm $0.2} \\
C_S\to \text{-0.24$\pm $0.6} \\
C'_S\to \text{-0.19$\pm $0.6} \\
\end{array}$  &  $0.44$  &  $-0.82$  &  $-1.02$  &  $1.27$  \\
\hline
\hline
\end{tabular}
\label{tab:NewLike}
\end{table*}

%%%%%%%%%%%%%%%%%%%%%%%%%%%%%%%%%%%%%%%%%%%%%%%%%	
\section{Methodology}\label{sec:method}
\subsection{Parameter Estimation}\label{sec:methodOptim}
%%%%%%%%%%%%%%%%%%%%%%%%%%%%%%%%%%%%%%%%%%%%%%%%%

The methodology adopted in this paper for the model selection is as follows:

\paragraph{Define Models:}  Considering the NP Wilson coefficients real, we take all possible combinations (511 in total) of the coefficients forming a predefined global set of different scenarios. Each scenario with a specific combination of coefficients thus constitutes a potential `model' to explain the experimental results. 
\paragraph{Numerical Optimization:} Next, for each such `model' $k$, as mentioned above, we perform a Frequentist statistical analysis optimizing a $\chi^2$ statistic which is a function of the Wilson coefficients. Whenever applicable, statistical (systematic) covariance matrices $V^{stat (syst)}$, are constructed by taking separate correlations. Theoretical uncertainties are propagated separately and are introduced in the $\chi^2$ in terms of a `theoretical' covariance matrix $V^{th}$. The effect of the interplay of the SM uncertainties and the NP parameters come in the fit at a higher order and are neglected without any loss of generality.
Following section \ref{sec:backexp}, we perform 4 types of fit for each `model': 
\begin{enumerate}[label=(\alph*)] 
	\item \textit{New} data with \textit{Likelihood} data for angular observables, a total of 214 observables.%, with constraints from \textit{LCSR and Lattice}.
	\item \textit{New} data with \textit{Moments} data for angular observables, a total of 258 observables.%, with constraints from \textit{LCSR and Lattice}.
	\item \textit{Old} data with \textit{Likelihood} data for angular observables, a total of 211 observables.%, with constraints from \textit{LCSR and Lattice}.
	\item \textit{Old} data with \textit{Moments} data for angular observables, a total of 255 observables.%, with constraints from \textit{LCSR and Lattice}.
\end{enumerate}
All fits are done in batch using \textit{Mathematica}\textsuperscript{\textcopyright} in the form of a package \cite{OptEx}. The chosen optimization method is `Differential Evolution', a stochastic parallel direct search evolution strategy \cite{Storn1997} \footnote{Capable of handling non-differentiable, nonlinear and multi-modal objective functions. Considerably faster than generic genetic algorithms and extremely able to find the global minimum.}.
\paragraph{Post-process:}\label{sec:normality} In the post-process for each fit, we obtain fit-quality using $p$-value and find outliers by constructing a `Pull' (related to Studentized residuals; for our purpose, the difference between the fitted and experimental results, normalized by the uncertainty of the data, including theory uncertainties \cite{CYRSP303,Demortier:2008}) for each data-point. We also check the normality of the `Pull'-distribution (i.e. consistency with a Gaussian of $\mu=0$ and $\sigma=1$) to ensure the applicability of the $\chi^2$ as the fit-statistic. We use the ``Cram\'{e}r-von Mises" criterion \cite{cramer1928} for the normality check. Scenarios not satisfying the normality criterion are dropped from the analysis.
\paragraph{Parameter-space:} Parameter uncertainties are obtained both from the Fisher matrix\footnote{In case of approximately Gaussian parameter-profile likelihoods, it is possible to obtain the `HESSE' errors \cite{James:1975dr}, which are, obviously, symmetric.} and the profile-likelihood curve\footnote{Range of the $1\sigma$ confidence level (CL) of the profile likelihoods of the said parameter. One and two dimensional profile likelihoods in this analysis will be depicted as 1-CL plots, closely following the PROB method followed in ref. \cite{Aaij:2016kjh}}

With the remaining scenarios, we perform a model-selection procedure for each data-set. In the following sub-section, we elaborate the methods used to do the multi-model selection procedure.

%%%%%%%%%%%%%%%%%%%%%%%%%%%%%%%%%%%%%%%%%%%%%%%%%
\subsection{Model Selection}\label{sec:methodModsel}
%%%%%%%%%%%%%%%%%%%%%%%%%%%%%%%%%%%%%%%%%%%%%%%%%

Following the `concept of parsimony' \cite{boxjenkins}, we need to optimize the dimension (measure of the degree of structure) of the model explaining our data. All model selection methods, to some extent, depend on the principle of parsimony \cite{breiman}. In statistical terms, this is expressed as a bias versus variance trade-off. In general, bias decreases and variance increases as the model-dimension increases.

\subsubsection{Cross-Validation:}\label{sec:methodXval}
As we have mentioned in the introduction, `cross-validation' is the most generally applicable, powerful, reliable, and computationally expensive method for model comparison. The most straightforward and the most expensive flavor of cross-validation is ``leave-one-out cross-validation" (LOOCV). In LOOCV, one of the data points is left out and the rest of the sample (``training set'') is optimized for a particular model. Then that result is used to find the predicted residual for the left out data point. This process is repeated for all data points and a mean-squared-error (MSE) is obtained using all those residuals. This process is repeated for all models. The models with the least MSE are the best ones.

\begin{figure*}[t]
	\caption{\small Comparison of the $CP$-averaged angular observables obtained in experiment, SM and from our fit results considering all the avilable inputs.}
	\label{fig:angobscp}
	\centering
	\subfloat[]{\includegraphics[width=0.24\textwidth]{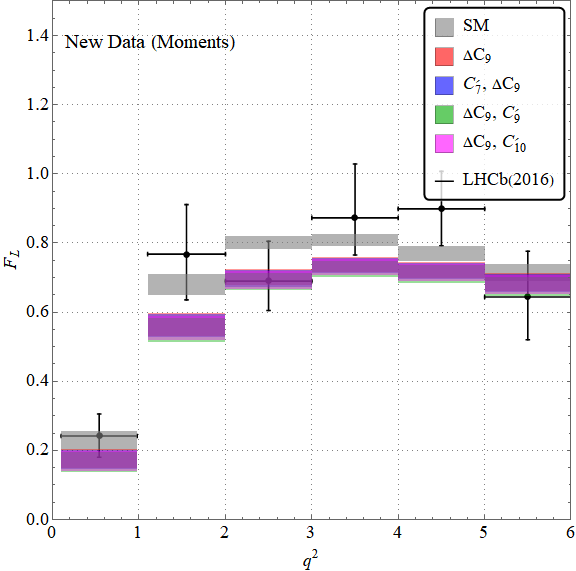}
		\label{fig:angobsmomFL}}~
	\subfloat[]{\includegraphics[width=0.24\textwidth]{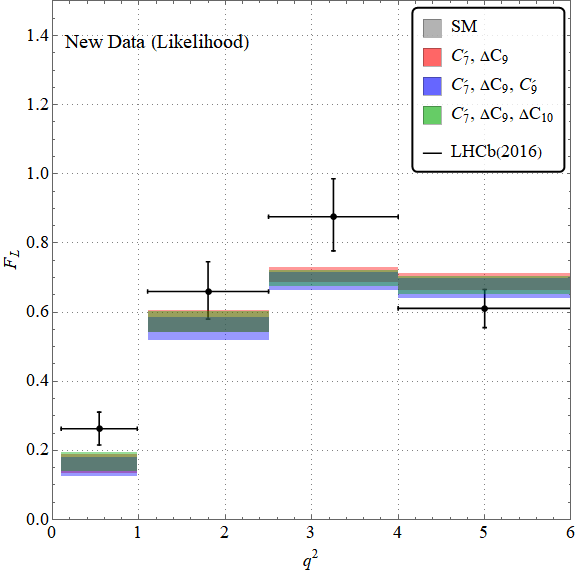}
		\label{fig:angobslikeFL}}~
	\subfloat[]{\includegraphics[width=0.24\textwidth]{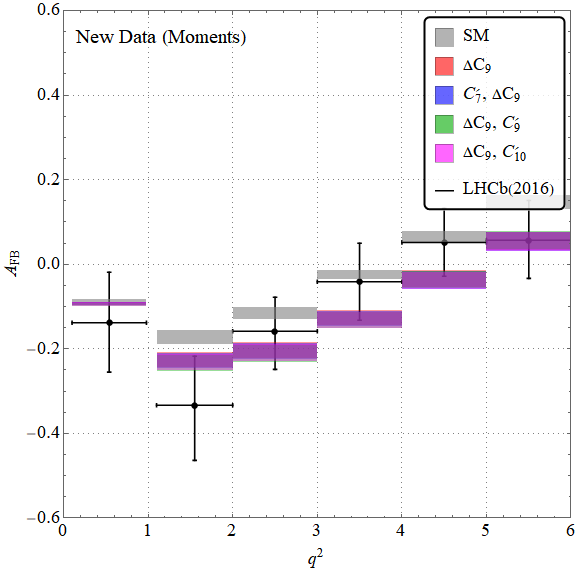}
		\label{fig:angobsmomAFB}}~
	\subfloat[]{\includegraphics[width=0.24\textwidth]{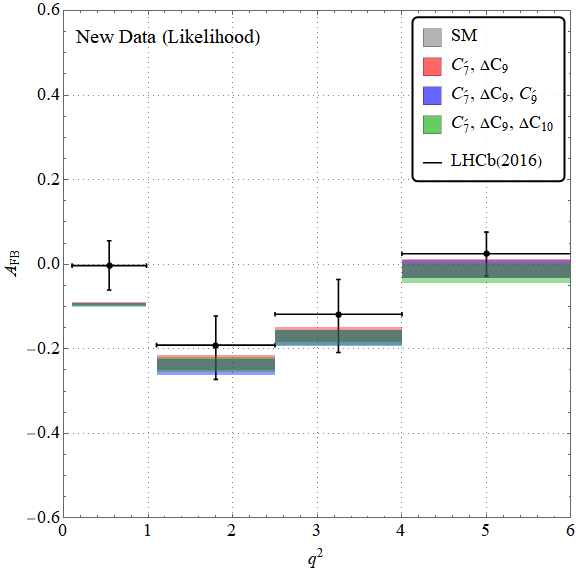}
		\label{fig:angobslikeAFB}}\\
	\subfloat[]{\includegraphics[width=0.24\textwidth]{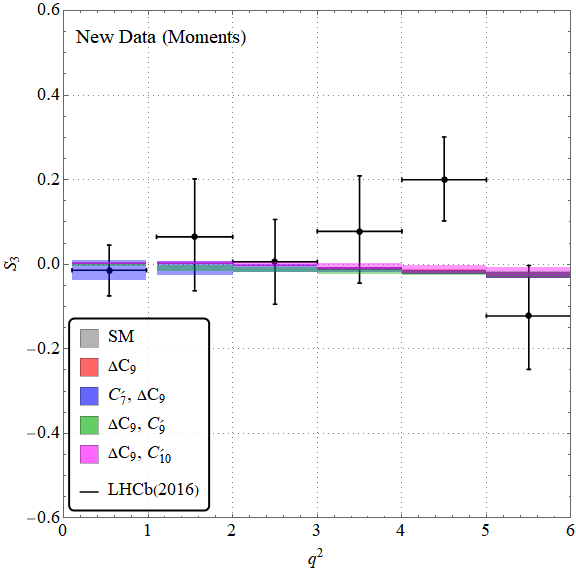}
		\label{fig:angobsmomS3}}~
	\subfloat[]{\includegraphics[width=0.24\textwidth]{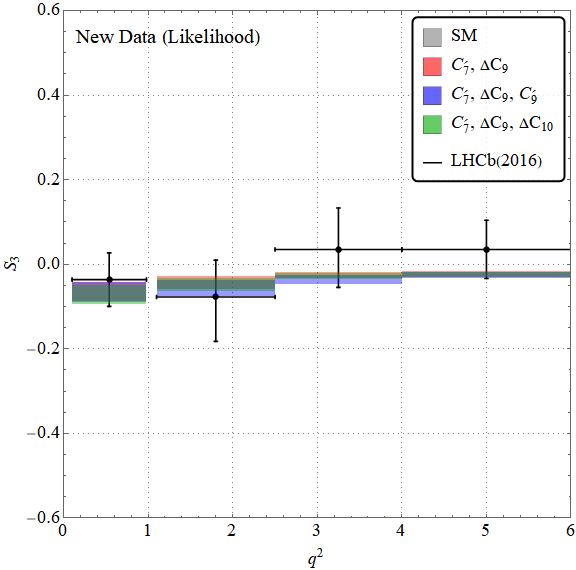}
		\label{fig:angobslikeS3}}~
	\subfloat[]{\includegraphics[width=0.24\textwidth]{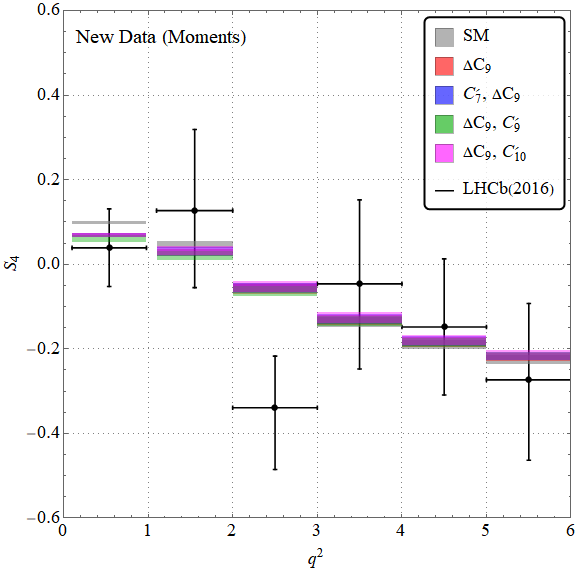}
		\label{fig:angobsmomS4}}~
	\subfloat[]{\includegraphics[width=0.24\textwidth]{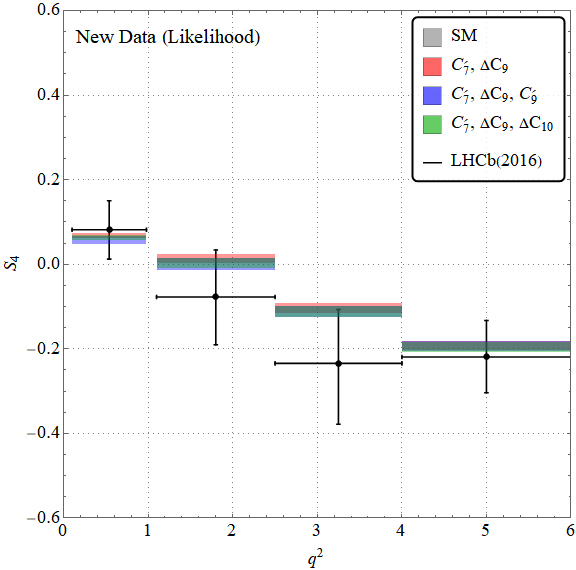}
		\label{fig:angobslikeS4}}\\
	\subfloat[]{\includegraphics[width=0.24\textwidth]{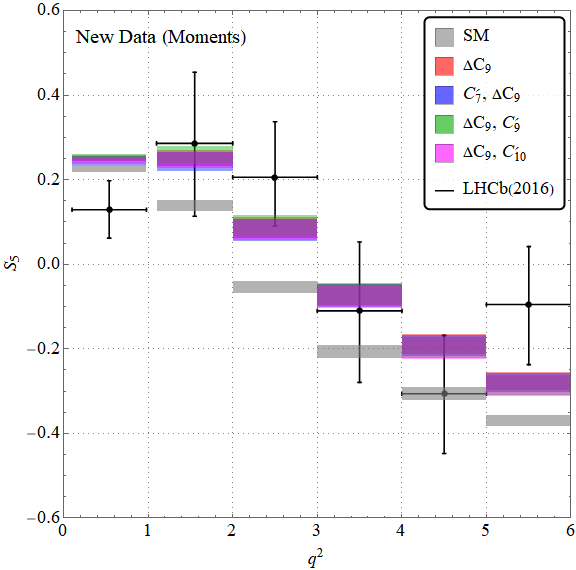}
		\label{fig:angobsmomS5}}~
	\subfloat[]{\includegraphics[width=0.24\textwidth]{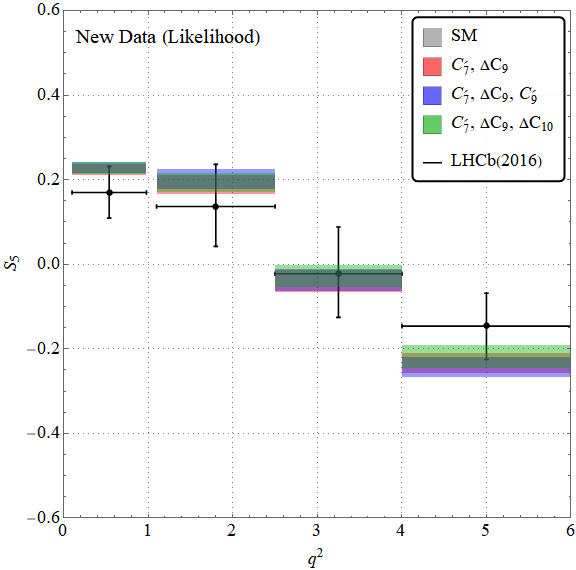}
		\label{fig:angobslikeS5}}
\end{figure*}

\begin{figure*}[htbp]
	\caption{\small Comparison of the optimized angular observables obtained in experiment, SM and from our fit results considering all the avilable inputs.}
	\label{fig:angobsopt}
	\centering
	\subfloat[]{\includegraphics[width=0.24\textwidth]{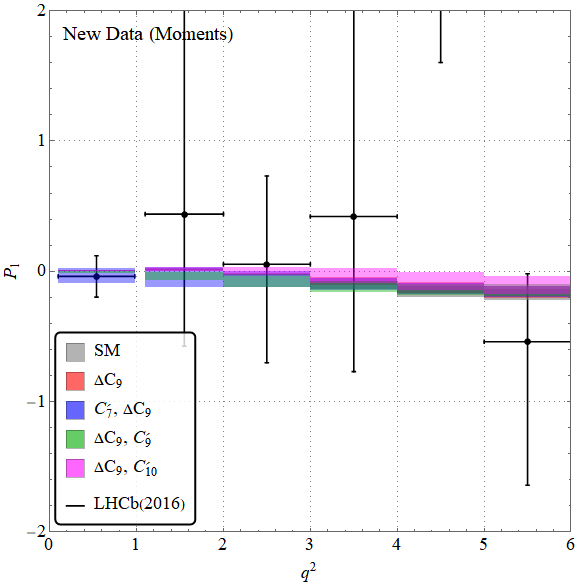}
		\label{fig:angobsmomP1}}~
	\subfloat[]{\includegraphics[width=0.24\textwidth]{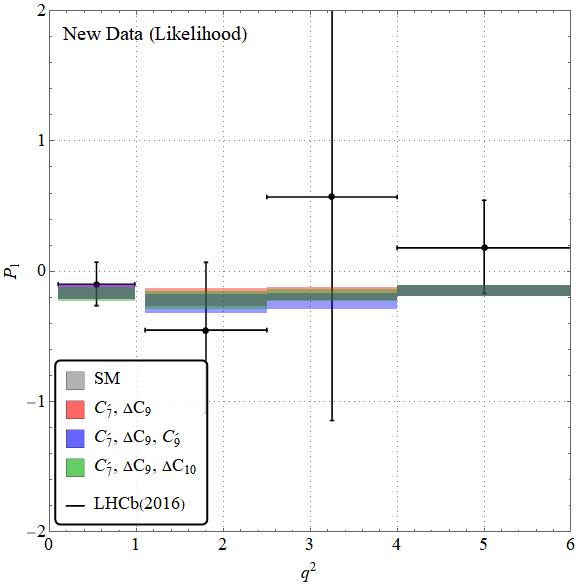}
		\label{fig:angobslikeP1}}~
	\subfloat[]{\includegraphics[width=0.24\textwidth]{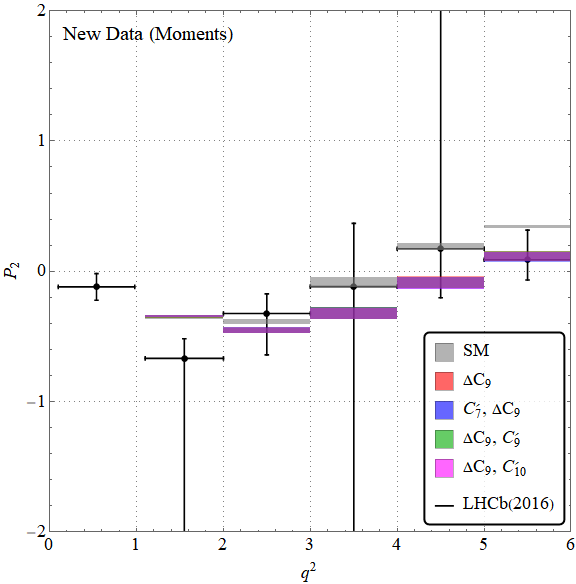}
		\label{fig:angobsmomP2}}~
	\subfloat[]{\includegraphics[width=0.24\textwidth]{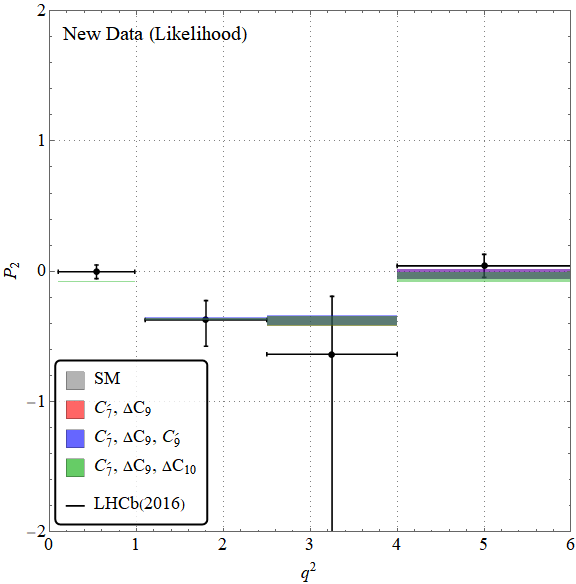}
		\label{fig:angobslikeP2}}\\
	\subfloat[]{\includegraphics[width=0.24\textwidth]{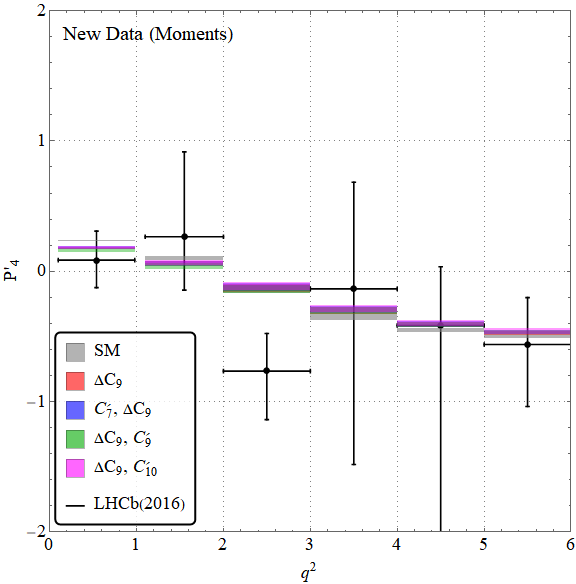}
		\label{fig:angobsmomP4pr}}~
	\subfloat[]{\includegraphics[width=0.24\textwidth]{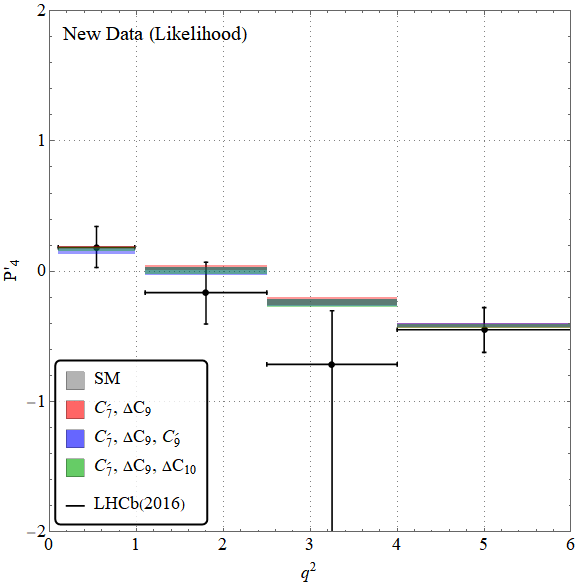}
		\label{fig:angobslikeP4pr}}~
	\subfloat[]{\includegraphics[width=0.24\textwidth]{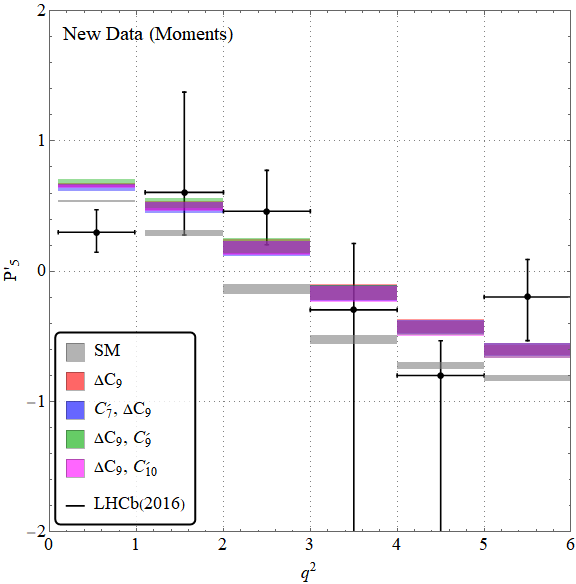}
		\label{fig:angobsmomP5pr}}~
	\subfloat[]{\includegraphics[width=0.24\textwidth]{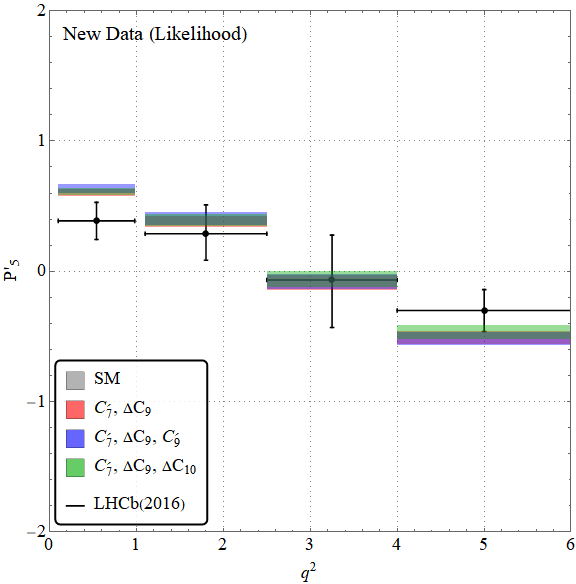}
		\label{fig:angobslikeP5pr}}
\end{figure*}

\subsubsection{Criteria from Information Theory:}\label{sec:methodAIC}
In addition to the extreme computational cost demanded by cross-validation methods, especially LOOCV, its applicability is questionable  to very small sample sizes \cite{BELEITES,Varoquaux}. Due to this reason, in our earlier works \cite{Bhattacharya:2016zcw,Bhattacharya:2018kig}, we have shown the importance and use of the information-theoretic criterion `AIC'\cite{akaike} and its second order variant `AIC$_c$'\cite{sugiura78}. It has been shown that minimizing AIC is asymptotically equivalent to cross-validation\cite{shibata}. For a detailed discussion on AIC$_c$, we point the reader to those papers and references therein. Here, let us reiterate the main important aspects of AIC$_c$ in with respect to model selection in the present work:
\paragraph{AIC$_c$:} If the full reality or truth is noted as $f$ and an approximating model in terms of probability distribution is $g$, then we can define a model selection criterion in terms of the $\chi^2_{min}$ (equivalent to the maximum point of the empirical log-likelihood function) in the parameter space:
\begin{align}
	{\rm AIC}_c = \chi^2_{min} + 2 K + \frac{2 K (K+1)}{n - K -1}\,
	\label{aicc}
\end{align}
where $n$ is the number of data points and $K$ is the number of estimable parameters\footnote{A more preferable way of calculating `number of estimable parameters' is to calculate the $p$-value of the fit from toy Monte-Carlo (MC) method. Under the assumption that the fit-statistic follows a $\chi^2$ distribution, this can give us the number of degrees of freedom, and thus the number of estimable parameters. Still, as we need the differences between the AIC$_c$ values instead of the absolute ones, the naive way of parameter counting works fine.}.

\paragraph{$w^{\Delta\text{AIC}_c}_i$:} The model which is the `closest' to the unknown reality generating the data should have the smallest value of AIC$_c$ among the considered models. Simple differences of them ($\Delta^{AIC}_i = {\rm AIC}^i_c - {\rm AIC}^{min}_c$) estimate the relative expected information loss between $f$ and $g_i$ allowing comparison and ranking of candidate models in increasing order of $\Delta^{AIC}_i$. Generally, the level of empirical support in favor of $g_i$ is considered substantial when $\Delta$AIC$_c$ is between 0 and 2 ($\Delta$AIC$_c \leq 4$ is considered to be a conservative and loose bound). We can also quantify the weight of evidence in favor of model $i$ by defining a set  of positive ``Akaike weights":
\begin{align}
	w^{\Delta\text{AIC}_c}_i = \frac{e^{(-\Delta^{AIC}_i / 2)}}{\sum_{r = 1}^R e^{(-\Delta^{AIC}_r / 2)}}\,,
\end{align}
adding up to $1$ \cite{Burnham}. As these depend on the entire set, adding or dropping a model during an analysis requires re-computation for all models in the new set.

In the present analysis, we have a unique opportunity to not only test the relative capability of MSE from cross-validation and $w^{\Delta\text{AIC}_c}_i$, but also the validity of the empirical rule of selecting models with $\Delta\text{AIC}_c \leq 2$. To that end, we first select a large number of competing models by using the conservative limit of $\Delta\text{AIC}_c \leq 4$, and then distribute them in the plane of MSE$_{\text{X-val}}$ vs. $w^{\Delta\text{AIC}_c}$ and check how they are clustered. Models with a low value of MSE$_{\text{X-val}}$ and a high value of $w^{\Delta\text{AIC}_c}$ are the undoubtedly the best ones to explain the data.

\begin{figure*}[htbp]
	\caption{\small Predicted values, and the correlation of different observables in a few selected scenarios.}
	\label{fig:corpltrad}
	\centering
	\subfloat[]{\includegraphics[width=0.32\textwidth]{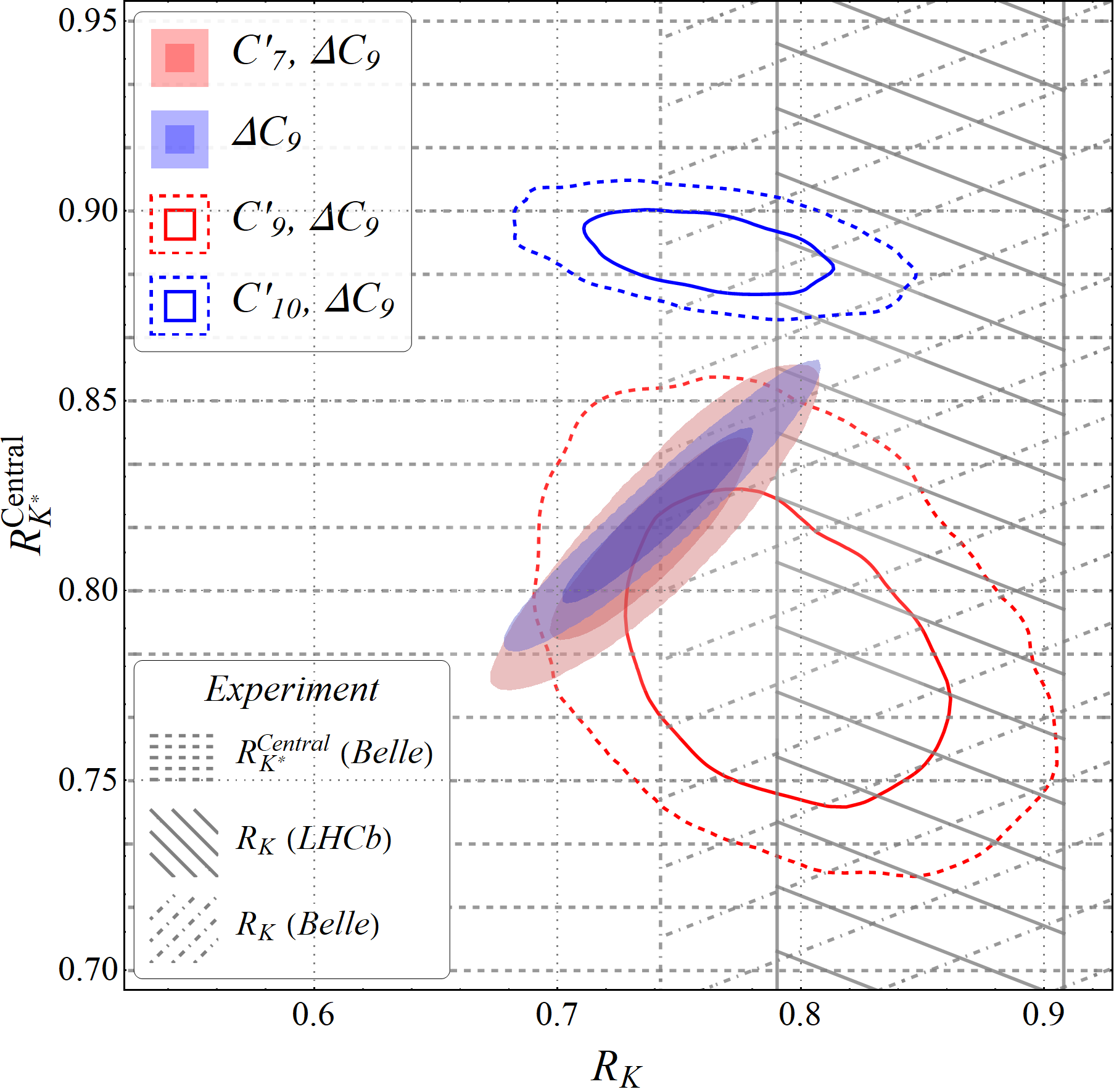}\label{fig:rkrkstcencor}}~
	\subfloat[]{\includegraphics[width=0.32\textwidth]{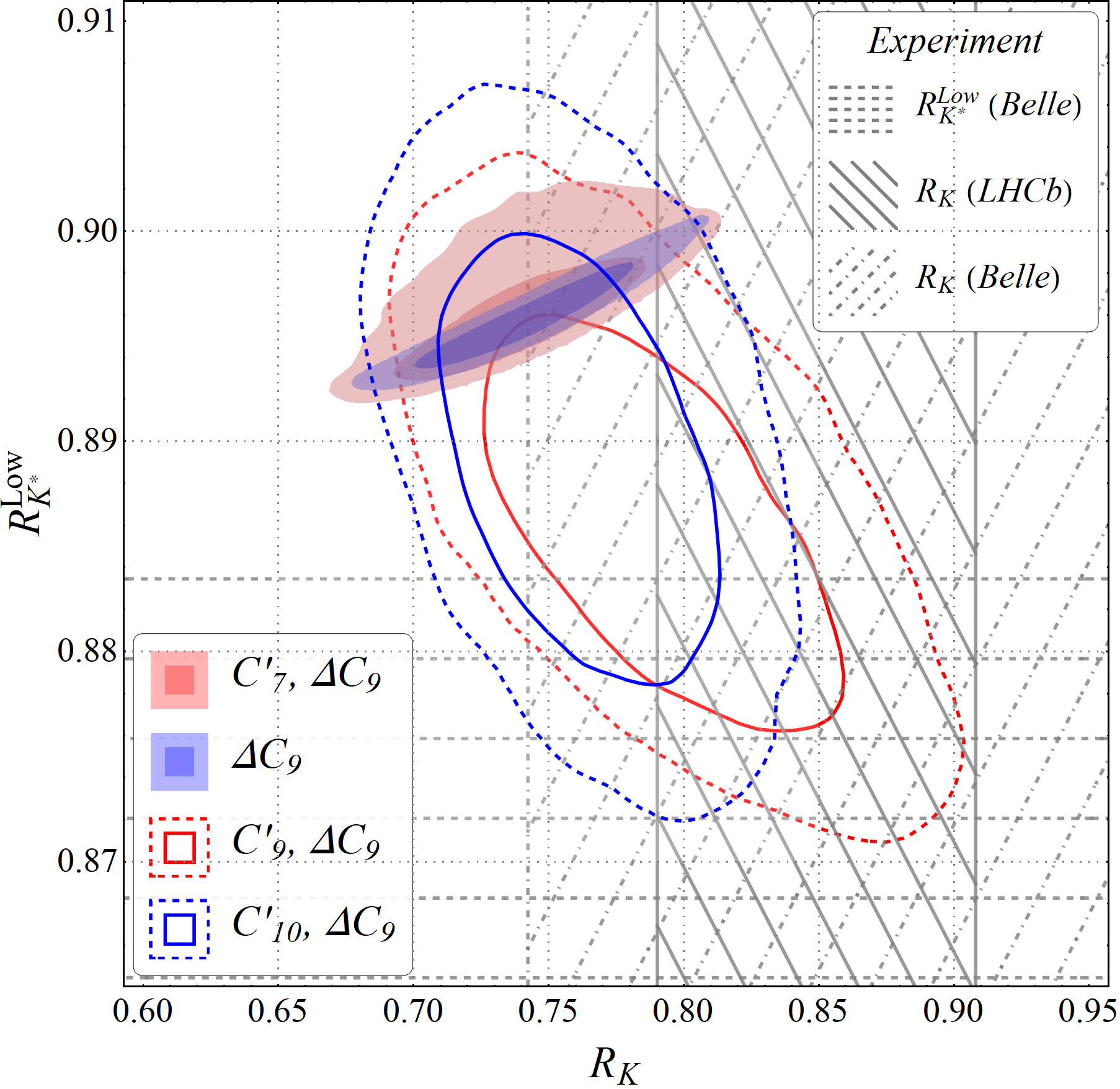}\label{fig:rkrkstLowcor}}~
	\subfloat[]{\includegraphics[width=0.32\textwidth]{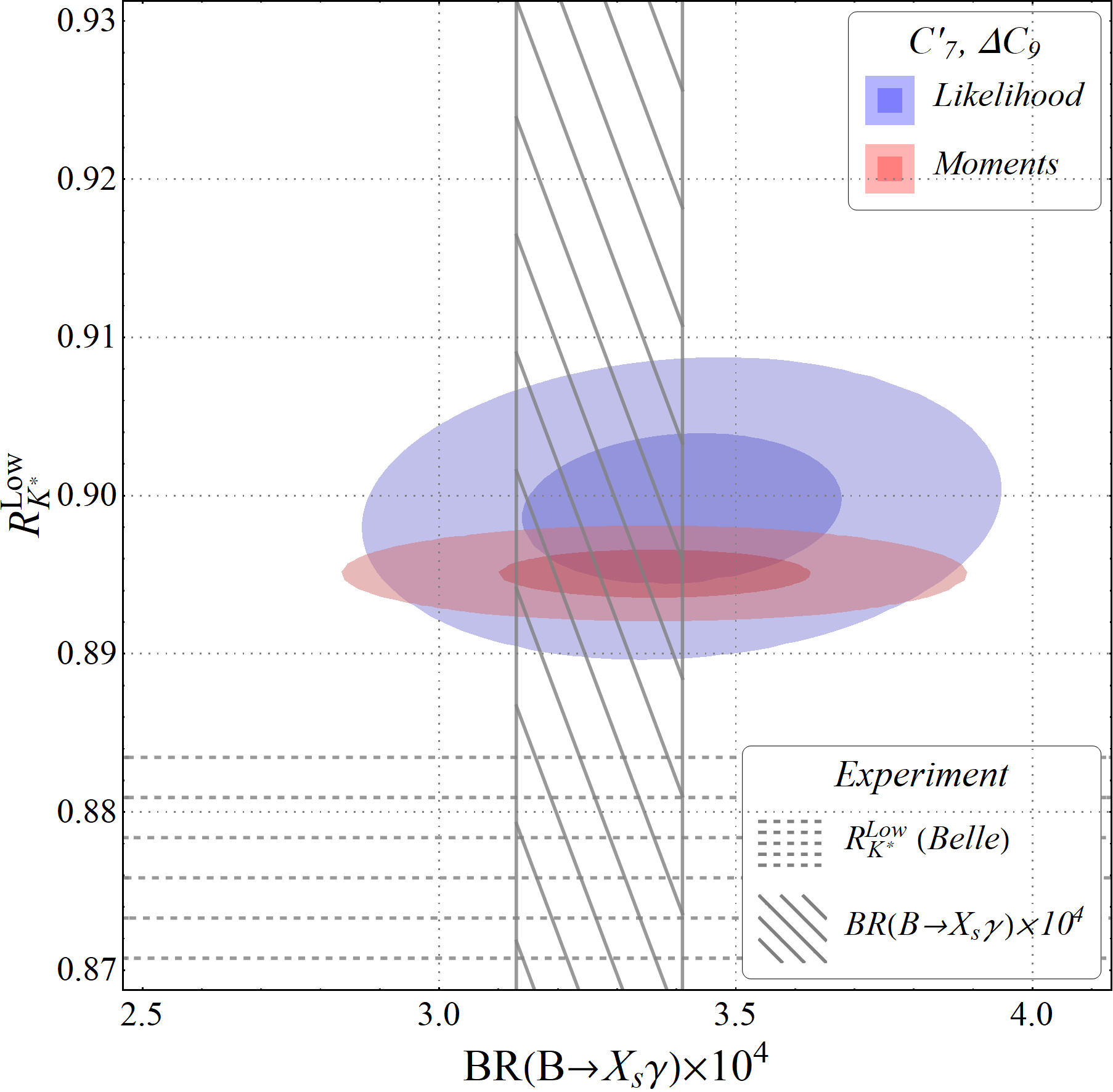}\label{fig:b2xsrkst}}\\
	\subfloat[]{\includegraphics[width=0.32\textwidth]{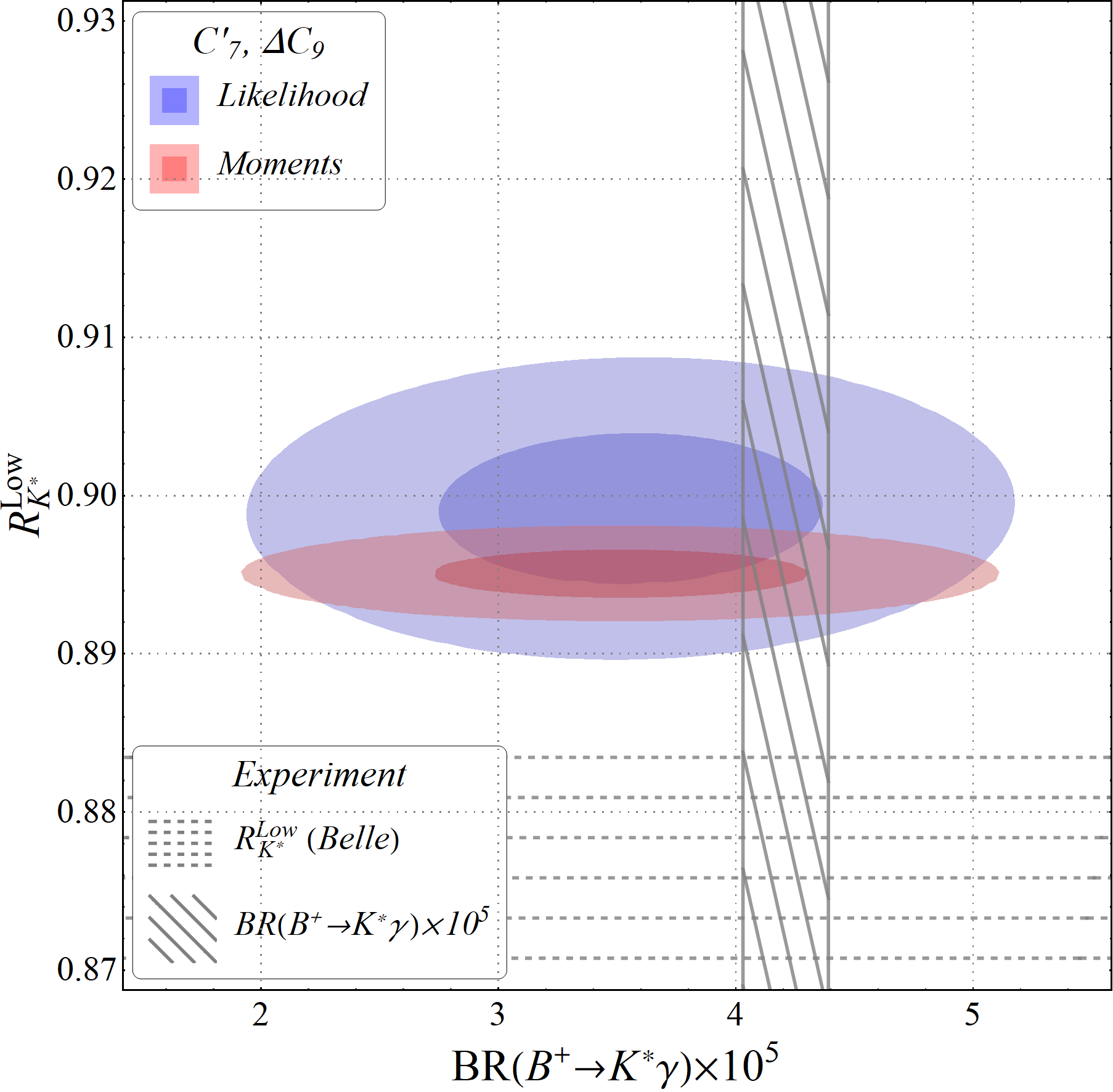}\label{fig:Bp2kgrkst}}~
	\subfloat[]{\includegraphics[width=0.32\textwidth]{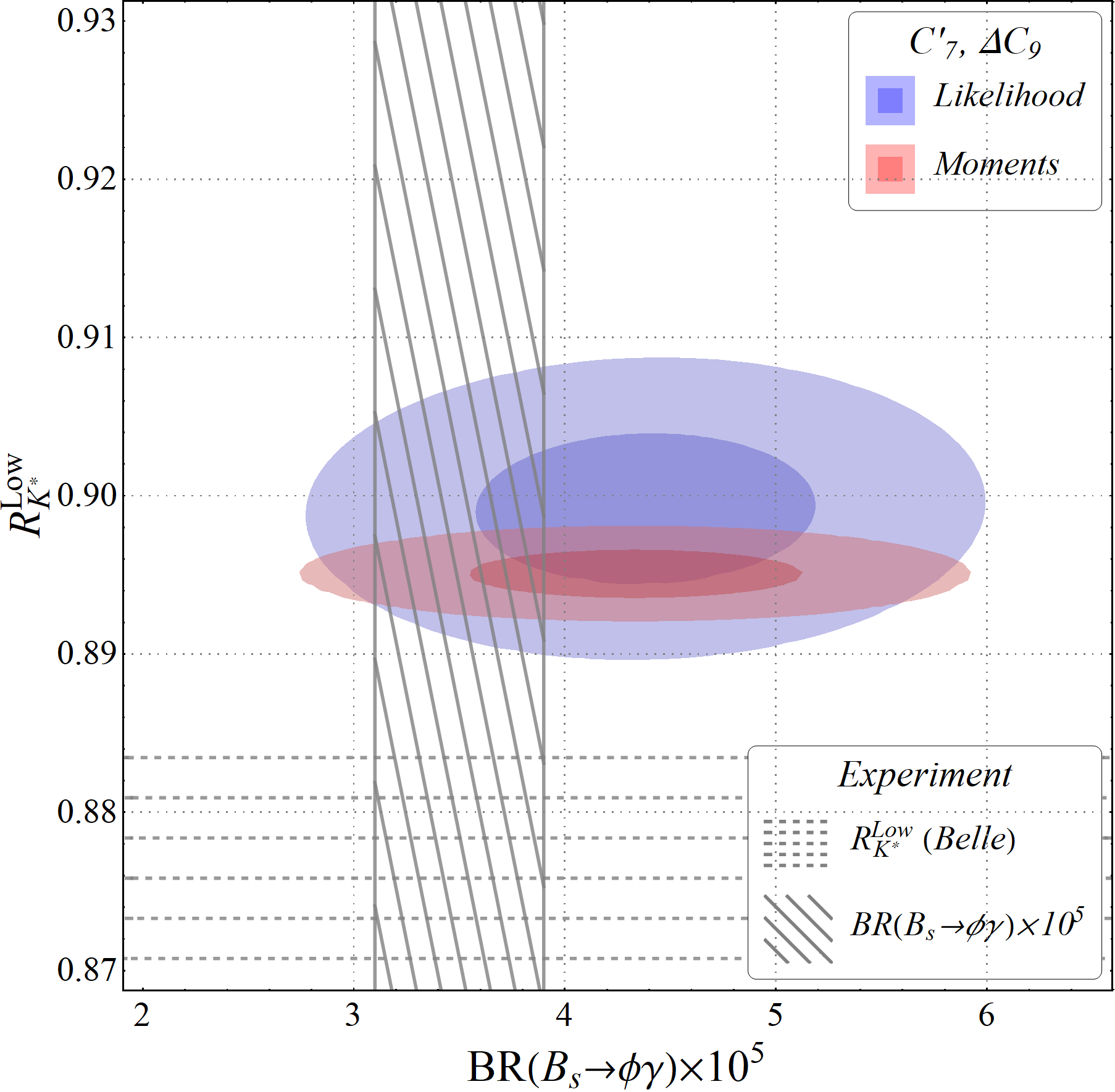}\label{fig:Bp2phirkst}}~
	\subfloat[]{\includegraphics[width=0.32\textwidth]{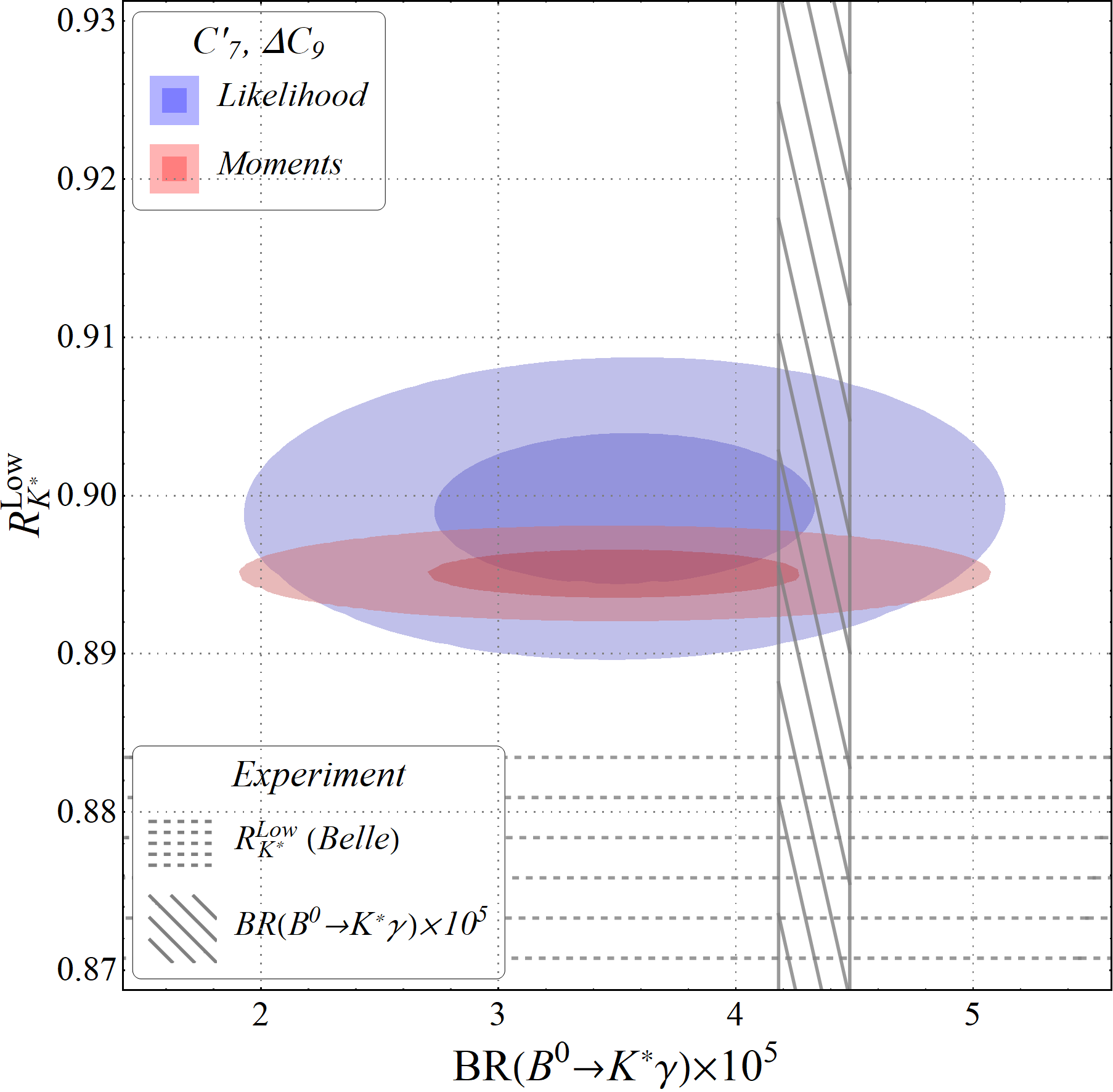}\label{fig:Bz2kgrkst}}~
\end{figure*}

\begin{figure*}[t]
\caption{\small $q^2$ distributions of a few angular observables.}
\label{fig:q2dist}
\centering
\subfloat[]{\includegraphics[width=0.3\textwidth]{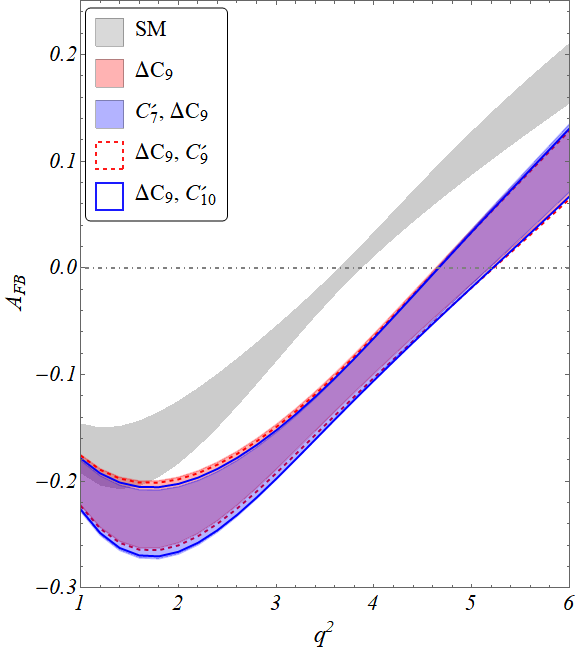}\label{fig:AFBq2plot}}~
\subfloat[]{\includegraphics[width=0.3\textwidth]{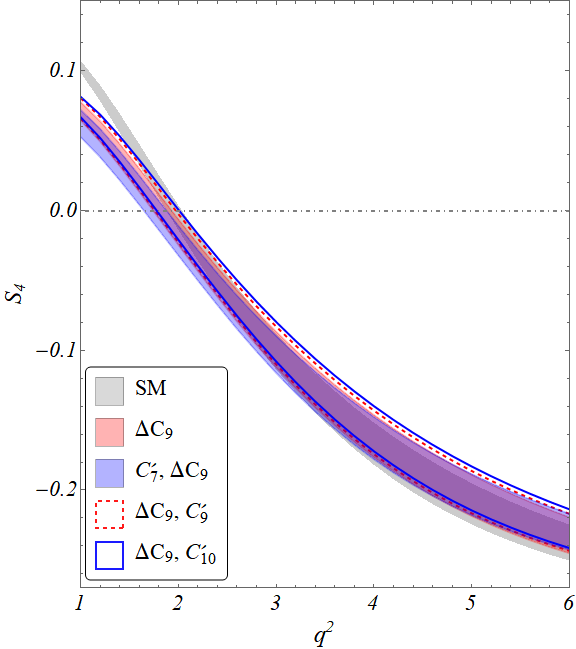}\label{fig:S4q2plot}}~
\subfloat[]{\includegraphics[width=0.3\textwidth]{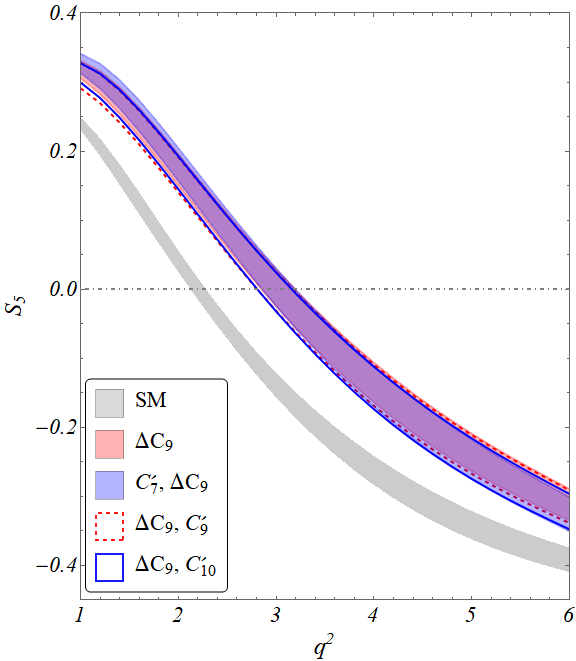}\label{fig:S5q2plot}}
\end{figure*}

%%%%%%%%%%%%%%%%%%%%%%%%%%%%%%%%%%%%%%%%%%%%%%%%%
\section{Results}\label{sec:res}
\subsection{Model Selection}\label{modelselection}
%%%%%%%%%%%%%%%%%%%%%%%%%%%%%%%%%%%%%%%%%%%%%%%%% 

As explained in section \ref{sec:method}, we perform the fit for four different sets. After applying the normality check (as explained in sec. \ref{sec:methodOptim}) on all the 511 models (thus ensuring only valid fits remain in our data-set), we pick out the large set of models with $\Delta\text{AIC}_c \leq 4$. The list of models, thus selected, are shown in figures \ref{fig:ModselNew} and \ref{fig:ModselOld}, which are based on the analysis of all the available data sets given in sec \ref{sec:backexp}.  Each point in these figures represents a `selected' model (i.e. a model for which $\Delta\text{AIC}_c \leq 4$). Among these, the indices (labels for points) for models selected by $\Delta\text{AIC}_c \leq 2$ are framed. As is evident from the figure (and explained earlier in section \ref{sec:methodXval}), the lower the MSE$_{\text{X-val}}$, the better the model. One can clearly see three separate clusters depending only on MSE$_{\text{X-val}}$, and we can safely label the lowest one as the cluster of the best models (from MSE$_{\text{X-val}}$) and discard the rest. Similarly, there are three clusters in the $w^{\Delta\text{AIC}_c}$ direction as well, where the cluster with the largest $w^{\Delta\text{AIC}_c}$ value contains only one model. All models with $\Delta\text{AIC}_c \leq 2$ lie in the two rightmost clusters. Following section \ref{sec:methodAIC}, we know that a model with a larger value of $w^{\Delta\text{AIC}_c}$ is more probable to explain the data. So, we put a commensurate bound on $w^{\Delta\text{AIC}_c}$. We note that out of the various possible combinations, only a few are `selected' by the combined criteria of MSE$_{\text{X-val}}$ and $\text{AIC}_c$. 

Figures \ref{fig:ModselNewMom} and \ref{fig:ModselNewLike} respectively compare the selected models in the analyses of moments and likelihood data on angular observables along with the new data on $R_{K^{(*)}}$. Similar comparisons are done with the old data on $R_{K^{(*)}}$ in figures \ref{fig:ModselOldMom} and \ref{fig:ModselOldLike}, respectively. We note that apart from the two-operator scenario $[{\cal O}^{\prime}_7, {\cal O}_9]$, the likelihood and moments data pick up very different combinations of operators after model selection. The likelihood data prefer scenarios with more operators than those that are required for moments data. This could be since most of the observables which are determined from an unbinned maximum likelihood fit are consistent with their respective SM predictions. A large value of a single WC may lead to a discrepancy between the measured values and the corresponding SM predictions. It is thus preferable to have simultaneous contributions from different operators. The same observation can be made from the comparison of the quality of fits with moments and likelihood data which are shown in table \ref{tab:NewMom} and \ref{tab:NewLike}, respectively. The values of the reduced $\chi^2$ and corresponding $p$-values, indicating the quality of the fit and the relative quality of the various models (scenarios) for a given set of data are provided in each corresponding table. We note that in general, the quality of the fits is better in the analysis with moments data compared to that with likelihood data. It can be seen that the ranking of the models depends not only on the quality of fit but also on the penalty function defined in eq. \ref{aicc}. This is one of the major advantage of AICc over data-fitting. Results of a similar analysis after dropping $R_{K^{(*)}}$ are presented in figure \ref{fig:ModselNewwoLFUV} in Appendix~\ref{app2}. We note that after dropping the LFUV observables from the fits for the respective cases, the multi-operator scenarios can not survive the competition from those scenarios with relatively less number of operators.   

As we know, the data from Belle has significant errors compared to those given by LHCb. To check whether this data has any influence over the selected scenarios, we have performed an analysis where we drop the $R_{K^{(*)}}$ data from Belle only. We note that the selected scenarios are the same as that given in figure \ref{fig:ModselNew}. Also, the quality of fits or the respective $p$-values do not change. The reason is straightforward: the Belle data being consistent with LHCb, and the extracted $R(K)$ and $R(K^*)$ in the selected models in figure \ref{fig:ModselNew} are consistent with both LHCb and Belle. Hence, the conclusions remain unchanged even after dropping Belle data. 

There are a few other relevant observations here as well. For example, both in the case of the `Old' and `New' data sets, the majority of the selected scenarios are with two, three, or four operators and all of them contain the operator corresponding to the WC $\Delta C_9$. Also, the only scenario with a single operator selected by the `New' data set (in particular the `Moments' dataset) has the WC $\Delta C_9$ (model 2). Model 2 is clearly the better option for all the datasets. However, for the `New Moments' dataset, scenarios with two operators corresponding to the WCs [$\Delta C_9$, $C'_9$] (model 18), [$\Delta C_9$, $C'_7$] (model 10), and [$\Delta C_9$, $C'_{10}$] (model 20) are other possible choices. An important feature of the data is that the `Likelihood' and `Moments' datasets pick completely different types of models. For likelihood, an explanation of the observed data with a single operator is less plausible.

We note that some models have low MSE scores but are rejected by AICc. Although leave-one-out-cross-validation (LOO-CV) is asymptotically equivalent to AIC, there are differences between the two. Theoretical considerations aside, AIC is just likelihood penalized by the degrees of freedom. Evidently, AIC accounts for uncertainty in the data (-2Log(L)) and assumes that more parameters lead to a higher risk of overfitting (2k). Cross-validation (CV) just looks at the test set performance of the model, with no further assumptions. There is no explicit measure of model complexity, unlike AIC. Clearly, AIC penalizes model complexity more than CV. The accepted practice in the literature is that if one cares mostly about making predictions and assumes that the test set(s) to be reasonably similar to the validation sets, one should go for CV (only with a large number of data) \cite{arlot2010}.

Following the reasoning in the paragraph above, it seems clear \textit{a priori} that models with low MSE but not selected by AICc may have more parameters. In our analysis, the models are organized in such a way that more complex models with a larger number of parameters have a higher index. A quick examination of the plots in figures \ref{fig:ModselNew} and \ref{fig:ModselOld} validates this assumption, as the left sides of the plots are littered with models of higher indices.

On another note, as the title suggests, this study clearly finds that under the simultaneous application of both AIC and CV, the models cluster in such a way that lets us carry out a tighter selection of models than would have been possible by the use of any one of these criteria. So the fact that some of the selected models by CV are further discarded by AIC due to relative complexity is actually quite a non-trivial finding.

The best fit values of the new WCs with the corresponding errors for the selected models in Figure \ref{fig:ModselNew} are given in tables  \ref{tab:NewMom} and \ref{tab:NewLike}. We note that for all the fit results, the allowed values of $\Delta C_9$ are negative and greater than one. Other WCs appearing along with $\Delta C_9$ as probable solutions have values $<< 1\ \text{or}\ -1$. The selected models with $C_7^{\prime}$ as one of the WCs will impact radiative decays. Hence, along with the allowed values for the new WCs, we have shown that in all the relevant models the branching fractions for $B\to K^*\gamma$, $B\to \phi\gamma$ and $B\to X_s\gamma$ are consistent with their respective measurements within 1$\sigma$ confidence interval in the relevant tables. The fitted values of the selected WCs remain almost unchanged in the analysis obtained after dropping $R_{K^{(*)}}$. The allowed parameter spaces of the respective WC's are similar in the analyses with old data-set which are given in the Appendix in tables \ref{tab:OldMom}, \ref{tab:OldLike}. However, note that the fit qualities are relatively better in the analysis with old data which is an indicator of relatively poor NP-description of the new LFUV observables along with the angular observables.    

We compare our fit results for various angular observables with their respective measured values and the SM predictions in figures \ref{fig:angobscp} and \ref{fig:angobsopt} for a set of selected models. We see that the data in a few bins are inconsistent with their corresponding SM predictions, particularly in the data set for the `Moment' analysis. For the `Likelihood' analysis, most of the measured values in different bins are consistent with their SM predictions, albeit with exceptions. Assuming the observed discrepancies are due to NP, most of them can be resolved by our selected models. For the likelihood data, our analysis shows that the angular observables obtained from the selected models are fully consistent with their SM predictions as well as measured values. On the contrary, the fitted angular observables in the selected models in the moment-data analysis are shifted from their respective SM predictions in a few bins. Moreover, in some of those, the fit results shift from their measured values as well. One can also notice the correlations among various angular observables in the presence of different new operators from these figures. For example, $A_{FB}$ and $F_L$ are positively correlated, $F_L$ and $S_5$ are negatively correlated, etc. For some of the bins of $F_L$, the values predicted by the selected models are shifted from respective measured values. Similar plots obtained in our analysis after dropping $R_{K^{(*)}}$ is provided in fig. \ref{fig:angobswolfuv} in the appendix.

In figure \ref{fig:corpltrad}, the predicted values of different observables and their respective correlations are shown for a few of the models (only one or two operator scenarios) selected in our analysis. We note that $R^{Central}_{K^*}$ and $R_K$ are correlated differently in different two operator scenarios. However, all of them satisfy the current experimental bounds on these observables. Interestingly, the single operator scenario with $\Delta C_9$ and the two operator scenario with $[\Delta C_9, C'_7]$ are unable to satisfy the current experimental bounds on $R^{Low}_{K^*}$. In these scenarios, satisfying the experimental bounds on $R_K$, it is hard to get a value of $R^{Low}_{K^*}$ below $0.85$. However, the scenarios with $[\Delta C_9, C'_9]$ and  $[\Delta C_9, C'_{10}]$ as WCs can explain the observed data on $R_K$ and $R_{K^*}$. From the respective correlations, one can also see that $R^{Low}_{K^*}$ and
 $R_K$ are negatively correlated and values of $R^{Low}_{K^*}$ lower than $0.88$ prefers $R_K > 0.8$. In figures \ref{fig:b2xsrkst}, \ref{fig:Bp2kgrkst}, \ref{fig:Bp2phirkst}, and \ref{fig:Bz2kgrkst}, we provide the predicted values of the branching fractions of different radiative decays and their correlations with $R^{Low}_{K^*}$ in scenarios corresponding to the WCs $[\Delta C_9, C'_7]$. We note that there are no noticeable correlations between these branching fractions and $R^{Low}_{K^*}$, or for that matter with $R_K$ and $R^{Central}_{K^*}$.

The $q^2$ distributions and the zero crossing of the angular observables $A_{FB}$, $S_4$ and $S_5$ corresponding to the SM and the selected models are shown in fig. \ref{fig:q2dist}. We have noted discrepancies between the $q^2$ distributions for the SM and the selected models in the cases of $A_{FB}$ and $S_5$ while $S_4(q^2)$ in the selected models are fully consistent with the SM. The $q^2$ distributions of these angular observables in different selected models overlap with each other. Hence it is hard to discriminate models from these distributions. In the Appendix in figure \ref{fig:zerocrossing}, we compare the values of $q^2$ at the zero crossing ($q_0^2$) between SM, our selected models and the measured values for the above mentioned observables.

The uncertainties of the observables, in terms of the parameters, are obtained by two different techniques for figures  \ref{fig:corpltrad} and \ref{fig:q2dist}. 
For figure \ref{fig:corpltrad}, which contains spaces for any 2 observables, we take, from our fit-results, all the information (best-fit values, uncertainties, and correlations) of only the NP parameters occurring in those observables, create a multivariate distribution out of those, and sample a large number of points ($\sim 5000$) from that distribution using Monte-Carlo. For each of those points, we get sets of values of the observables, which in turn lets us draw the $1 ~(39.35\%)$ and $2 ~(63.21\%) ~\sigma$ contours (for 2-observable plots) from these data-sets.

For figure \ref{fig:q2dist}, we are dealing with only one observable at a time and that too, for a specific value of $q^2$. Taking the $1 \sigma ~(68\%)$ confidence levels for the marginal likelihoods around the central (for SM/nuisance parameters) or best-fit values (for NP) of the parameters as uncertainties and the corresponding correlations between them, we propagate the uncertainties to get the central values and uncertainties of the observables at some $q^2$. Doing this for many points over the allowed $q^2$ range and interpolating between them gives us the plots of $q^2$ distribution.

We should also mention here that the scenario $\Delta C_9 = -\Delta C_{10}$ that arises, for example, in some Leptoquark models (among other model-dependent origins) does not pass the criteria of $\Delta\text{AIC}_c \leq 4$. We hence do not display or discuss this scenario.

%%%%%%%%%%%%%%%%%%%%%%%%%%%%%%%%%%%%%%%%%%%%%%%%%%%%
\section{Summary}\label{sec:summ}
%%%%%%%%%%%%%%%%%%%%%%%%%%%%%%%%%%%%%%%%%%%%%%%%%%%%

In this article, we have analyzed the semileptonic $b\to s \ell\ell$ decays in a model independent framework with the relevant dimension six effective operators invariant under the strong and electromagnetic gauge groups. Our chosen set of operators does not include the four quark operators, chromomagnetic operators and tensor operators. Different possible combinations of all the effective operators have been considered, and following the statistical tools like cross-validation and the small-sample-corrected Akaike Information Criterion ($AIC_c$), we have found out the combinations which best explain the available data. We have provided separate analyses for the data on angular observables obtained from an unbinned maximum likelihood fit and that due to the principal moments of the angular distribution in $B \to K^*\mu^+\mu^-$ decay. 

Among all the possible combinations, a relatively small number of one, two and three-operator scenarios satisfy the criterion of a selected `best' model. All the selected scenarios contain a left-handed quark current with vector muon coupling as an operator (${\cal O}_9$). This is also the only one-operator scenario that survives the exclusion test in our search for the `best' model(s). We have noted differences between the selected models in the analysis with angular data from likelihood fit and those from the principal moments analysis. The $R_{K^{(*)}}$, along with the angular observables associated with $B \to K^*\mu^+\mu^-$ decays, have played an important role in this selection. In the analysis with the new data on $R_{K^{(*)}}$, the scenarios with three, four and five operators are selected. This could be due to the fact that the tension between the updated measured values of $R_K$ and $R_{K^*}$ and their respective SM predictions have reduced in comparison to that for their old experimental measurements. Therefore, in order to explain all the data simultaneously, simultaneous contributions from different operators are required. We have noticed changes in the selected scenarios when we drop $R_{K^{(*)}}$ from the list of inputs. We have performed the analysis with and without the 2019 updates on $R_{K^{(*)}}$ from Belle and have compared them. We have noticed changes in the allowed parameter spaces for the Wilson coefficients of the selected scenarios. 

We have compared our fit results for the angular observables with the corresponding SM predictions and the measured values in different bins for a few selected models. While our fit results are fully consistent with both the measured values and the respective SM predictions in the analysis of likelihood data, in the analysis of moment data, there are discrepancies between our fitted results and the respective SM predictions and the measured values in some bins.
For some of the selected scenarios, we have studied the correlations between different observables, which show that the operator ${\cal O}_9$  and the combination of ${\cal O}_9$ and ${\cal O}'_7$ (flavor changing electromagnetic dipole operator) can not explain all the available data on $R_{K^{(*)}}$ simultaneously. In particular, they have difficulty in explaining the observed results of $R_{K^*}$ in the low $q^2$ bin ($q^2 \in \ [0.045, 1.1]\ {\it GeV}^2$).

We have studied the NP effects in $R_{K^{(*)}}$ only, and noticed that the operator with a left-handed quark current with an axial-vector muon coupling (${\cal O}_{10}$) is the only one-operator scenario that can explain the data. Also, the parameter space for the corresponding Wilson-coefficient $\Delta C_{10}$, allowed by $R_{K^{(*)}}$, is tightly constrained by the measured values of $Br(B_s \to \mu^+\mu^-)$. However, there are a few two-operator scenarios which have the potential to explain the current observation. Those operators are obtained from possible combinations of  ${\cal O}_9$, ${\cal O}_{10}$ and operators like right-handed quark current with vector or axial-vector muon couplings  (${\cal O}'_9$, ${\cal O}'_{10}$). In the two-operator scenarios, the allowed parameter spaces for $\Delta C_{10}$ or/and $C'_{10}$ can comfortably explain the observed data on $Br(B_s \to \mu^+\mu^-)$.

{\bf Note}:The complete list of all models used in this analysis is too long to include in this draft. We have added an ancillary file named ``\texttt{models.json}" along with this draft (to be found within the `arXiv' source file). This file contains all combinations of WCs, relating them with their corresponding indices in our analysis. 

\begin{acknowledgments}
	This project, S.N., and S.K.P. are supported by the Science and Engineering Research Board, Govt. of India, under the grant CRG/2018/001260.
\end{acknowledgments}

\appendix

\begin{table*}[t]
	%\centering
	\small
	\caption{Fit-qualities, model selection criteria, parameter estimates and effects on radiative decays for the `best' selected models with the `Old' data-set, with the `Moments' estimate of the angular observables. Selected models are obtained from fig. \ref{fig:ModselOldMom}. Last four columns showcase the deviations (in units of `$\sigma$') between the experimental value of the radiative decays and the corresponding value obtained with the fit results.}
	\begin{tabular}{*{10}{c}}
		\hline
		\hline
		$\text{Model}$  &  $\left.\chi _{\text{Min}}^2\right/$  &  $\text{p-val}$  &  $\omega ^{\text{$\Delta $AIC}_c}$  &  $\text{MSE}_{X-\text{val}}$  &  $\text{Parameter}$  &  \multicolumn{4}{c|}{Deviation in $\sigma$} \\
		\cline{7-10}
		$\text{Index}$  &  $\text{DOF}$  &  $\text{($\%$)}$  &  $\text{($\%$)}$  &  $\text{}$  &  $\text{Values}$  &  $\text{B$\to $}X_s\gamma$  &  $B^+\to K^*\gamma$  &  $\text{$\Delta $B}^0\to K^*\gamma$  &  $\text{$\Delta $B}_s\to \phi \gamma$  \\
		\hline
		$2$  &  $\text{245.67/254}$  &  $63.5$  &  $5.$  &  $0.918$  &  $\begin{array}{l}
		\Delta C_9\to \text{-1.26$\pm $0.14} \\
		\end{array}$  &  $-$  &  $-$  &  $-$  &  $-$  \\
		\cline{6-6}
		$10$  &  $\text{244.92/253}$  &  $63.1$  &  $2.6$  &  $0.916$  &  $\begin{array}{l}
		C'_7 \to \text{0.013$\pm $0.015} \\
		\Delta C_9\to \text{-1.3$\pm $0.15} \\
		\end{array}$  &  $0.32$  &  $-0.87$  &  $-1.06$  &  $1.22$  \\
		\cline{6-6}
		$19$  &  $\text{245.42/253}$  &  $62.2$  &  $2.$  &  $0.926$  &  $\begin{array}{l}
		\Delta C_9\to \text{-1.22$\pm $0.16} \\
		\Delta C_{10}\to \text{0.061$\pm $0.123} \\
		\end{array}$  &  $-$  &  $-$  &  $-$  &  $-$  \\
		\cline{6-6}
		$21$  &  $\text{245.48/253}$  &  $62.1$  &  $2.$  &  $0.923$  &  $\begin{array}{l}
		\Delta C_9\to \text{-1.27$\pm $0.15} \\
		C_S\to \text{-0.021$\pm $0.026} \\
		\end{array}$  &  $-$  &  $-$  &  $-$  &  $-$  \\
		\cline{6-6}
		$22$  &  $\text{245.51/253}$  &  $62.$  &  $1.9$  &  $0.923$  &  $\begin{array}{l}
		\Delta C_9\to \text{-1.27$\pm $0.15} \\
		C'_S\to \text{0.02$\pm $0.026} \\
		\end{array}$  &  $-$  &  $-$  &  $-$  &  $-$  \\
		\cline{6-6}
		$18$  &  $\text{245.55/253}$  &  $62.$  &  $1.9$  &  $0.915$  &  $\begin{array}{l}
		\Delta C_9\to \text{-1.25$\pm $0.14} \\
		C'_{9}\to \text{0.067$\pm $0.195} \\
		\end{array}$  &  $-$  &  $-$  &  $-$  &  $-$  \\
		\cline{6-6}
		$20$  &  $\text{245.59/253}$  &  $61.9$  &  $1.9$  &  $0.92$  &  $\begin{array}{l}
		\Delta C_9\to \text{-1.26$\pm $0.14} \\
		C'_{10}\to \text{-0.03$\pm $0.109} \\
		\end{array}$  &  $-$  &  $-$  &  $-$  &  $-$  \\
		\hline
		\hline
	\end{tabular}
	\label{tab:OldMom}
\end{table*}
\begin{table*}[htbp]
\small
\caption{Same as table \ref{tab:OldMom}, but with the `Likelihood' estimate of the angular observables. Selected models are obtained from fig. \ref{fig:ModselOldLike}.}
\begin{tabular}{*{10}{c}}
\hline
\hline
$\text{Model}$  &  $\left.\chi _{\text{Min}}^2\right/$  &  $\text{p-val}$  &  $\omega ^{\text{$\Delta $AIC}_c}$  &  $\text{MSE}_{X-\text{val}}$  &  $\text{Parameter}$  &  \multicolumn{4}{c}{Deviation in $\sigma$} \\
\cline{7-10}
$\text{Index}$  &  $\text{DOF}$  &  $\text{($\%$)}$  &  $\text{($\%$)}$  &  $\text{}$  &  $\text{Values}$  &  $\text{B$\to $}X_s\gamma$  &  $B^+\to K^*\gamma$  &  $\text{$\Delta $B}^0\to K^*\gamma$  &  $\text{$\Delta $B}_s\to \phi \gamma$  \\
\hline
$10$  &  $\text{213.78/209}$  &  $39.6$  &  $5.3$  &  $0.973$  &  $\begin{array}{l}
C'_7\to \text{0.028$\pm $0.015} \\
\Delta C_9\to \text{-1.37$\pm $0.14} \\
\end{array}$  &  $0.37$  &  $-0.85$  &  $-1.04$  &  $1.24$  \\
\cline{6-6}
$2$  &  $\text{217.19/210}$  &  $35.2$  &  $2.7$  &  $0.989$  &  $\begin{array}{l}
\Delta C_9\to \text{-1.28$\pm $0.13} \\
\end{array}$  &  $-$  &  $-$  &  $-$  &  $-$  \\
\cline{6-6}
$49$  &  $\text{213.2/208}$  &  $38.8$  &  $2.5$  &  $0.981$  &  $\begin{array}{l}
C'_7\to \text{0.029$\pm $0.015} \\
\Delta C_9\to \text{-1.4$\pm $0.14} \\
C_S\to \text{-0.028$\pm $0.019} \\
\end{array}$  &  $0.38$  &  $-0.85$  &  $-1.04$  &  $1.25$  \\
\cline{6-6}
$50$  &  $\text{213.23/208}$  &  $38.7$  &  $2.5$  &  $0.981$  &  $\begin{array}{l}
C'_7\to \text{0.029$\pm $0.015} \\
\Delta C_9\to \text{-1.4$\pm $0.14} \\
C'_S\to \text{0.028$\pm $0.019} \\
\end{array}$  &  $0.38$  &  $-0.85$  &  $-1.04$  &  $1.25$  \\
\cline{6-6}
$47$  &  $\text{213.65/208}$  &  $37.9$  &  $2.$  &  $0.976$  &  $\begin{array}{l}
C'_7\to \text{0.029$\pm $0.015} \\
\Delta C_9\to \text{-1.39$\pm $0.15} \\
\Delta C_{10}\to \text{-0.042$\pm $0.117} \\
\end{array}$  &  $0.38$  &  $-0.85$  &  $-1.04$  &  $1.25$  \\
\hline
\hline
\end{tabular}
\label{tab:OldLike}
\end{table*}

\section{Best fit values of the new WCs in the analysis with old dataset}
The selected models in the analysis with old datasets are shown in figure \ref{fig:ModselOld}. The best fit values of the corresponding WCs along with their respective errors are given in tables \ref{tab:OldMom} and \ref{tab:OldLike}, respectively.

%\begin{figure*}[t]
%	\caption{\small Same as fig. \ref{fig:ModselNew}, but for the fit we have dropped the Belle data on $R_K$ and $R_{K^*}$.}
%	\label{fig:Modselwobelle}
%	\centering
%	\subfloat[Moments ]{\includegraphics[width=\columnwidth]{modsel_moment_wobelle.png}
%		\label{fig:Modselmomentswobelle}}~~
%	\subfloat[Likelihood ]{\includegraphics[width=\columnwidth]{modsel_likelihood_wobelle.png}
%		\label{fig:ModselLikewobelle}}
%\end{figure*}

\begin{figure*}
	\caption{\small Results similar to fig. \ref{fig:ModselNew} but with $R_K$ and $R_{K^*}$ dropped.}
	\label{fig:ModselNewwoLFUV}
	\centering
	\subfloat[New data (Moments; w/o $R_{K^{(*)}}$)]
	{\includegraphics[width=\columnwidth]{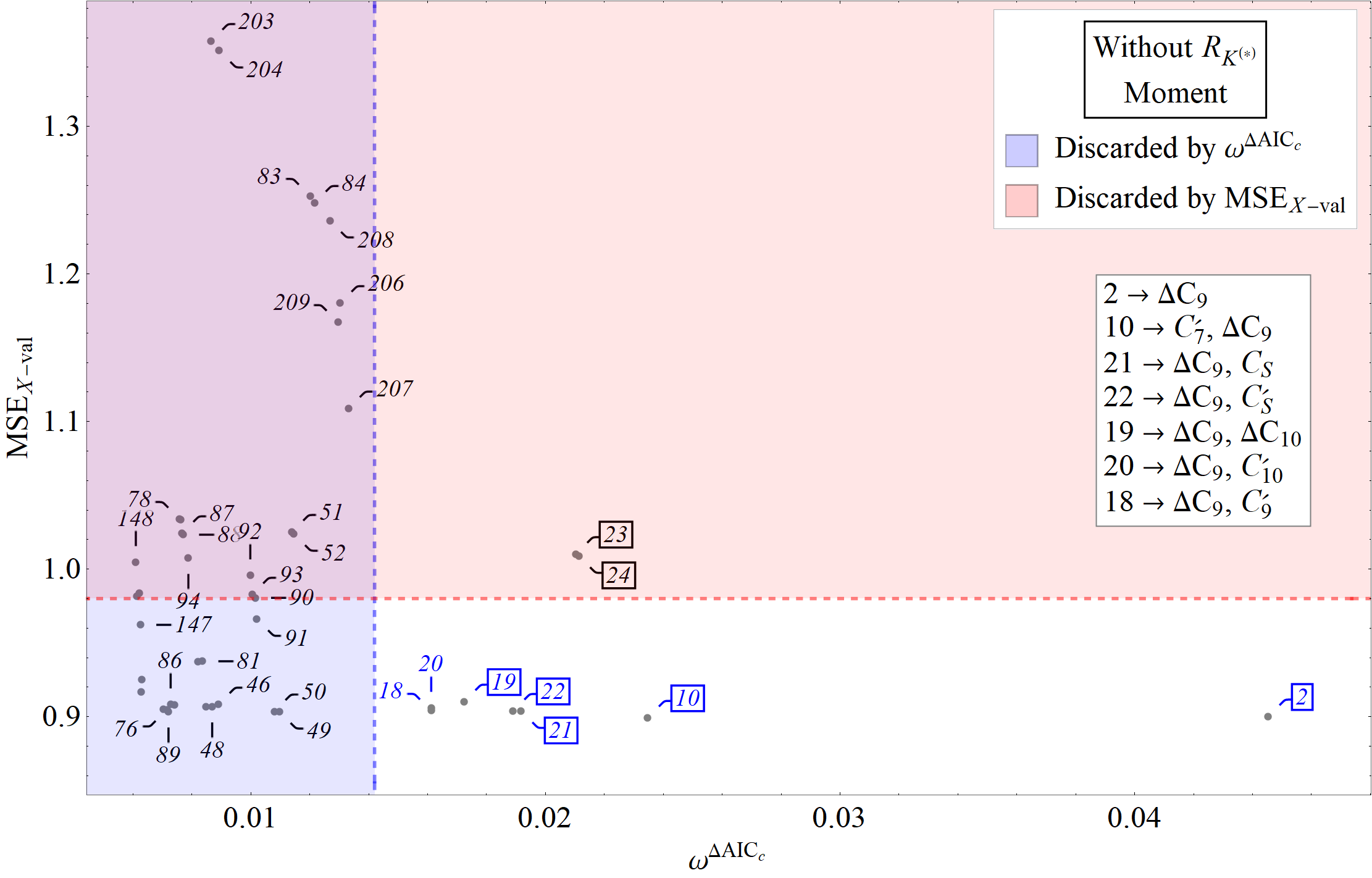}
		\label{fig:ModselNewMomwoLFUV}}~~
	\subfloat[New data (Likelihood; w/o $R_{K^{(*)}}$)]
	{\includegraphics[width=\columnwidth]{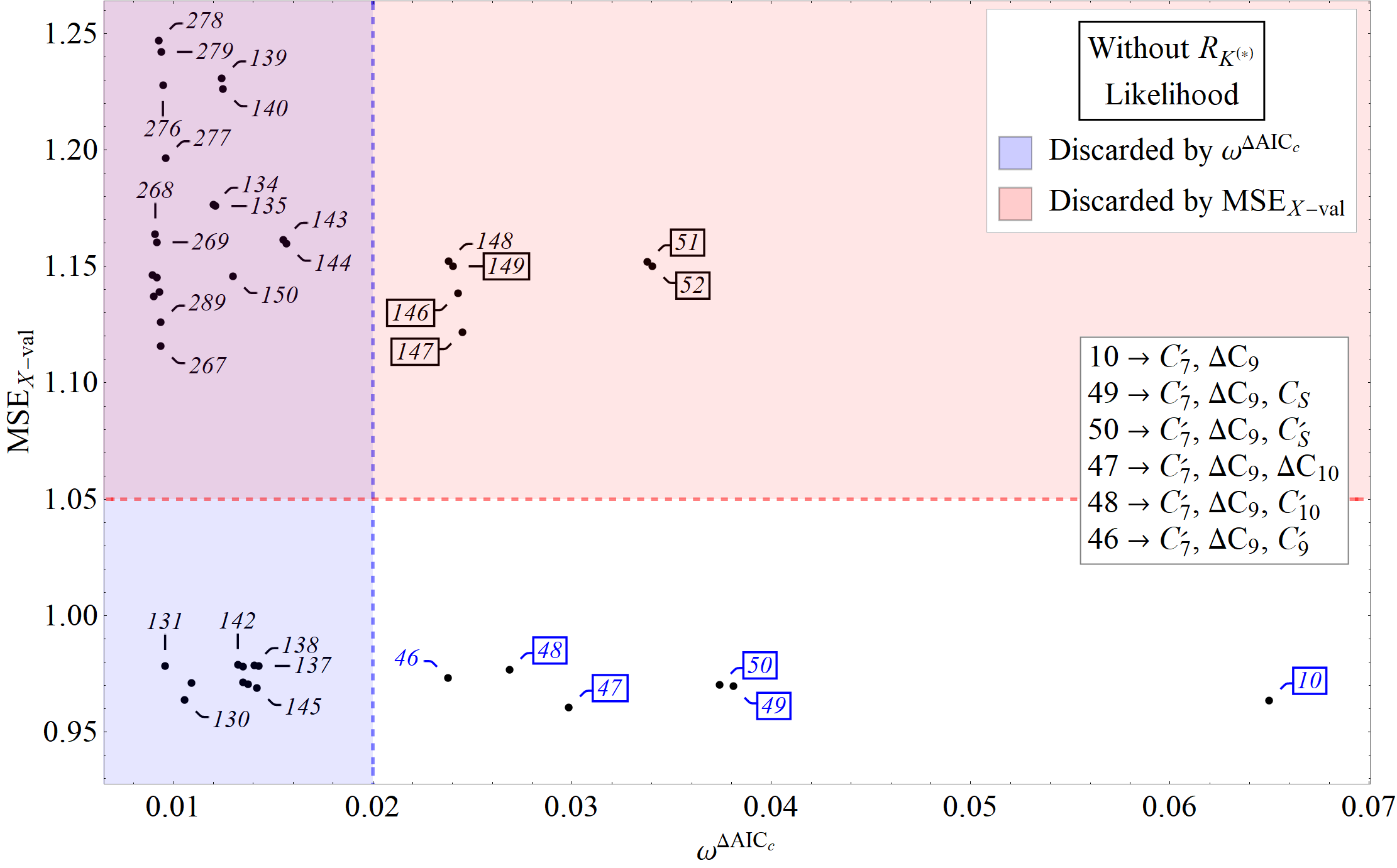}
		\label{fig:ModselNewLikewoLFUV}}
\end{figure*}

\section{Results for model-selection with datasets other than "New-moments" and "New-likelihood"}\label{app2}

\begin{figure*}[htb]
 	\caption{\small Comparison of the $CP$-averaged angular observables in different bins which are obtained in experiment, SM and from our fit results after dropping $R_{K^{(*)}}$.}
 	\label{fig:angobswolfuv}
 	\centering
 	\subfloat[]{\includegraphics[width=0.2\textwidth]{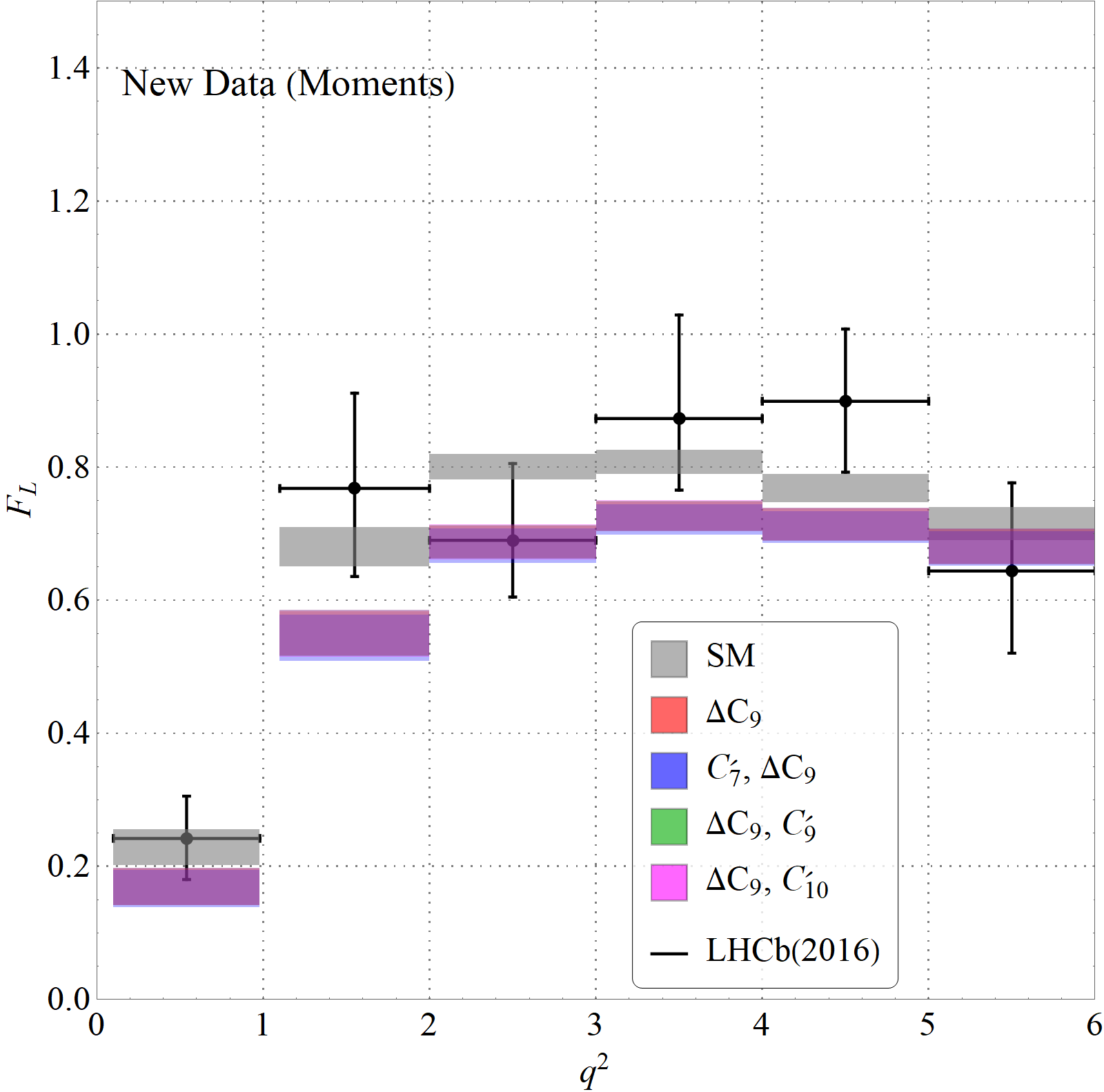}
 		\label{fig:angobsmomFLwolfuv}}
 	\subfloat[]{\includegraphics[width=0.2\textwidth]{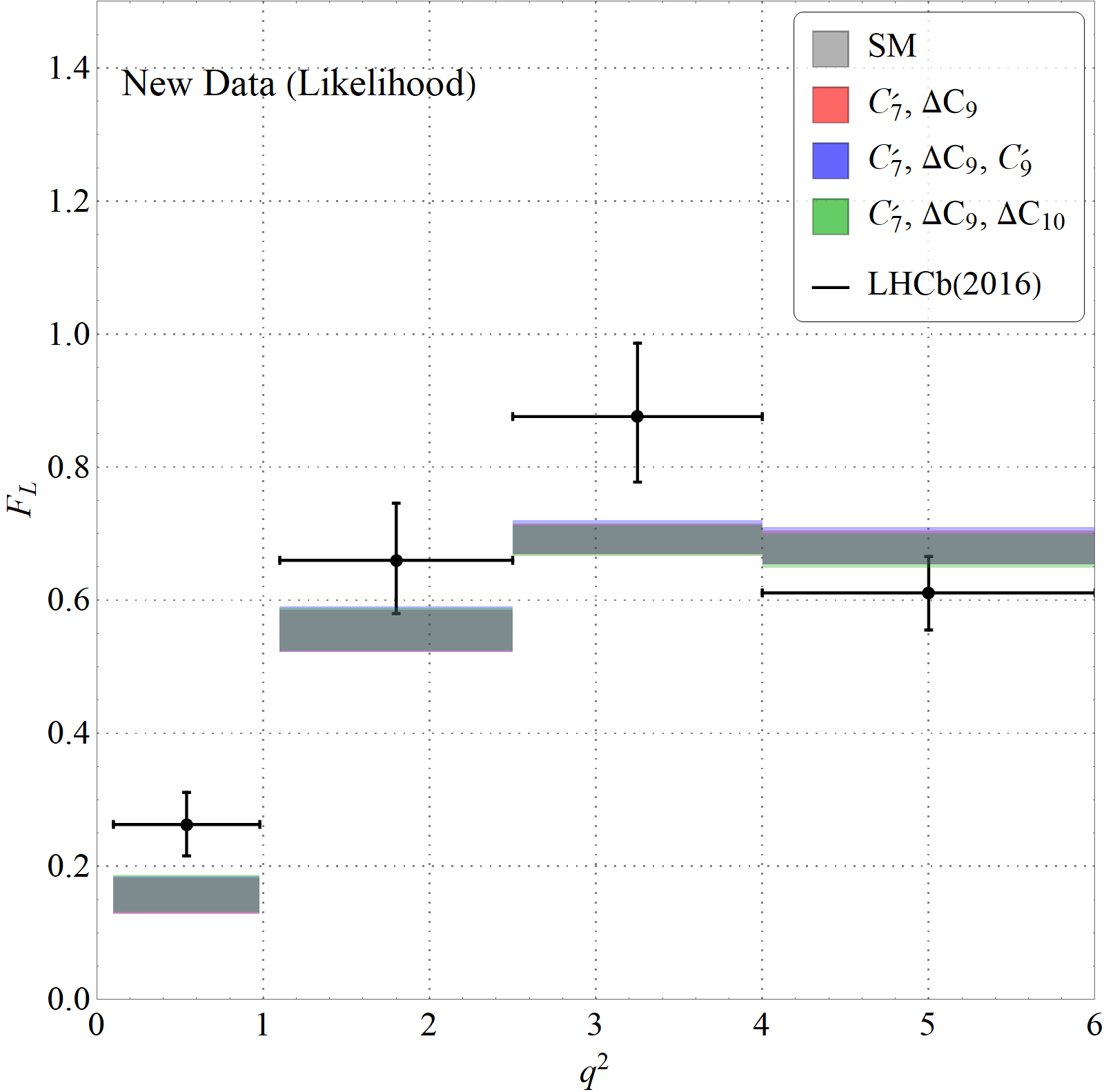}
 		\label{fig:angobslikeFLwolfuv}}
 	\subfloat[]{\includegraphics[width=0.2\textwidth]{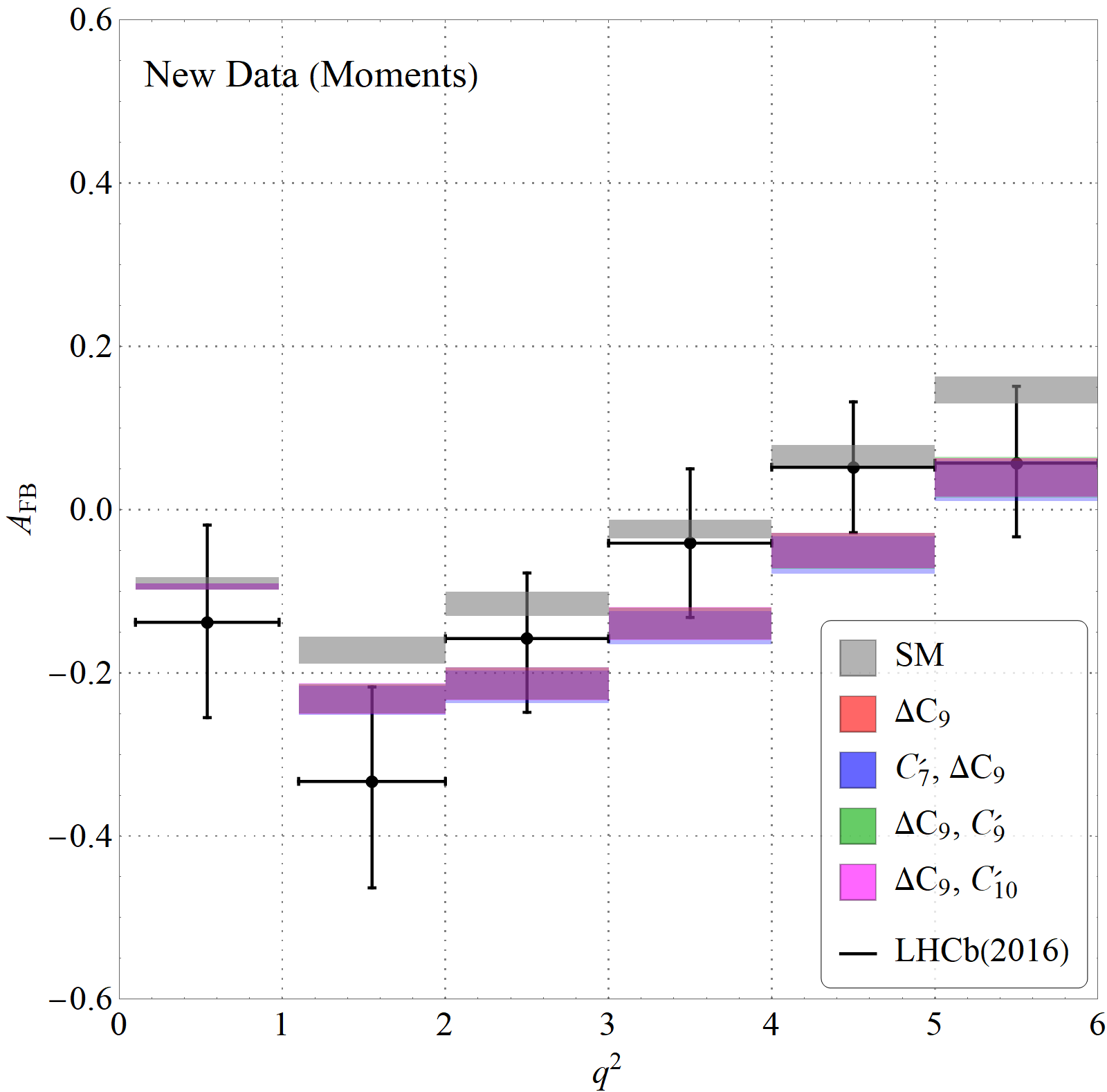}
 		\label{fig:angobsmomAFBwolfuv}}
 	\subfloat[]{\includegraphics[width=0.2\textwidth]{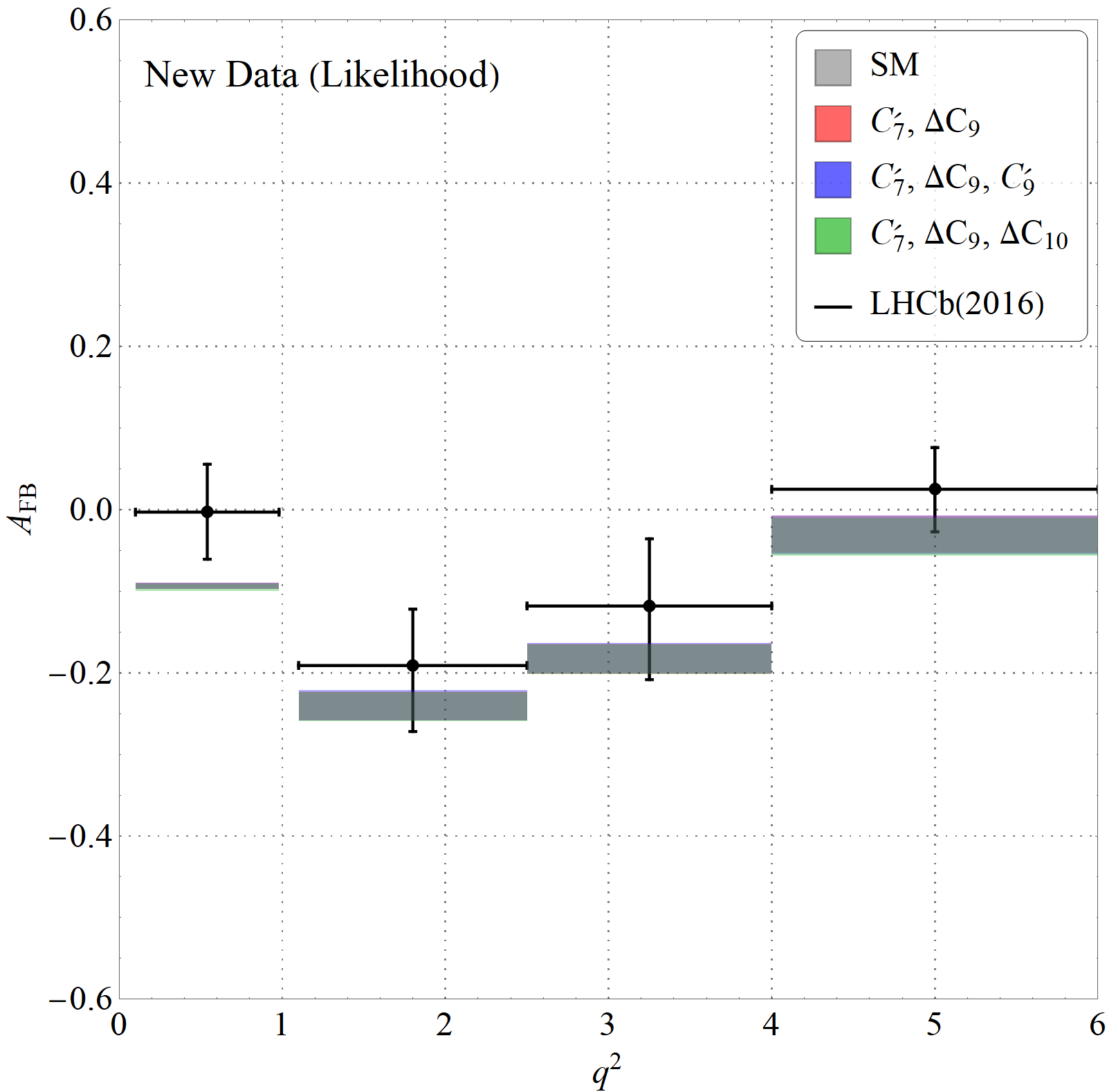}
 		\label{fig:angobslikeAFBwolfuv}}
 	\subfloat[]{\includegraphics[width=0.2\textwidth]{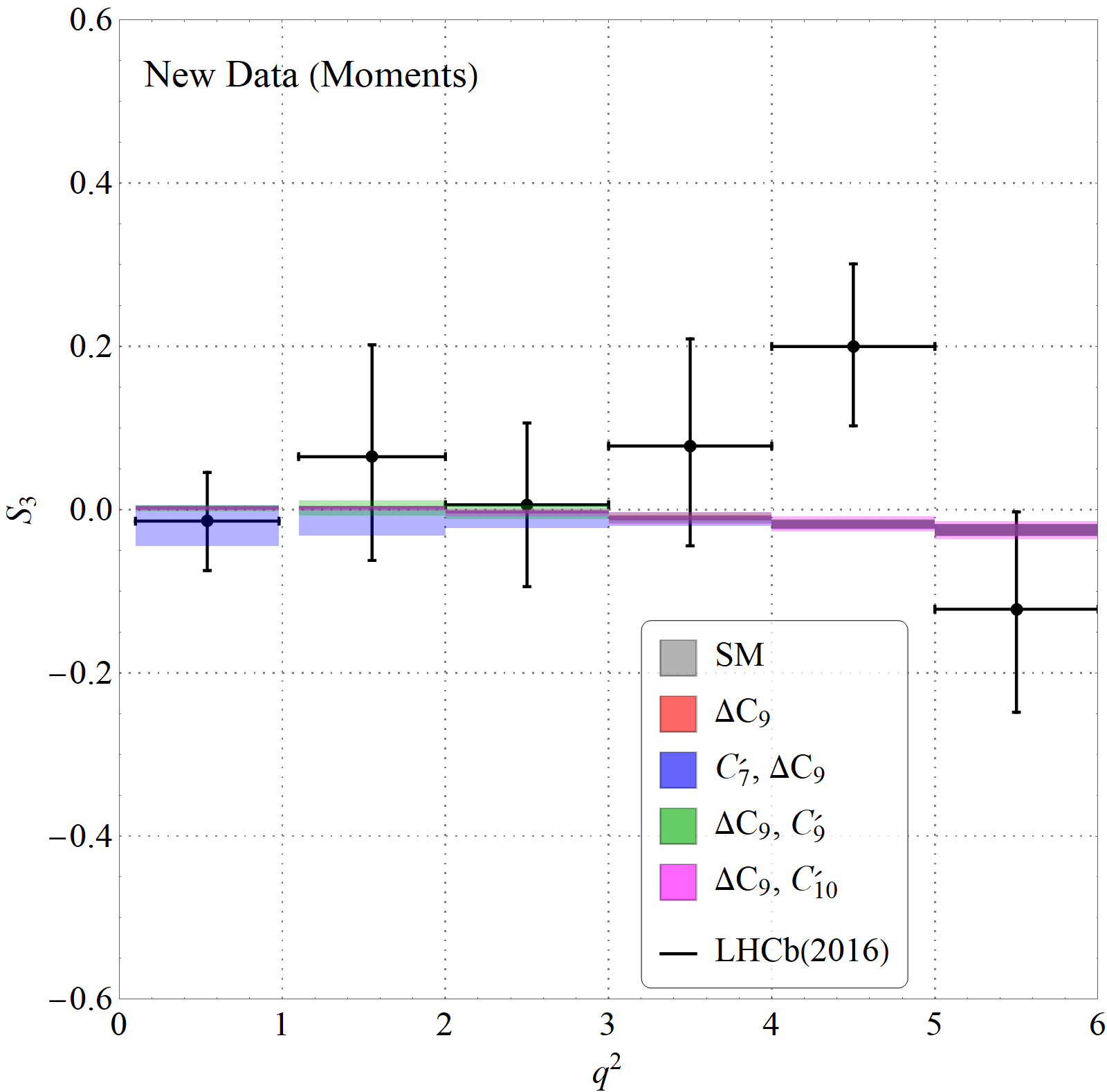}
 		\label{fig:angobsmomS3wolfuv}}\\
 	\subfloat[]{\includegraphics[width=0.2\textwidth]{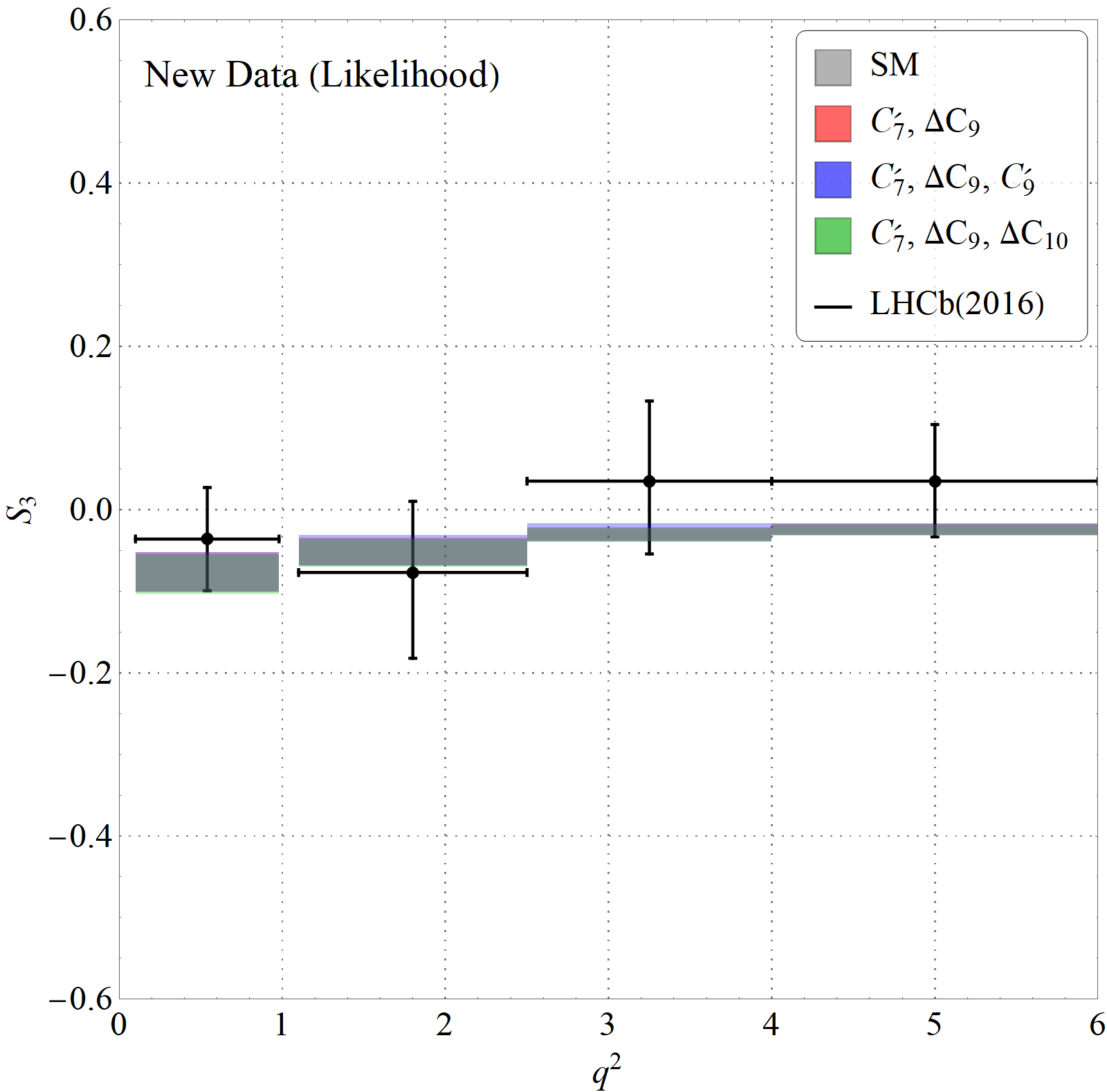}
 		\label{fig:angobslikeS3wolfuv}}
 	\subfloat[]{\includegraphics[width=0.2\textwidth]{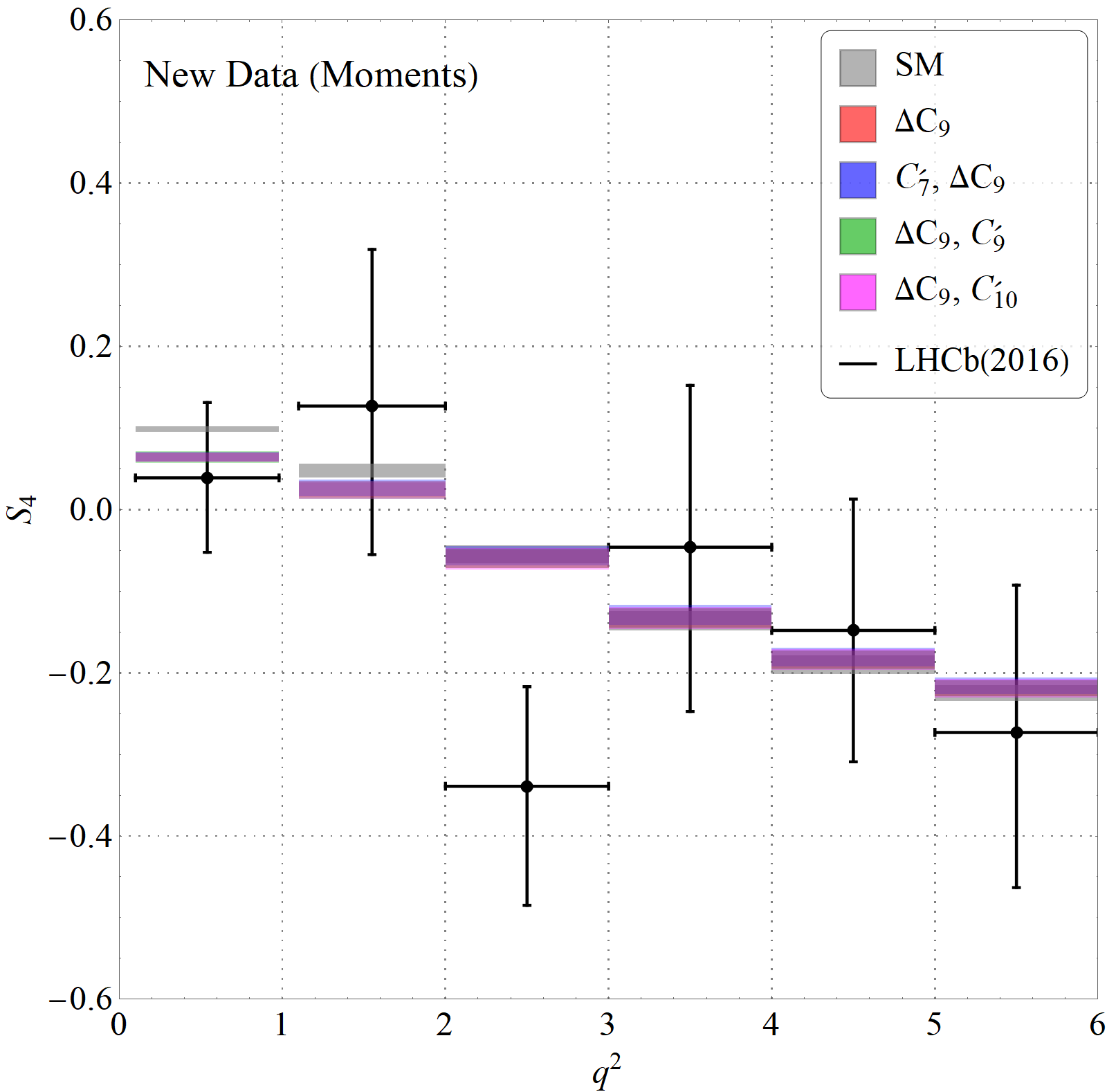}
 		\label{fig:angobsmomS4wolfuv}}
 	\subfloat[]{\includegraphics[width=0.2\textwidth]{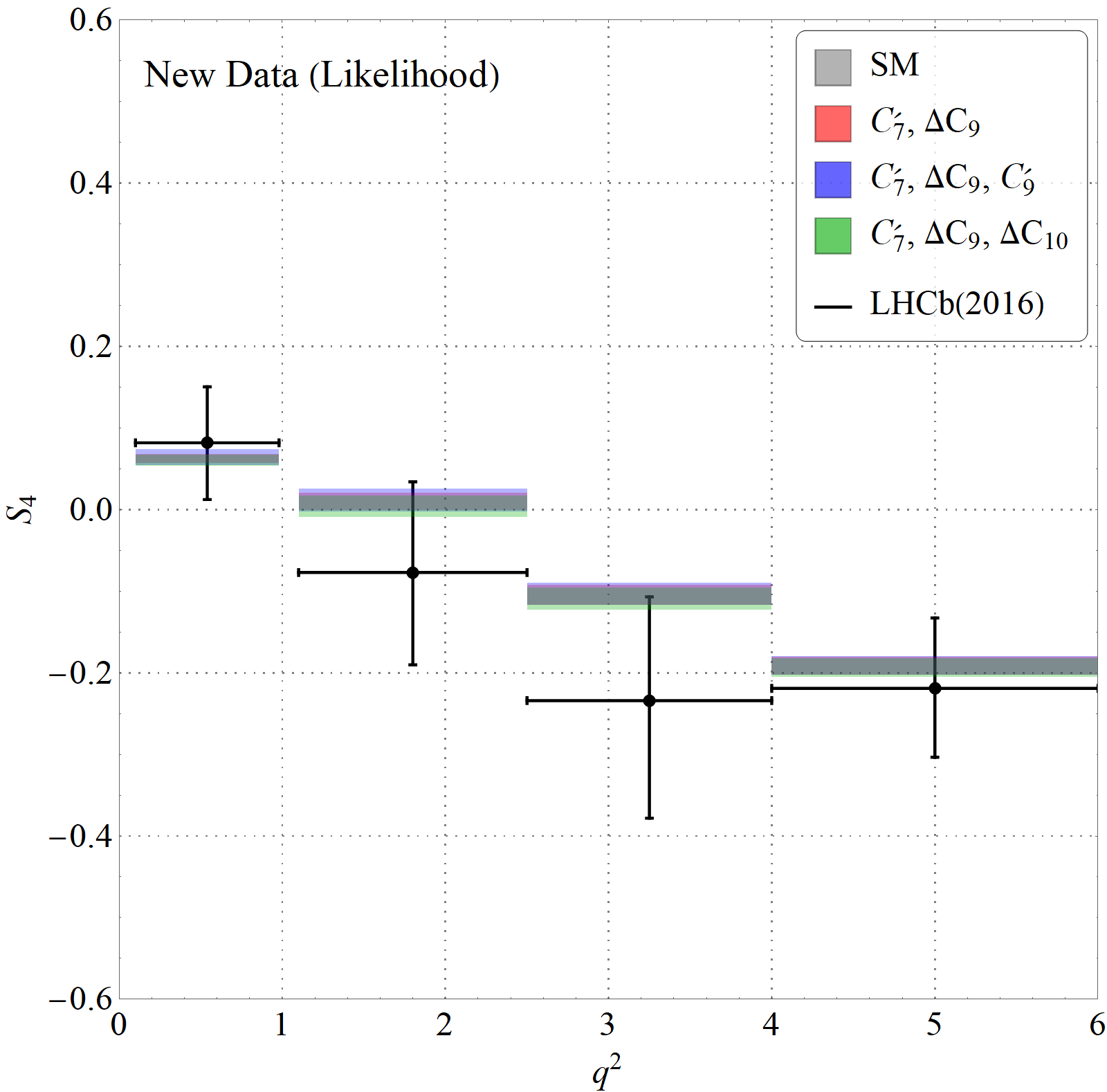}
 		\label{fig:angobslikeS4wolfuv}}
 	\subfloat[]{\includegraphics[width=0.2\textwidth]{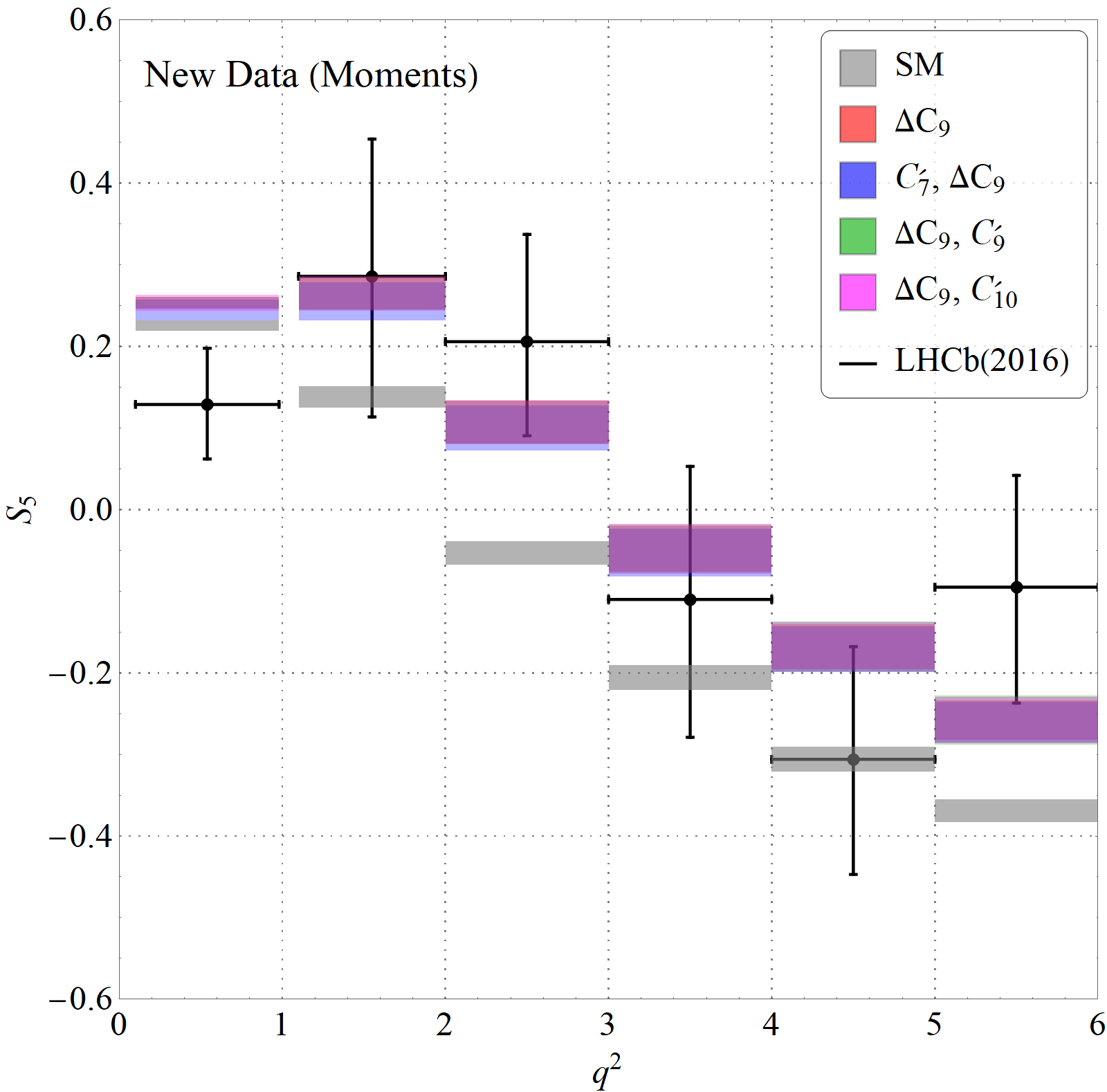}
 		\label{fig:angobsmomS5wolfuv}}
 	\subfloat[]{\includegraphics[width=0.2\textwidth]{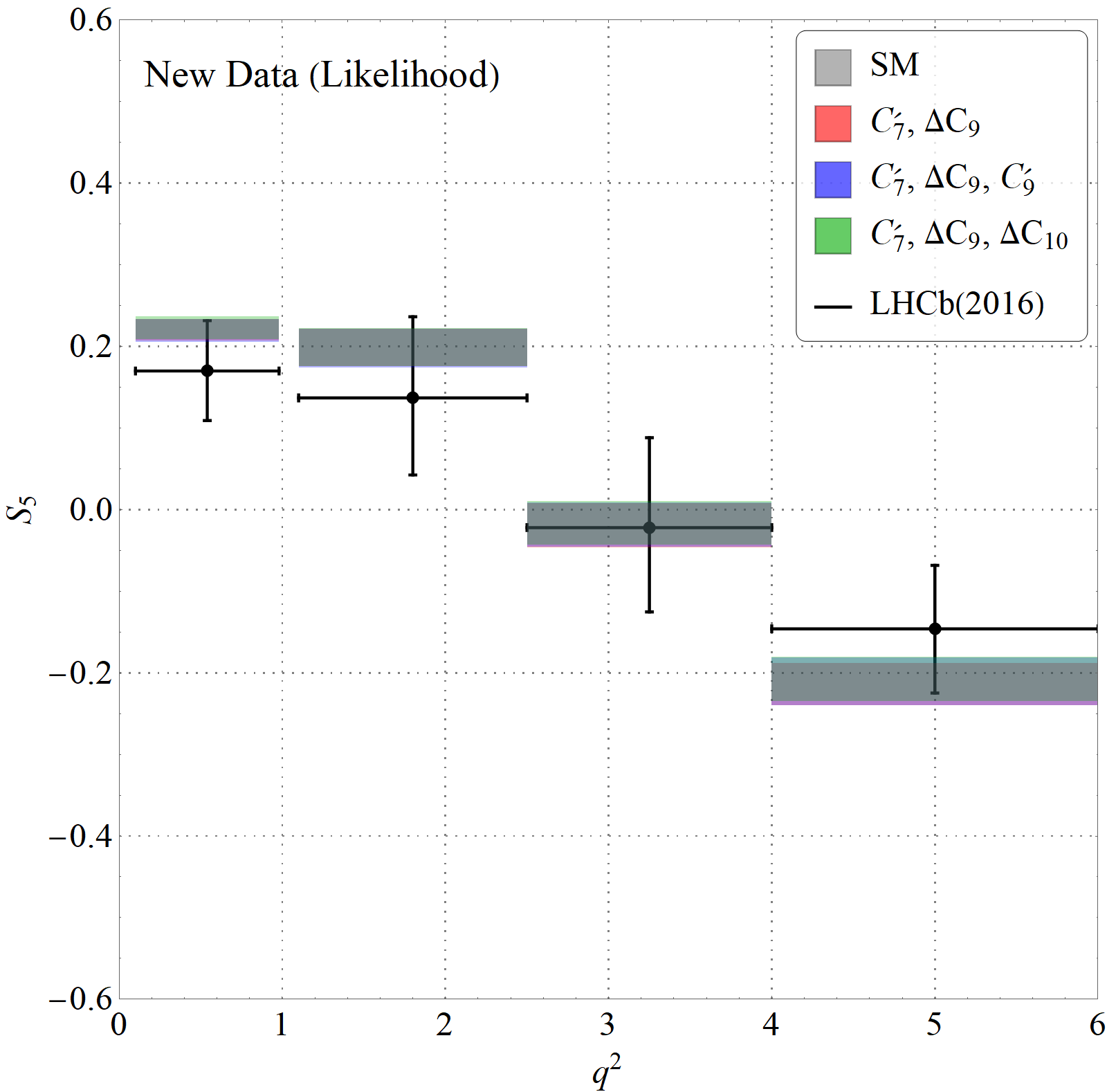}
 		\label{fig:angobslikeS5wolfuv}}\\
 	\subfloat[]{\includegraphics[width=0.2\textwidth]{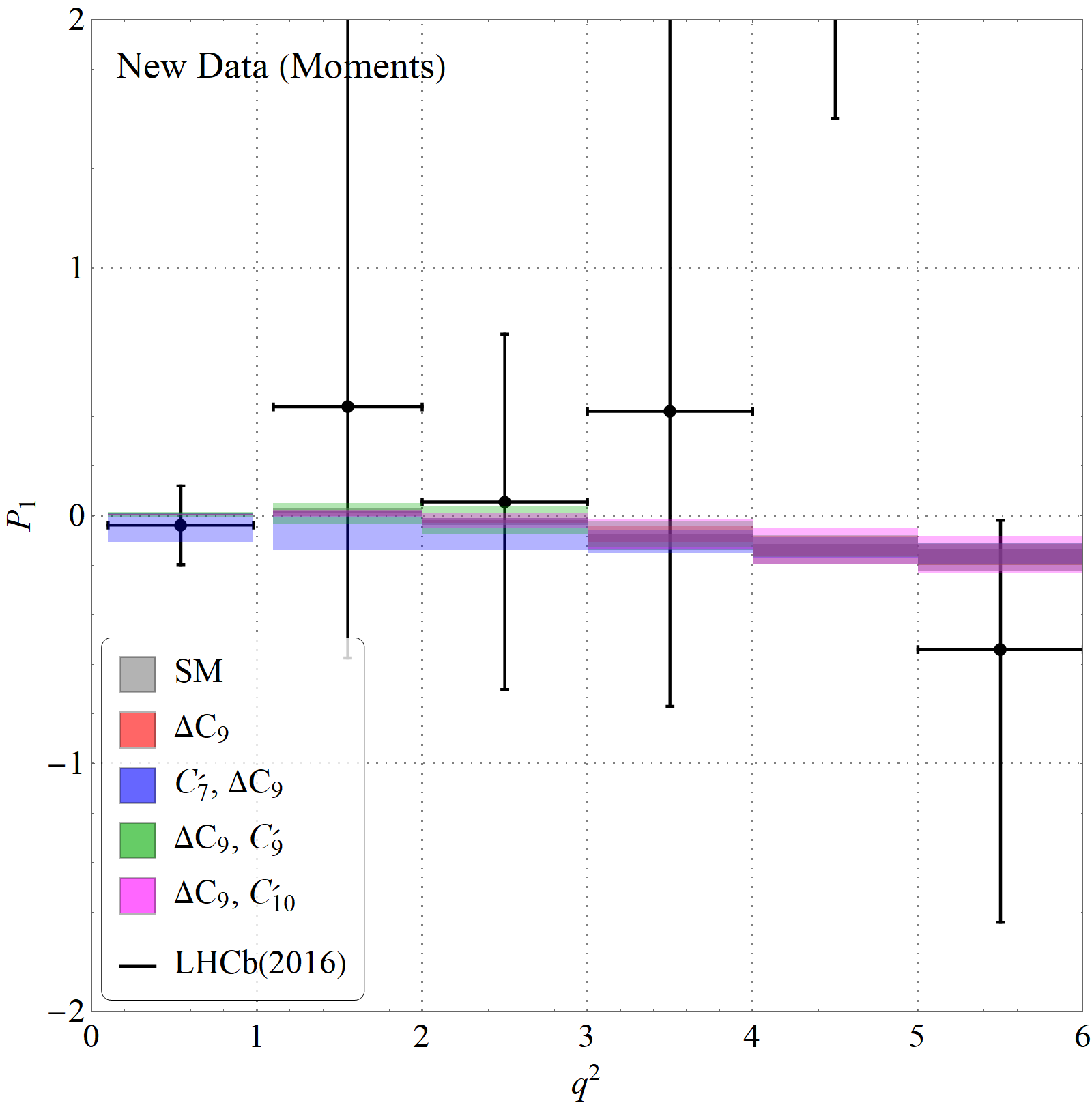}
 		\label{fig:angobsmomP1wolfuv}}
 	\subfloat[]{\includegraphics[width=0.2\textwidth]{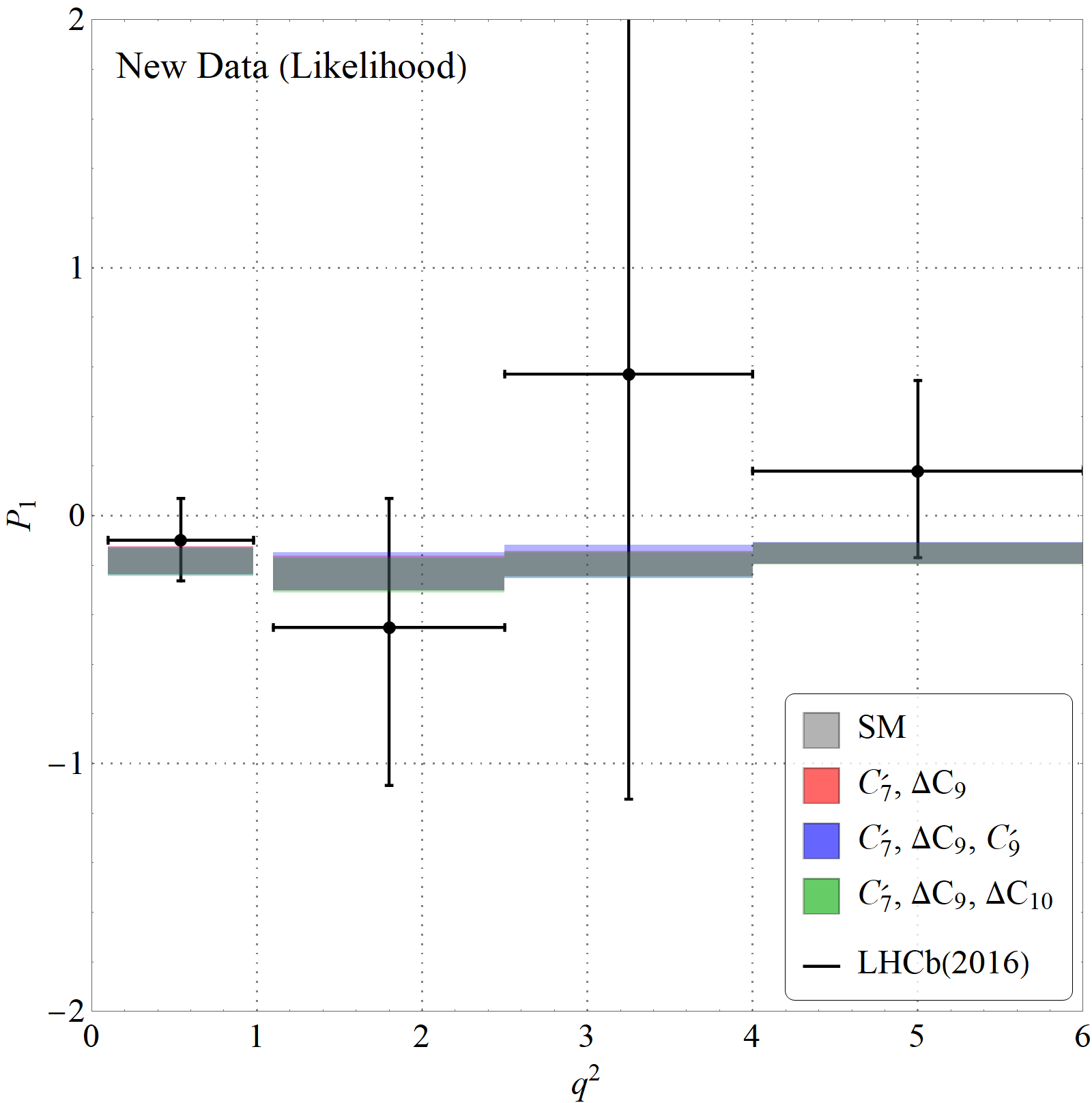}
 		\label{fig:angobslikeP1wolfuv}}
 	\subfloat[]{\includegraphics[width=0.2\textwidth]{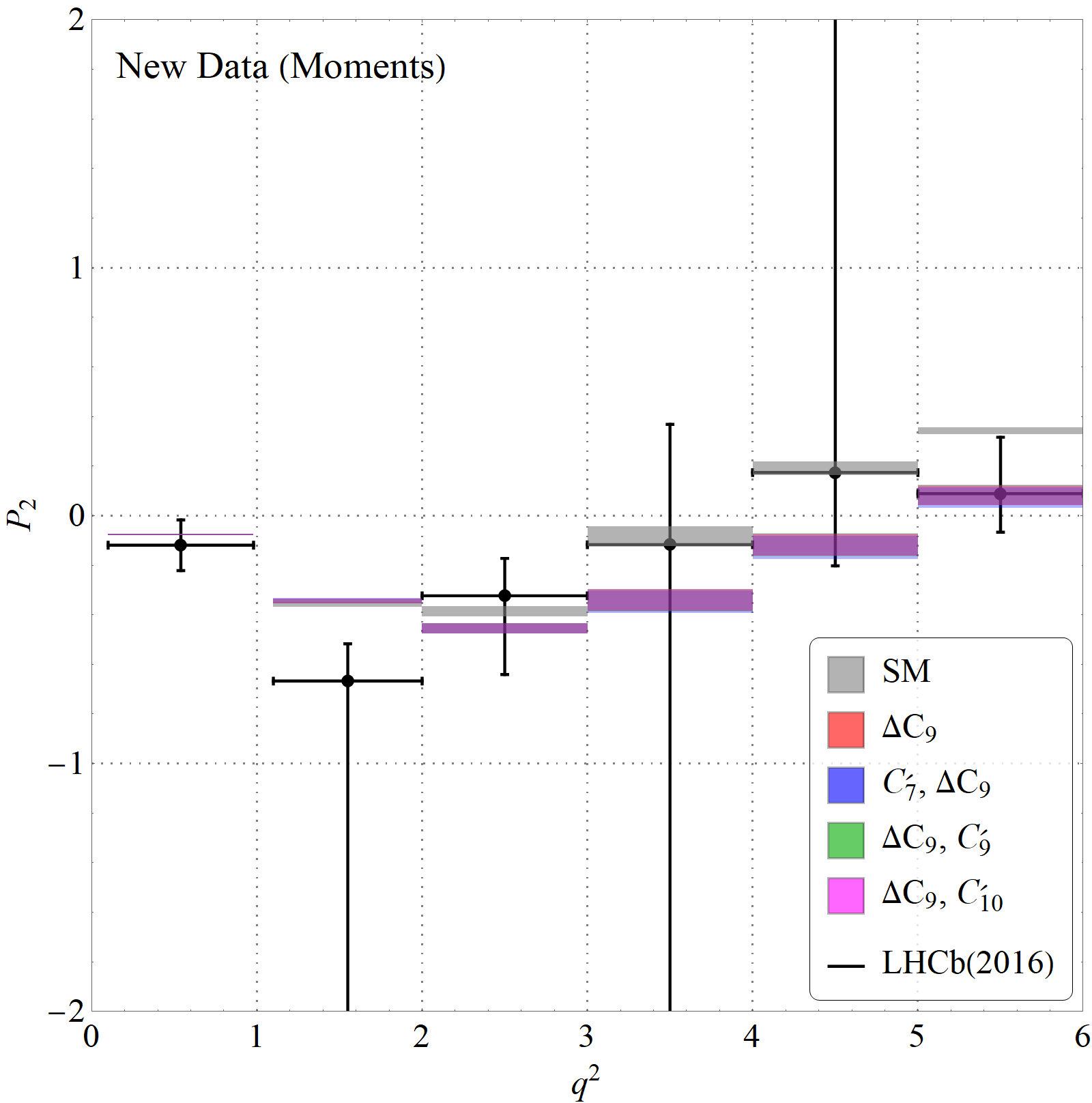}
 		\label{fig:angobsmomP2wolfuv}}
 	\subfloat[]{\includegraphics[width=0.2\textwidth]{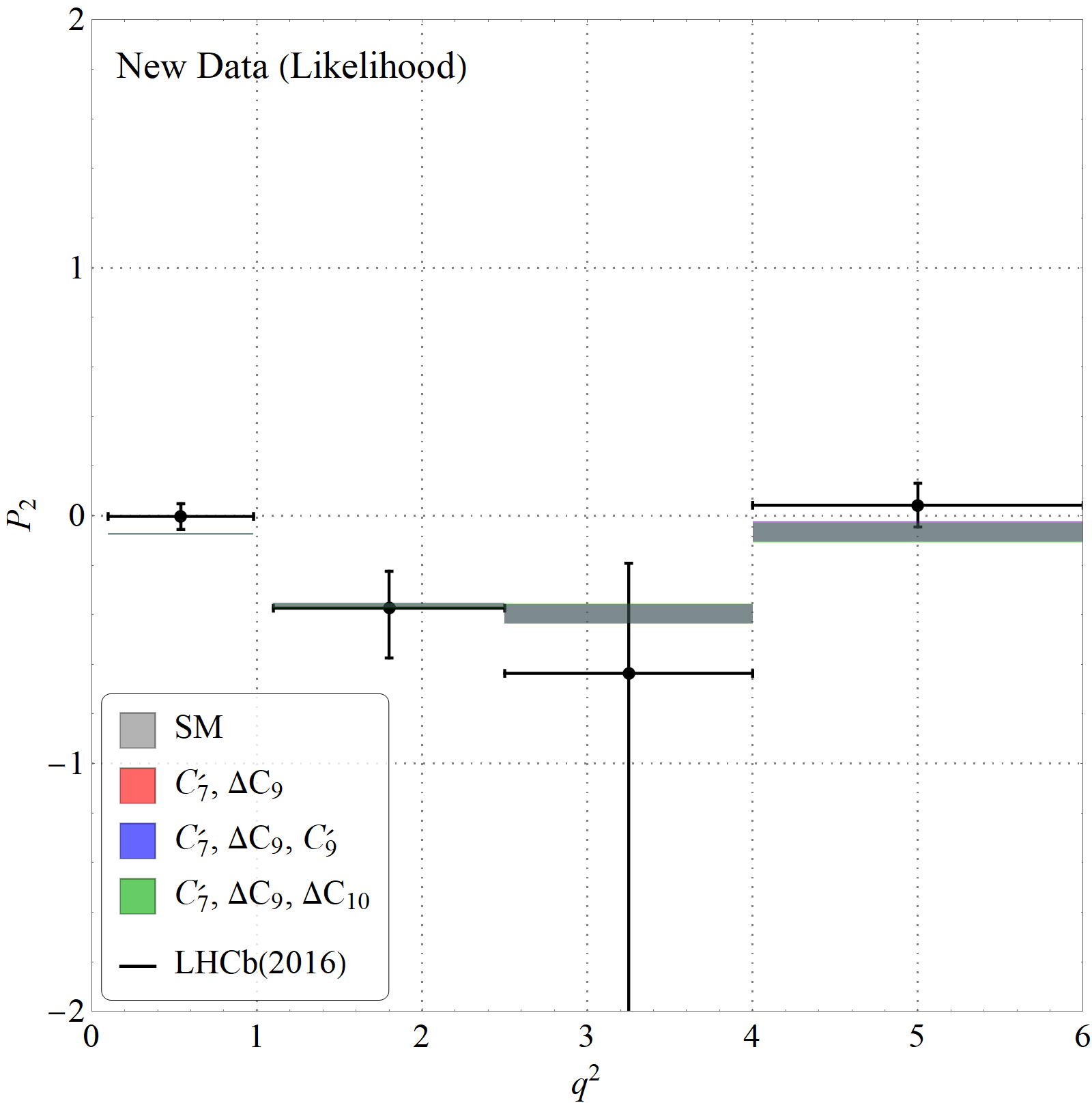}
 		\label{fig:angobslikeP2wolfuv}}
 	\subfloat[]{\includegraphics[width=0.2\textwidth]{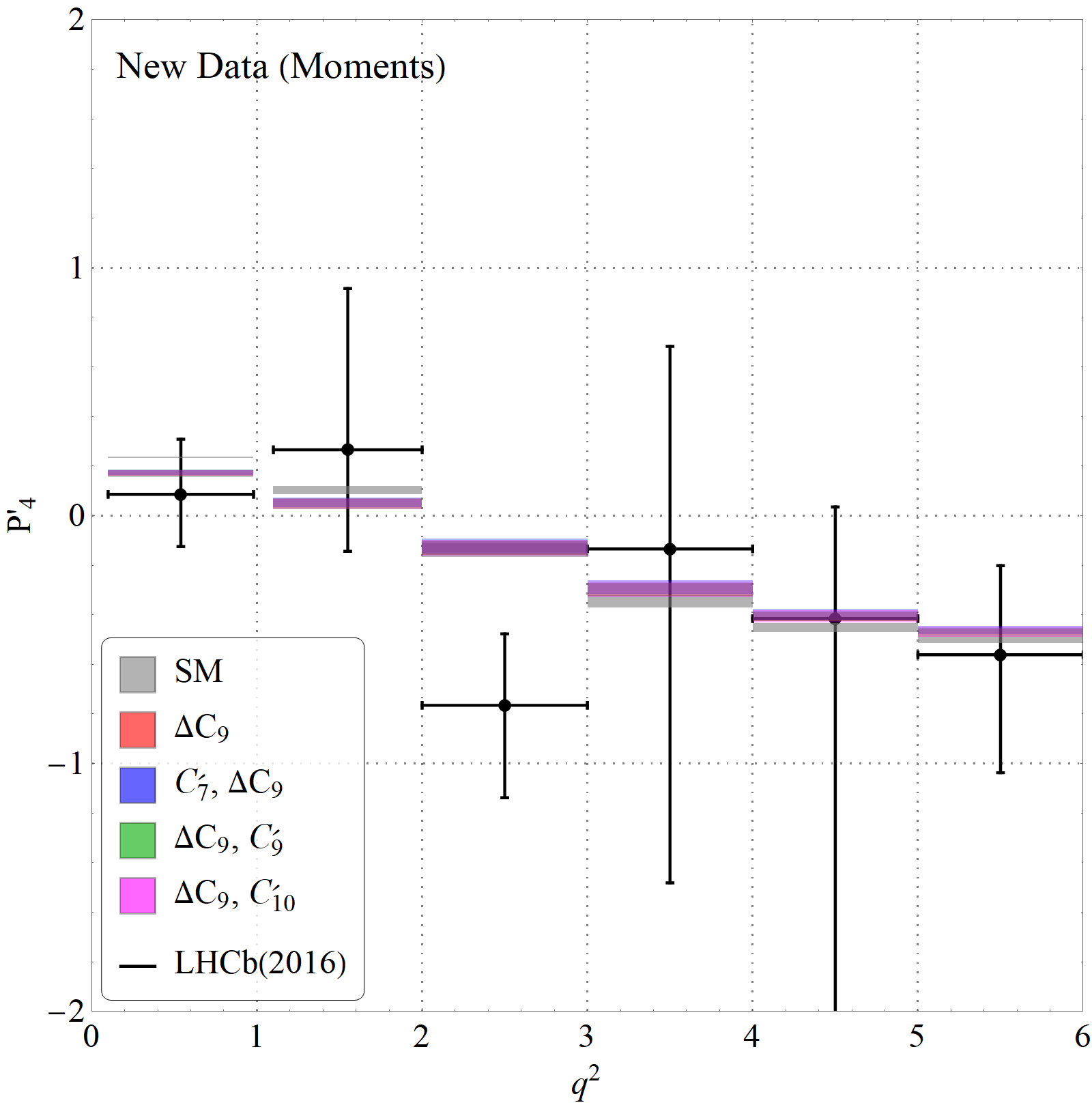}
 		\label{fig:angobsmomP4prwolfuv}}\\
 	\subfloat[]{\includegraphics[width=0.2\textwidth]{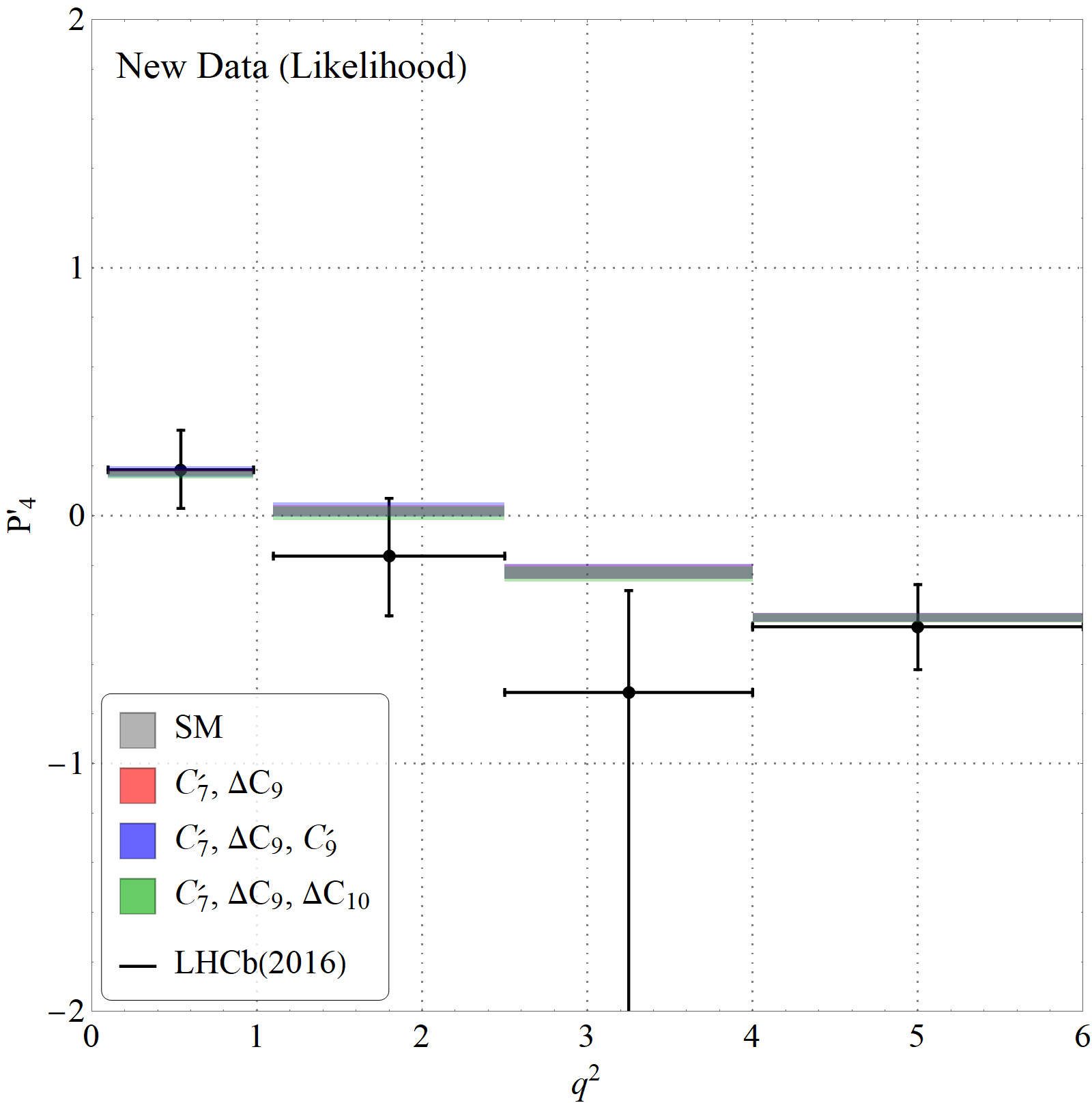}
 		\label{fig:angobslikeP4prwolfuv}}
 	\subfloat[]{\includegraphics[width=0.2\textwidth]{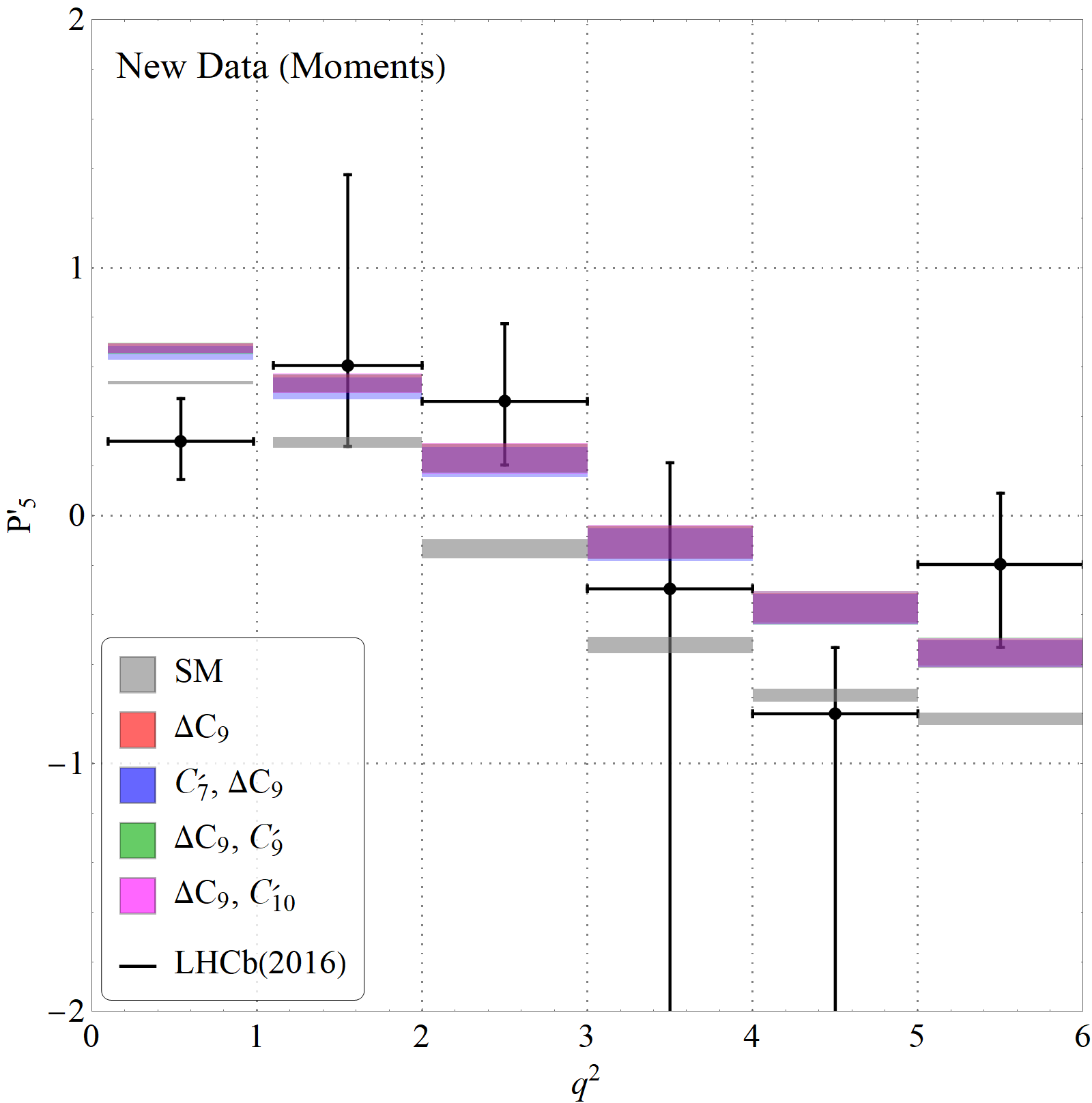}
 		\label{fig:angobsmomP5prwolfuv}}
 	\subfloat[]{\includegraphics[width=0.2\textwidth]{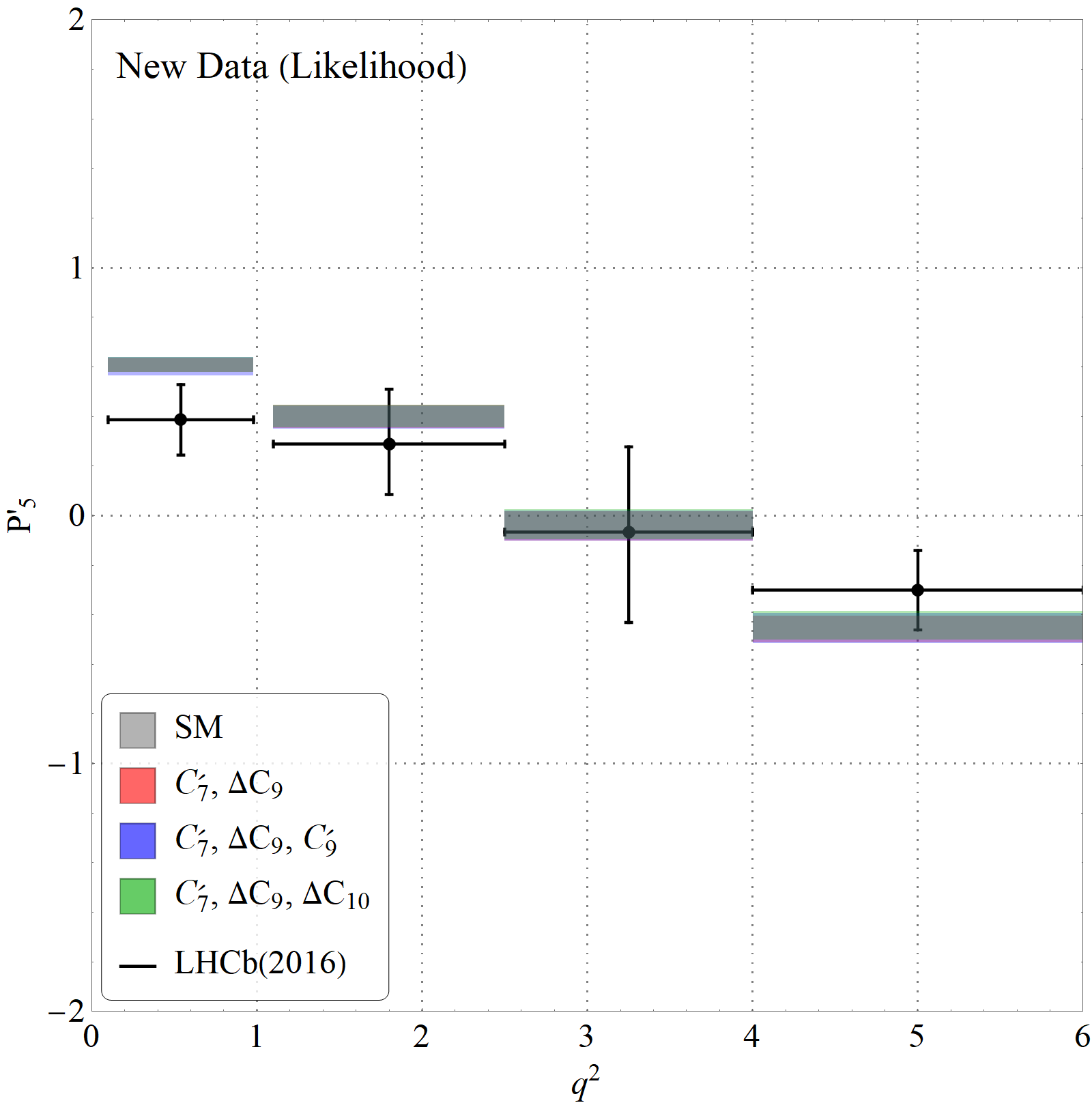}
 		\label{fig:angobslikeP5prwolfuv}}
 \end{figure*}

 A comparison of the selected models in figures \ref{fig:ModselNew} and \ref{fig:ModselNewwoLFUV} is helpful in understanding the impact of the new $R_{K^{(*)}}$ measurements by LHCb and Belle on the process of model selection. With more precise measurements on $R_K$ and $R_{K^*}$, the consistency between the data and the corresponding SM measurements may increase further. Therefore, it is important to gain insights on the probable NP effects in the angular observables. In figure \ref{fig:ModselNewwoLFUV}, we provide the results for the model selection corresponding to the "New-moments" and the "New-likelihood" datasets after dropping the LFUV observables $R_K$ and $R_{K^*}$. The selected scenarios can be compared with the one given in figure \ref{fig:ModselNew}. We note that in both the likelihood and moments data under the given selection setup, the number of selected scenarios reduce after we drop $R(K)$ and $R(K^*)$ from the inputs. A comparison of between figures \ref{fig:ModselNewMomwoLFUV} and \ref{fig:ModselNewMom} indicates that the three-operator scenarios become less favourable. Similarly, from figures \ref{fig:ModselNewLikewoLFUV} and \ref{fig:ModselNewLike} we see that the four and five-operators scenarios are less favourable in the analysis without $R(K)$ and $R(K^*)$. As explained in the main text, the explanation of new data prefers NP scenarios with more than two operators like three, four or five-operator scenarios. In particular, we have noted that the fit qualities improve once we drop these LFUV observables from the fits in general, which is at par with our expectations.  For comparison with the `New' data-set, figure \ref{fig:angobswolfuv} lists the angular observables from SM, experiment, and our fit results.          

\begin{figure*}[t]
	\caption{\small  Allowed parameter space for $\Delta C_9$ in one-operator scenario and allowed NP parameter spaces and their respective correlation for some selected two-operator scenarios.}
	\label{fig:paramspace1D2D}
	\centering
	\subfloat[]{\includegraphics[width=0.49\textwidth]{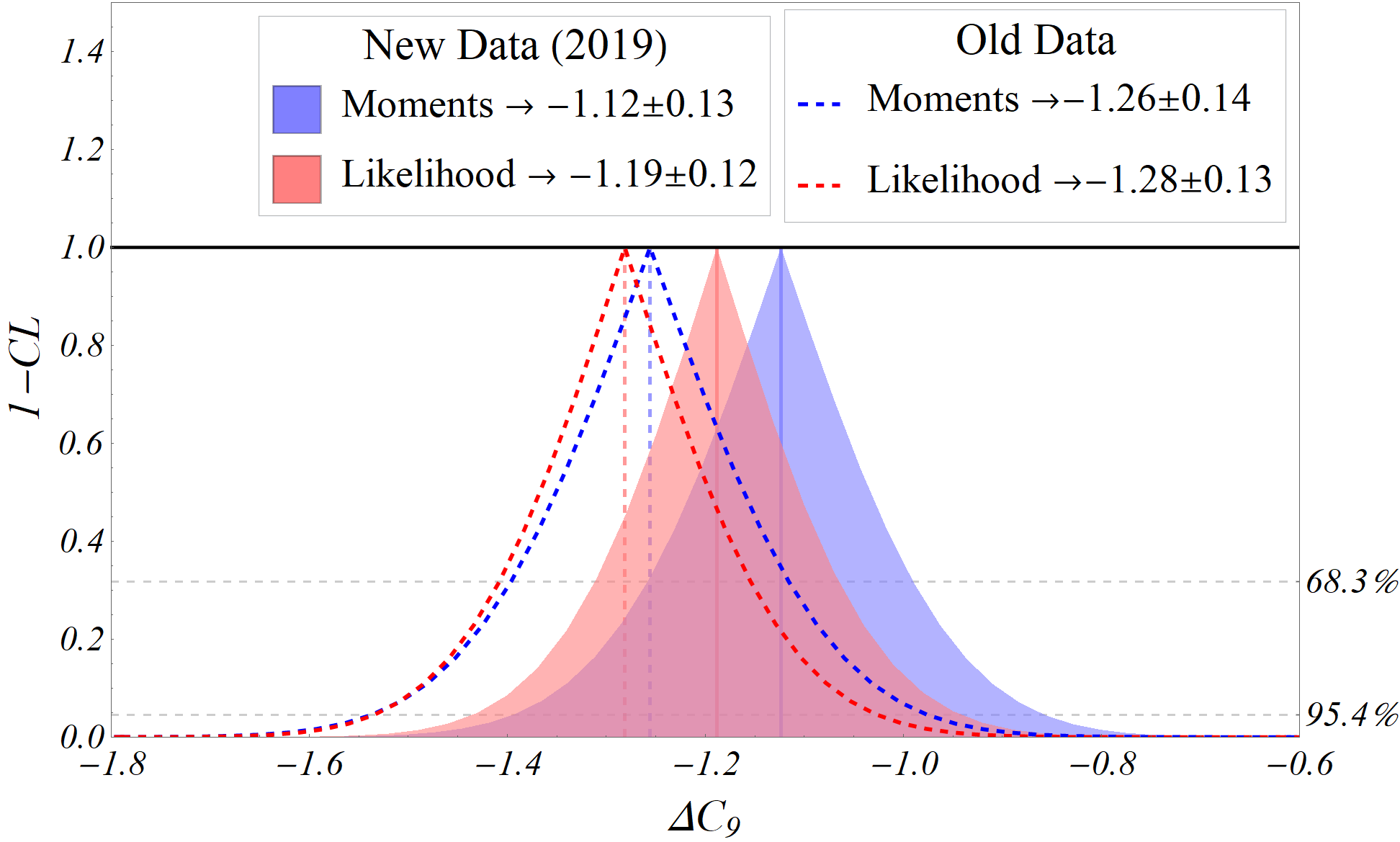}\label{fig:dc91DPlot}}\\
	\subfloat[]{\includegraphics[width=0.32\textwidth]{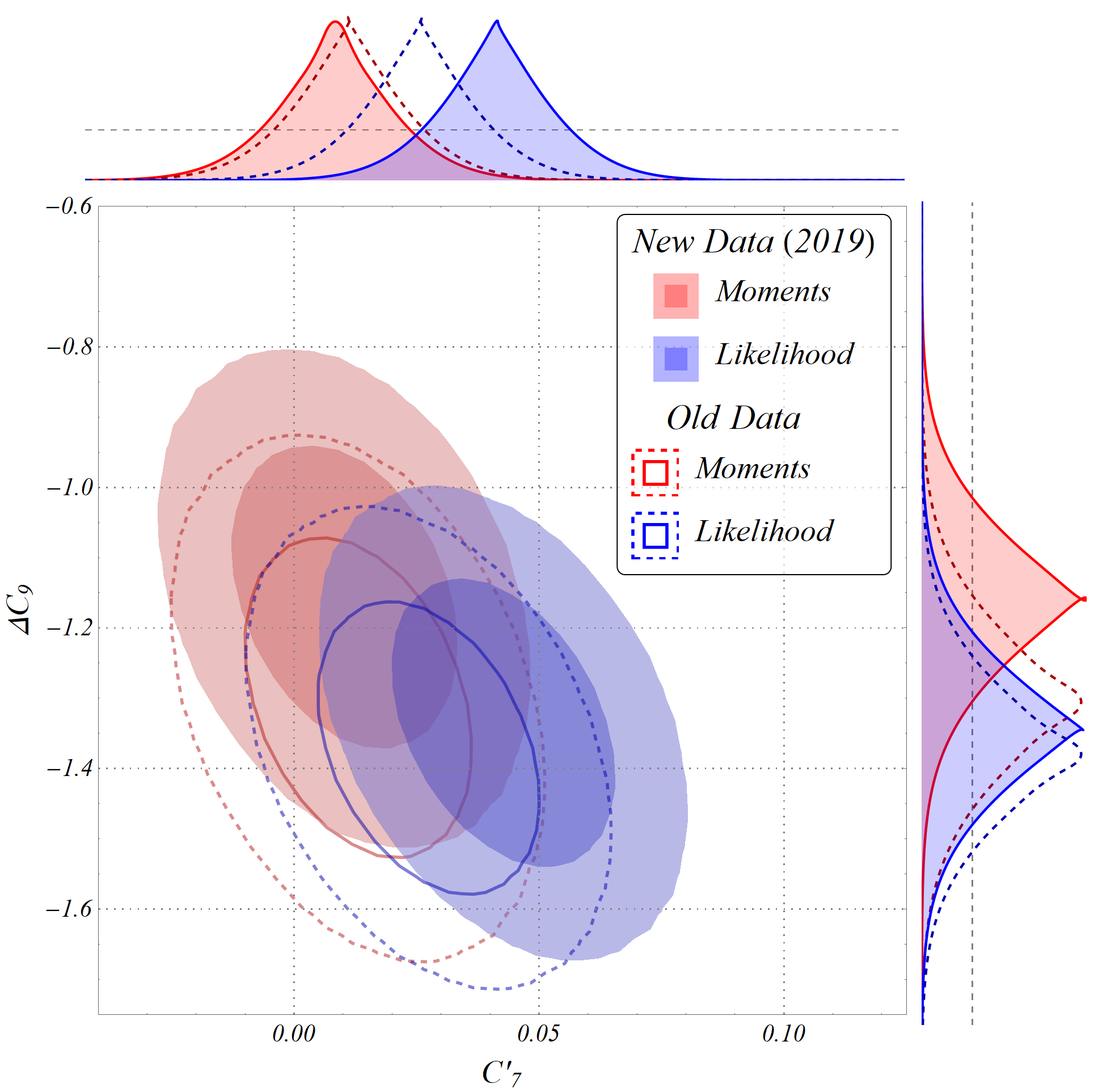}\label{fig:c7prdc9plt}}~~
	\subfloat[]{\includegraphics[width=0.32\textwidth]{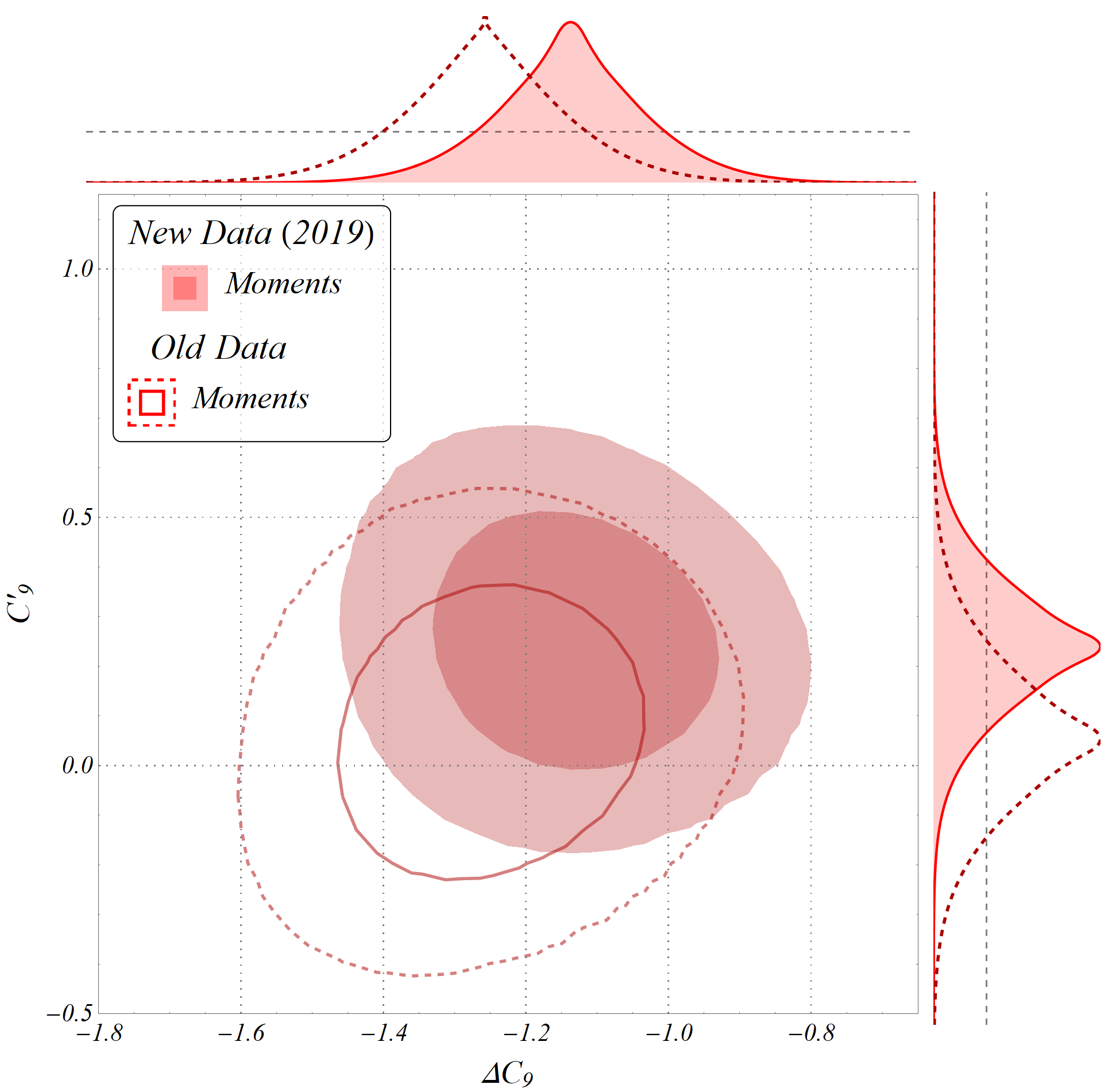}\label{fig:c9prdc9plt}}~~
	\subfloat[]{\includegraphics[width=0.32\textwidth]{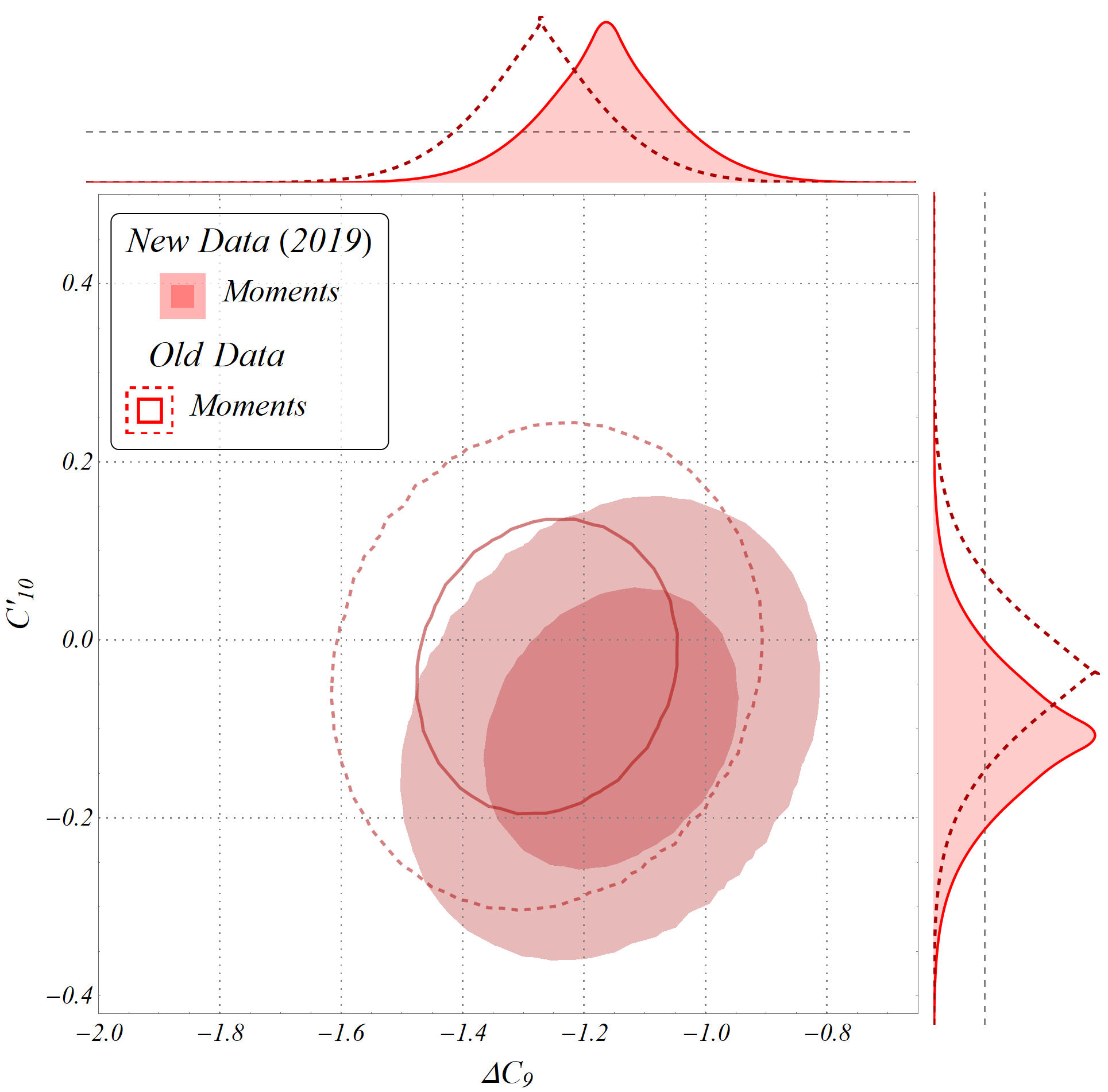}\label{fig:c10prdc9plt}}
\end{figure*}

\begin{figure*}
	\caption{\small Zero crossing values in the $q^2$ distributions of the angular observables shown in fig. \ref{fig:q2dist}. $q_0^2$ represents the value of $q^2$ at the zero crossing.}
	\label{fig:zerocrossing}
	\centering
	\subfloat[]{\includegraphics[width=0.45\textwidth]{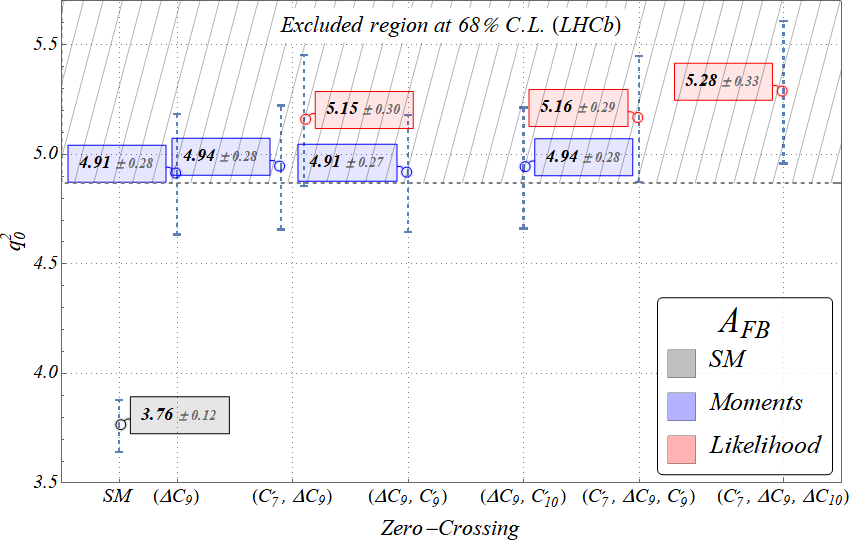}\label{fig:AFBZCPlot}}~~
	\subfloat[]{\includegraphics[width=0.45\textwidth]{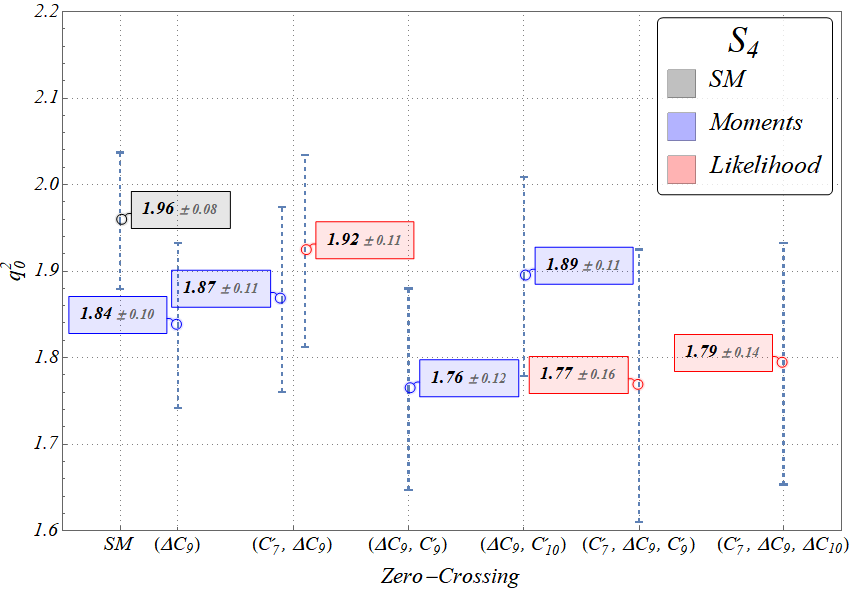}\label{fig:S4ZCPlot}}\\
	\subfloat[]{\includegraphics[width=0.45\textwidth]{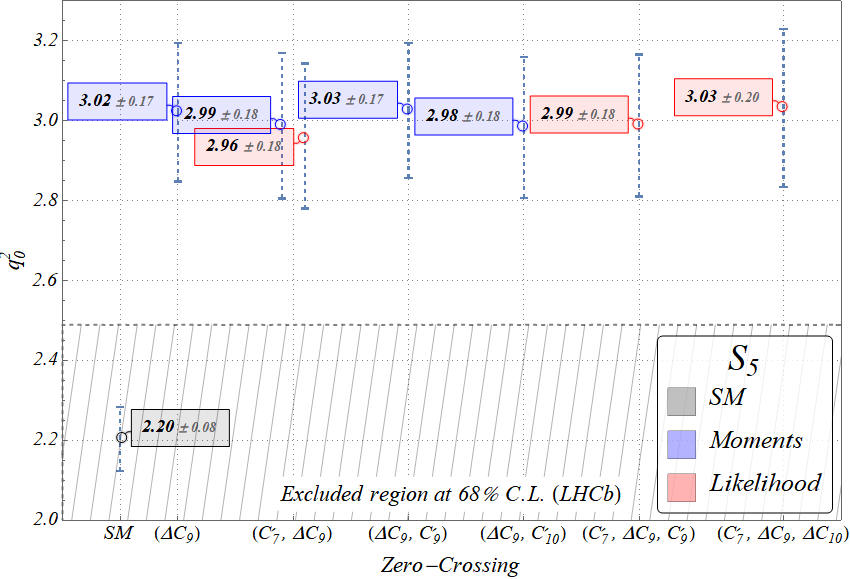}\label{fig:S5ZCPlot}}
\end{figure*}

Figure \ref{fig:paramspace1D2D} depicts the allowed parameter spaces of the most commonly occurring one and two parameter scenarios selected from different types of fits and data-sets. For one operator scenario (${\cal O}_9$), the allowed parameter space of the corresponding WC $\Delta C_9$, is shown in figure \ref{fig:dc91DPlot}. For the two operator scenarios, figures \ref{fig:c7prdc9plt}, \ref{fig:c9prdc9plt} and \ref{fig:c10prdc9plt} shows the correlations between the WCs. We note that the allowed values of $C'_7$ and $\Delta C_9$ have reasonably small ranges and they are negatively correlated. The corresponding value is $-0.316$. Large (negative) values of $\Delta C_9$ prefers large positive values of $C'_7$. The other two plots show the allowed parameter spaces and the correlations of $\Delta C_9$ with $C'_9$ with $C'_{10}$, respectively. The fitted values of $C'_9$ and $C'_{10}$ have large errors. Guessing the exact correlations between them from the figures alone is therefore difficult. However, one can see that in the analysis with new data, the value of the correlation between $\Delta C_9$ and $C'_{10}$ ( + 0.24) is greater than that between $\Delta C_9$ and $C'_9$ which is $- 0.11$. 

In figure \ref{fig:zerocrossing} we compare the values of $q^2$ at the zero crossing ($q_0^2$) between SM, our selected models and the measured values for the above mentioned observables. We note that in our selected models, the $q_0^2$ for all these three observables are in good agreement with the corresponding measured values.

\newpage
\bibliography{ref_SSSA}

\end{document}